\documentclass[aip, reprint, graphicx]{revtex4-1}

\usepackage{amsmath,amssymb,amsfonts,amsthm,bm}
\usepackage{mathtools}
\usepackage{tikz-cd} 							
\usepackage{dsfont} 	    
\usepackage{algorithm, algorithmicx, algpseudocode}
\usepackage{bbm} 
\usepackage{hyperref} 
\usepackage{subcaption}
\usepackage{upgreek}

\newtheorem{definition}{Definition}
\newtheorem{theorem}{Theorem}
\newtheorem{lemma}{Lemma}
\newtheorem{prop}{Proposition}

\usepackage[utf8]{inputenc}
\usepackage[T1]{fontenc}

\hypersetup{%
  colorlinks=true,
  linkcolor=blue,
  linkbordercolor=red,
  citecolor=red,
}

\newcommand{\bmsf}[1]{\boldsymbol{\mathsf{#1}}}        
\newcommand{\msf}[1]{\mathsf{#1}}      
\algdef{SE}[DOWHILE]{Do}{doWhile}{\algorithmicdo}[1]{\algorithmicwhile\ #1}%

\draft 

\begin{document}


\title{A self-consistent Hamiltonian model of the ponderomotive force and its structure preserving discretization} 

\newcommand{\Oden}{Oden Institute for Computational Engineering and Sciences, The University of Texas at Austin, Austin, Texas 78712, USA}
\newcommand{\IFS}{Institute for Fusion Studies, The University of Texas at Austin, Austin, Texas 78712, USA}
\newcommand{\NMPP}{Max Planck Institute for Plasma Physics, Division of Numerical Methods in Plasma Physics, Garching, Bavaria 85748, Germany}
\newcommand{\Phys}{Department of Physics, The University of Texas at Austin, Austin, Texas 78712, USA}
\newcommand{\TUM}{Technical University of Munich, Department of Mathematics, Garching, Bavaria 85748, Germany}

\author{William Barham}
\email[]{william.barham@utexas.edu}
\affiliation{\Oden}
\affiliation{\IFS}

\author{Yaman Güçlü}
\affiliation{\NMPP}

\author{Philip J. Morrison}
\email[]{morrison@physics.utexas.edu}
\affiliation{\Phys}
\affiliation{\IFS}

\author{Eric Sonnendrücker}
\affiliation{\NMPP}
\affiliation{\TUM}

\date{\today}

\begin{abstract}
In the presence of an inhomogeneous oscillatory electric field, charged particles experience a net force, averaged over the oscillatory timescale, known as the ponderomotive force. We derive a one-dimensional Hamiltonian model which self-consistently couples the electromagnetic field to a plasma which experiences the ponderomotive force. We derive a family of structure preserving discretizations of the model of varying order in space and time using conforming and broken finite element exterior calculus spectral element methods. In all variants of our discretization framework, the method is found to conserve the Casimir invariants of the continuous model to machine precision and the energy to the order of the splitting method used. 
\end{abstract}


\maketitle 

\tableofcontents


\section{Introduction}


A well known laser-plasma interaction model gives rise to a nonlinear polarization of the electromagnetic medium. In the presence of an inhomogeneous oscillatory electric field, charged particles experience a net force, averaged over the oscillatory timescale, known as the ponderomotive force. Starting from the two-fluid equations, we derive a Hamiltonian model in which the ponderomotive force provides a current source for Maxwell's equations. The asymptotic assumptions under which this model is derived correspond to what is frequently called the ``linear regime'' in laser plasma interaction models \cite{641305}. Our model is possibly the simplest self-consistent model of the ponderomotive force, and closely resembles other fluid models of the ponderomotive force in the linear regime \cite{10.1063/1.860884}, although the model therein is stated in terms of the vector potential whereas our model is stated in terms of the electromagnetic fields. Our model is perhaps the first to elucidate the Hamiltonian structure of the 1D linear regime, and provides an alternative approach to its derivation based on applying the asymptotics directly to the energy functional and Poisson bracket rather than the equations of motion. This self-consistent model of the ponderomotive force leads to a nonlinear polarization of the electromagnetic medium; we adapted ideas from a modeling framework for kinetic theories in nonlinear media to build the model studied herein \cite{morrison_gauge_free_lifting}. We call our model the 1D ponderomotive Maxwell system. 


The simple Poisson bracket one finds for the 1D ponderomotive Maxwell system makes this model straightforward to discretize in a structure preserving manner: its Poisson bracket is a direct sum of the Poisson brackets for Maxwell and acoustic wave equations, as well as a coupling bracket. Each of these brackets is field-free, i.e.\ the Poisson bivector is independent of the dynamical fields. Hence, antisymmetry is a sufficient condition for the Jacobi identity to be satisfied \cite{pjm82}, and a structure preserving discretization is easily accomplished using finite element exterior calculus (FEEC) \cite{ArnoldDouglasN2010Feec}. In particular, as the model studied in this paper is a nonlinear wave equation, we study the model using a variety of FEEC which is built from broken elements (i.e. the method uses discontinuous shape functions which are local to each element) \cite{conga_pic_1, conga_pic_2, campos-pinto2016:CONGA, campos-pinto_and_guclu_2021, güçlü2022broken}. These are referred to variously as CONGA (COnforming/Non-conforming GAlerkin) or broken-FEEC methods. We consider both standard and broken-FEEC methods in our numerical studies. In general, self-consistent models of the interaction of a plasma with the electromagnetic field come from field-dependent Poisson brackets making structure preserving discretizations rare. For example, Hamiltonian structure preserving discretizations of the Vlasov-Maxwell system and 2D vorticity dynamics rely on particle based discretizations: Hamiltonian structure preserving discretizations of the Vlasov-Maxwell system have been derived based on a particle based representation of the distribution function \cite{GEMPIC, 10.1063/1.4976849}, and the reduction of 2D vorticity to finite dimensional Hamiltonian mechanics of point vorticies is classical \cite{kirchhoff_point_vortex}. The nonlinearities in the ponderomotive Maxwell model are all contained in the Hamiltonian leaving the Poisson bracket extremely simple. The amenability of this model to structure preserving discretization points to a broad class of nonlinear-wave models likewise admitting a similar numerical treatment. 


This paper is organized as follows. In section~\ref{sec:direct_derivation}, the 1D ponderomotive Maxwell model is derived from the two-fluid Maxwell-Euler system via a traditional order-by-order asymptotic approach. In section~\ref{sec:hamiltonian_derivation}, the Hamiltonian structure of the 1D ponderomotive Maxwell system is derived by applying the same asymptotic considerations as before to the Hamiltonian structure of the two-fluid Maxwell-Euler system. In section~\ref{sec:ponder_maxwell_discretization}, a Hamiltonian structure preserving discretization of the 1D ponderomotive Maxwell system is derived. Finally, in section~\ref{sec:numerical_results}, the numerical results from simulating the 1D ponderomotive Maxwell system using the algorithm derived in~\ref{sec:ponder_maxwell_discretization} are presented. Appendix~\ref{appendix:legendre_transform_EM} elucidates the Hamiltonian structure of Maxwell's equations in nonlinear media. Appendix~\ref{appendix:1d_spectral_element_feec} provides the details of the 1D FEEC spectral element method used in section~\ref{sec:ponder_maxwell_discretization}. Appendix~\ref{appendix:disc_functional_derivatives} provides a technical result used in the discretization procedure. 

\section{Direct derivation from the two-fluid Maxwell-Euler model} \label{sec:direct_derivation}
In the presence of a inhomogeneous oscillatory electric field, charged particles experience a net force, averaged over the oscillatory timescale known as the ponderomotive force given by
\begin{equation}
	\bm{F}_p = - \frac{q^2}{4 m \omega_0^2} \nabla | \bm{E} |^2 \,,
\end{equation}
where $m$ and $q$ are the particles' mass and charge, and $\omega_0$ and $\bm{E}$ are the frequency and amplitude of the oscillatory field. In an appropriate asymptotic regime, we shall find that the envelope of the oscillatory electromagnetic fields evolve according to the macroscopic Maxwell's equations 
\begin{equation}
	\partial_t \bm{D} = c \nabla \times \bm{B}
	\quad \text{and} \quad
	\partial_t \bm{B} = - c \nabla \times \bm{E} \,,
\end{equation}
where $\bm{D} = \bm{E} + \bm{P}$ with some to be determined self-consistent polarization. In this section, we elucidate the asymptotic derivation of such a model which we call the ponderomotive Maxwell equation beginning from a two-fluid model coupled to Maxwell's equations, and then provide a self-consistent Hamiltonian formulation of the same model. 

The parent model from which we derive our asymptotic model, which we call the ponderomotive Maxwell system, is the two-fluid model. In Gaussian units, this model may be written:
\begin{equation}
	\begin{aligned}
		\partial_t \bm{E} - c\nabla \times \bm{B} &= -4 \pi (q_e n_e \bm{v}_e + q_i n_i \bm{v}_i) \\
		\partial_t \bm{B} + c\nabla \times \bm{E} &= 0 \\
		\partial_t \bm{v}_\alpha + \frac{1}{2} \nabla | \bm{v}_\alpha |^2 - 
                &\bm{v}_\alpha \times \nabla \times \bm{v}_\alpha \\
		&= \frac{q_\alpha}{m_\alpha} \left( \bm{E} 
                + \frac{1}{c} \bm{v}_\alpha \times \bm{B} \right) + \frac{\nabla p_\alpha}{n_\alpha} \\
		\partial_t n_\alpha + \nabla \cdot ( n_\alpha \bm{v}_\alpha ) &= 0 \,,
	\end{aligned}
\end{equation}
where $\alpha = e, i$, and the thermodynamic pressure, $p_\alpha = n_\alpha^2 (U_\alpha)_{n_\alpha}$, is prescribed by an internal energy, $U_\alpha(n_\alpha)$, which depends only on density. As a matter of notation convention, $(U_\alpha)_{n_\alpha} = \partial U_\alpha/ \partial n_\alpha$. Using the chain rule, one may show that
\begin{equation}
    \frac{\nabla p_\alpha}{n_\alpha} = \nabla \left( (n_\alpha U_\alpha)_{n_\alpha} \right).
\end{equation}
We assume $q_i = - Z q_e$ where $q_e$ is the charge of an electron. Such an isothermal two-fluid model describes a plasma which has reached local thermodynamic equilibrium and which possesses a single temperature. We shall assume that all processes occur on a timescale faster than that at which the temperature changes so that the assumptions of this model are not violated. A more complete model would include an entropy advection law and an entropy dependent internal energy. While we retain a thermodynamic pressure for the purposes of deriving the 1D ponderomotive Maxwell system in this section and section \ref{sec:hamiltonian_derivation}, we will neglect it for simplicity in our discrete model in sections \ref{sec:ponder_maxwell_discretization} and \ref{sec:numerical_results}. 

\subsection{The Two-fluid system: longitudinal/transverse split}
The problem studied in this paper possesses a distinguished direction, namely the direction of laser propagation, which we call $\hat{z}$. The plane transverse to the direction of laser propagation is denoted with the symbol ``$\perp$''. The system written in this longitudinal/transverse split takes the following form:
\begin{equation}
	\begin{aligned}
		\partial_t \bm{E}_\perp - c \hat{z} \times \left( \partial_z \bm{B}_\perp - \nabla_\perp B_z \right) 
				&= -4 \pi \sum_\alpha q_\alpha n_\alpha \bm{v}_{\alpha,\perp} \\
		\partial_t \bm{B}_\perp + c \hat{z} \times \left( \partial_z \bm{E}_\perp - \nabla_\perp E_z \right) &= 0 \\
		\partial_t E_z - c\,\text{curl}_\perp(\bm{B}_\perp) &= -4 \pi \sum_\alpha q_\alpha n_\alpha v_{\alpha, z} \\
		\partial_t B_z + c\,\text{curl}_\perp(\bm{E}_\perp) &= 0
	\end{aligned}
\end{equation}
for Maxwell's equations, 
\begin{equation}
	\partial_t n_\alpha + \nabla_\perp \cdot ( n_\alpha \bm{v}_{\alpha,\perp} ) + \partial_z (n_\alpha v_{\alpha,z}) = 0
\end{equation}
for the continuity equations, and 
\begin{equation}
	\begin{aligned}
		\partial_t \bm{v}_{\alpha, \perp} 
			&+ \frac{1}{2} \nabla_\perp \left( | \bm{v}_{\alpha,\perp} |^2 + v_{\alpha,z}^2
			+ ( n_\alpha U_\alpha(n_\alpha) )_{n_\alpha} \right)
				\\
				&= v_{\alpha,z} \left( \nabla_\perp v_{\alpha,z} -  \partial_z \bm{v}_{\alpha,\perp} 
					+ \hat{z} \times \frac{q_\alpha}{m_\alpha c} \bm{B}_\perp \right) \\
				&+ \hat{z} \times \bm{v}_{\alpha,\perp} \left( \text{curl}_\perp( \bm{v}_{\alpha,\perp} ) 
					- \frac{q_\alpha}{m_\alpha c} B_z \right) 
                    + \frac{q_\alpha}{m_\alpha} \bm{E}_\perp \\
		\partial_t v_{\alpha,z}
			&+ \frac{1}{2} \partial_z \left( | \bm{v}_{\alpha,\perp} |^2 + v_{\alpha,z}^2
			+ ( n_\alpha U_\alpha(n_\alpha) )_{n_\alpha} \right)
			- \frac{q_\alpha}{m_\alpha} E_z \\
				&= - \bm{v}_{\alpha, \perp} \cdot \left( \nabla_\perp v_{\alpha,z} 
					- \partial_z \bm{v}_{\alpha,\perp} + \hat{z} \times \frac{q_\alpha}{m_\alpha c} \bm{B}_\perp \right)
	\end{aligned}
\end{equation}
for the momentum equations, where we define
\begin{equation}
	\nabla_\perp f = 
	\begin{pmatrix}
		\partial_x f \\
		\partial_y f \\
		0
	\end{pmatrix}
	\quad \text{and} \quad
	\text{curl}_\perp ( \bm{A} ) = \partial_x A_y - \partial_y A_x.
\end{equation}

\subsection{Laser-plasma interaction---1D linear regime}
We consider here what is frequently called the linear regime \cite{641305} in laser-plasma literature. In this regime, we assume the transverse electromagnetic fields scale as $\bm{E}_\perp, \bm{B}_\perp \sim \epsilon \ll 1$, the longitudinal electromagnetic fields scale as $E_z, B_z \sim \epsilon^2$, gradients are small in the transverse direction $\nabla_\perp \sim \epsilon^2$ while $\partial_z \sim O(1)$, and we expand the fluid variables in powers of $\epsilon$: e.g. $n_\alpha = \sum_k n_{\alpha}^{(k)}$ where $n_{\alpha}^{(k)} \sim \epsilon^k$. We further assume that all transverse fields are such that
\begin{equation}
	\bm{A}_\perp(\bm{x}, t) = \bm{A}_{\perp0}(\bm{x}, t) \cos \left( \omega_0 t + k_0 z \right) \,,
\end{equation}
where $\omega_0 \sim \epsilon^{-1}$ and $\partial_t \bm{A}_{\perp0} \sim \epsilon$. 

We first consider the continuity and momentum equations order by order. At leading order, $O(\epsilon^0)$, we find that the fluid variables become decoupled from the electromagnetic fields as the electromagnetic fields are asymptotically smaller and we find that
\begin{equation}
	n_\alpha^{(0)} = n_{0\alpha},
	\quad
	\bm{v}_{\alpha, \perp}^{(0)} = 0,
	\quad \text{and} \quad
	v_{\alpha,z}^{(0)} = 0.
\end{equation} 
At $O(\epsilon^1)$, we have
\begin{equation}
	\partial_{{t}} {\bm{v}}_{\alpha, \perp}^{(1)}  = \frac{q_\alpha}{ m_\alpha } \bm{E}_\perp,
	\quad
	v_{\alpha, z}^{(1)}  = 0,
	\quad \text{and} \quad
	n_{\alpha}^{(1)} = 0.
\end{equation}
Hence, due to the oscillatory character of the transverse electromagnetic fields, we find
\begin{equation}
	\bm{v}_{\alpha, \perp}^{(1)} =  \frac{q_\alpha}{ m_\alpha \omega_0} \bm{E}_0 \sin \left( \omega_0 t + k_0 z \right).
\end{equation}
The transverse momentum equation along with Faraday's equation at $O(\epsilon^1)$ also tells us 
\begin{equation}
	\hat{z} \times \partial_z \bm{v}_{\alpha, \perp}^{(1)} - \frac{q_\alpha}{m_\alpha c} \bm{B}_\perp = O(\epsilon^2). 
\end{equation}
Finally, at $O(\epsilon^2)$, we find that
\begin{equation}
	\begin{aligned}
		\partial_{t} \bm{v}_{\alpha,\perp}^{(2)} &= 0 \\
		\partial_t v_{\alpha,z}^{(2)} + \frac{1}{2} \partial_z \left( | \bm{v}_{\alpha,\perp}^{(1)} |^2 + \left( n_\alpha \left( U_\alpha(n_\alpha)  \right)_{n_\alpha} \right) \right) 
			&= \frac{q_\alpha}{m_\alpha} E_z \\
		\partial_{t} {n}_{\alpha}^{(2)} + \partial_z ( {n}_{\alpha}^{(0)} v_{\alpha,z}^{(2)} ) &= 0.
	\end{aligned}
\end{equation}

Because of the asymptotic ordering, Maxwell's equations in the transverse direction become
\begin{equation}
    \begin{aligned}
        &\partial_t \bm{E}_{0\perp} \cos(\omega_0 t + k_0 z) \\
            &\hspace{3em}- c \hat{z} \times \partial_z \bm{B}_{0\perp} \cos(\omega_0 t + k_0 z) \\
            &\hspace{6em}= -4 \pi \sum_\alpha q_\alpha n_\alpha^{(0)} \bm{v}_{\alpha,\perp}^{(1)} \\
        &\partial_t \bm{B}_{0\perp} + c \hat{z} \times \partial_z \bm{E}_{0\perp} = 0.
    \end{aligned}
\end{equation}
We can see that we have simply obtained evolution equations for the envelopes of the transverse fields. Additional simplification is needed in order to address the current source in the transverse Amp{\`e}re's Law, however this is deferred to later when the model is described as a Hamiltonian system. What we will find is that the current source in the transverse equation yields a nonlinear polarization of the medium. To leading order, the longitudinal Maxwell equations become
\begin{equation}
	\begin{aligned}
		\partial_t E_z &= -4 \pi \sum_\alpha q_\alpha n_\alpha^{(0)} v_{\alpha, z}^{(2)} \\
		\partial_t B_z &= 0.
	\end{aligned}
\end{equation}
Hence, the longitudinal field is electrostatic. The longitudinal and transverse fields have decoupled (except through the current sources in Amp{\`e}re's Law). 

While this direct asymptotic derivation reveals many of the characteristics of the desired model, the full details of the reduced model are more easily obtained through consideration of the Hamiltonian formulation of the model. This approach ensures energy conservation in the reduced model, and better motivates the intricate manner in which the plasma and the field  couple in this asymptotic regime. 

\section{Hamiltonian formulation of the 1D ponderomotive Maxwell system} \label{sec:hamiltonian_derivation}
As we saw in the previous section, the model truncates at $O(\epsilon^2)$ yielding a complete set of equations. We now derive the Hamiltonian structure of this model beginning from the two-fluid Maxwell-Euler equations which is known to be Hamiltonian \cite{PhysRevA.25.2437}. 

\subsection{Deriving the Hamiltonian}
In order to add special relativistic effects to the two-fluid Maxwell-Euler model, only the kinetic energy needs to be modified:
\begin{multline}
	H[\bm{p}_{\alpha,\perp}, \bm{E}_\perp, \bm{B}_\perp, E_z, B_z, p_{\alpha,z}, n_\alpha] = \\
	\sum_\alpha \int n_\alpha \left( m_\alpha (\gamma_\alpha - 1) c^2 + U_\alpha(n_\alpha) \right) \mathsf{d}^3 \bm{x} \\
		+ \frac{1}{8 \pi} \int \left( | \bm{E}_\perp |^2 + E_z^2 + | \bm{B}_\perp |^2 + B_z^2 \right) \mathsf{d}^3 \bm{x} \,,
\end{multline}
where $\bm{p}_\alpha = \gamma_\alpha m_\alpha \bm{v}_\alpha$ is the relativistic momentum, $(\bm{p}_{\alpha, \perp}, p_{\alpha,z})$ are its transverse and longitudinal components, and
the Lorentz factor is defined
\begin{equation}
	\gamma_\alpha = \left( 1 + \frac{ | \bm{p}_{\alpha,\perp} |^2 + p_{\alpha,z}^2 }{(m_\alpha c)^2} \right)^{1/2}.
\end{equation}
For the purposes of this paper, we need only weakly relativistic effects. Assuming $|\bm{p}_\alpha| \ll m_\alpha c$, we obtain the lowest order relativistic correction to the classical Hamiltonian:
\begin{multline}
	H[\bm{p}_{\alpha,\perp}, \bm{E}_\perp, \bm{B}_\perp, E_z, B_z, p_{\alpha,z}, n_\alpha] = \\
		\sum_\alpha \int n_\alpha 
                \bigg[ U_\alpha(n_\alpha) 
                    + \frac{1}{2 m_\alpha} \bigg( | \bm{p}_{\alpha, \perp} |^2 
                    + p_{\alpha,z}^2 \\
			        - \frac{1}{4 m_\alpha^2 c^2} \left( | \bm{p}_{\alpha,\perp} |^4 + 2 | \bm{p}_{\alpha, \perp} |^2 p_{\alpha,z}^2 + p_z^4 \right) \bigg) 
                \bigg] \mathsf{d}^3 \bm{x} \\
		+ \frac{1}{8 \pi} \int \left( | \bm{E}_\perp |^2 + E_z^2 + | \bm{B}_\perp |^2 + B_z^2 \right) \mathsf{d}^3 \bm{x}.
\end{multline}
We will find that the inclusion of this leading order relativistic correction of the Hamiltonian results in the electromagnetic medium having an intensity dependent index of refraction.  

In order to obtain the Hamiltonian for the reduced model, we make the following substitutions:
\begin{multline} \label{eq:leading_order_expressions}
	\bm{E}_\perp = \bm{E}_{0\perp} \cos(\theta),
	\quad
	\bm{B}_\perp = \bm{B}_{0\perp} \cos(\theta),
	\quad
	B_z = O(\epsilon^3), \\
	\bm{p}_{\alpha, \perp} = \frac{q_\alpha}{\omega_0} \bm{E}_0 \sin \left( \theta \right) + O(\epsilon^3), 
	\quad
	p_z = p_z^{(2)} + O(\epsilon^3), \\
	n_\alpha = n_\alpha^{(0)} + n_\alpha^{(2)} + O(\epsilon^3) \,,
\end{multline}
where $\theta = \omega_0 t + k_0 z$. At leading order, the relativistic transverse momentum equation is
\begin{equation}
    \partial_t \bm{p}_{\alpha, \perp} = q_\alpha \bm{E}_\perp \,,
\end{equation}
which implies the expression for $\bm{p}_{\alpha, \perp}$ in equation \eqref{eq:leading_order_expressions}. Further, by assuming gradients in the transverse plane are small, we effectively assume the fields are functions of the longitudinal coordinate and time alone. 

Retaining only up to $O(\epsilon^4)$ and averaging over $\theta$, the kinetic energy becomes
\begin{multline}
	H_{KE} = \sum_\alpha \int \bigg[ \frac{n_\alpha^{(0)}}{2 m_\alpha} p_{\alpha,z}^2 \\
            + \frac{q_\alpha^2 n_\alpha^{(0)} }{4 m_\alpha \omega_0^2} \left( 1 + \frac{n_\alpha^{(2)}}{n_\alpha^{(0)}} - \frac{3 q_\alpha^2}{16 c^2 m_\alpha^2 \omega_0^2} | \bm{E}_{0\perp} |^2 \right) | \bm{E}_{0\perp} |^2 \bigg] \mathsf{d} z.
\end{multline}
We can see that the longitudinal momentum reduces to the purely classical expression (so that we may let $p_{\alpha,z} = m_\alpha v_{\alpha,z}$) and the transverse momentum contributes a polarization to the transverse electromagnetic field. The longitudinal electromagnetic energy becomes
\begin{equation}
	H_{EM,z} = \frac{1}{8\pi} \int E_z^2 \mathsf{d} z.
\end{equation}
The longitudinal magnetic energy is a lower order contribution. Finally, the transverse electromagnetic energy becomes
\begin{equation}
	H_{EM,\perp} = \frac{1}{16\pi} \int \left( | \bm{E}_{0\perp} |^2 + | \bm{B}_{0\perp} |^2 \right) \mathsf{d} z.
\end{equation}

In averaging over $\theta$ in the transverse electromagnetic energy, we have picked up an additional factor of $1/2$. This seemingly spurious factor of $1/2$ in the energy cancels out with a compensating factor which appears in the Poisson bracket. This is most easily seen by considering the phase space Lagrangian formulation of the Maxwell subsystem. Explicitly, if $\bm{E}_\perp = \bm{E}_{0\perp} \sin(\theta)$ and $\bm{A}_\perp = \bm{A}_{0\perp} \sin(\theta)$ where $\bm{A}_\perp$ is the vector potential for the transverse field and $\dot{\theta} = \omega_0 = \text{const}$, then we find that
\begin{widetext}
\begin{equation} \label{eq:phase_space_lagrangian}
    \frac{1}{2\pi} \int_0^{2 \pi} \int_\Omega \bm{E}_\perp \cdot \dot{\bm{A}}_\perp \mathsf{d} z \mathsf{d} \theta 
    = \frac{1}{2\pi} \int_0^{2 \pi} \int_\Omega \bigg( \bm{E}_{0\perp} \cdot \dot{\bm{A}}_{0\perp} \cos^2(\theta) 
    - \omega_0 \bm{E}_{0\perp} \cdot \bm{A}_{0\perp} \sin(\theta) \cos(\theta) \bigg) \mathsf{d} z \mathsf{d} \theta 
    = \frac{1}{2} \int_\Omega \bm{E}_{0\perp} \cdot \dot{\bm{A}}_{0\perp} \mathsf{d} z.
\end{equation}
\end{widetext}
Hence, the canonical symplectic form picks up a factor of $1/2$. The canonical Poisson bracket, being the inverse of the symplectic form, is then multiplied by two. 

It is awkward to perform this averaging directly in the Poisson bracket. It would be interesting to study a more complex model with a nonlinear WKB ansatz which retains the dynamics of the phase variable, but this is beyond the scope of this paper. Previous work has been done in this vein starting from the phase-space Lagrangian formulation in order to study acoustic waves \cite{10.1063/1.5099383}. 

Hence, we find that that the Hamiltonian for the 1D ponderomotive Maxwell system is 
\begin{widetext}
\begin{multline}
	H[v_{\alpha,z}, n_\alpha^{(2)}, E_z, \bm{E}_{0\perp}, \bm{B}_{0\perp}] = \frac{1}{2} \sum_\alpha \int \left( m_\alpha n_\alpha^{(0)} v_{\alpha,z}^2 + n_\alpha^{(0)} \overline{U}_\alpha(n_\alpha^{(2)}) \right) \mathsf{d} z 
                + \frac{1}{8 \pi} \int E_z^2 \mathsf{d} z \\
			+ \frac{1}{16\pi} \int \left[ 
			\left( 1 + \frac{4 \pi q_\alpha^2 n_\alpha^{(0)} }{m_\alpha \omega_0^2} \left( 1 + \frac{n_\alpha^{(2)}}{n_\alpha^{(0)}} 
				- \frac{3 q_\alpha^2}{16 c^2 m_\alpha^2 \omega_0^2} | \bm{E}_{0\perp} |^2 \right) \right) | \bm{E}_{0\perp} |^2 
                + | \bm{B}_{0\perp} |^2 \right] \mathsf{d} z
\end{multline}
\end{widetext}
where $\overline{U}_\alpha$ is the modified internal energy we obtain upon truncating $U_\alpha$ to the appropriate asymptotic order.

\subsection{Derivatives of the Hamiltonian}
In order to conform to an existing Hamiltonian modeling paradigm for nonlinear polarization of the electromagnetic field \cite{morrison_gauge_free_lifting}, we seek a functional $K$ such that
\begin{multline} \label{eq:abstract_hamiltonian}
    H = K - \int \bm{E}_{0\perp} \cdot \frac{\delta K}{\delta \bm{E}_{0\perp} } \mathsf{d} z
		\\
    + \frac{1}{16\pi} \int ( | \bm{E}_{0\perp} |^2 + | \bm{B}_{0\perp} |^2 )\mathsf{d} z \,,
\end{multline}
where $\delta K/\delta \bm{E}_\perp$ is the functional derivative, a concept of fundamental importance in Hamiltonian field theories \cite{pjm82}. See equation \eqref{eq:functional_derivative} for a definition of the functional derivative. Small modifications must be made to the Hamiltonian modeling framework for nonlinear polarization \cite{morrison_gauge_free_lifting} due to the additional factor of $1/2$ appearing in the electromagnetic energy from the averaging procedure. 

The functional $K$ satisfying equation \eqref{eq:abstract_hamiltonian} is
\begin{widetext}
\begin{equation}
	K = \frac{1}{2} \sum_\alpha \int \bigg[ m_\alpha n_\alpha^{(0)} (v_{\alpha,z}^{(2)})^2 + E_z^2 + n_\alpha^{(0)} \overline{U}_\alpha(n_\alpha^{(2)})
			- \frac{q_\alpha^2 n_\alpha^{(0)} }{4 m_\alpha \omega_0^2} \left( 1 + \frac{n_\alpha^{(2)}}{n_\alpha^{(0)}} 
				- \frac{q_\alpha^2}{16 c^2 m_\alpha^2 \omega_0^2} | \bm{E}_{0\perp} |^2 \right) | \bm{E}_{0\perp} |^2 \bigg] \mathsf{d} z.
\end{equation}
Hence, we define the displacement field via
\begin{equation}
    \bm{D}_{0\perp} = \bm{E}_{0\perp} - 8 \pi \frac{\delta K}{\delta \bm{E}_{0\perp}}
    = \left[ 1 + \sum_\alpha
	\frac{4 \pi q_\alpha^2 n_\alpha^{(0)} }{m_\alpha \omega_0^2} \left( 1 + \frac{n_\alpha^{(2)}}{n_\alpha^{(0)}} 
	- \frac{q_\alpha^2}{8 c^2 m_\alpha^2 \omega_0^2} | \bm{E}_{0\perp} |^2 \right) \right] \bm{E}_{0\perp}.
\end{equation}
\end{widetext}
We have defined $\bm{D}_{0\perp} = \bm{E}_{0\perp} - 8\pi \delta K/\delta \bm{E}_{0\perp}$ rather than the expected expression $\bm{D}_{0\perp} = \bm{E}_{0\perp} - 4\pi \delta K/\delta \bm{E}_{0\perp}$ to compensate for the additional factor of $1/2$ in the electromagnetic Hamiltonian. The additional factor of $1/2$ in the energy is canceled by a corresponding multiplicative factor of $1/2$ in the symplectic form. Since $\bm{D}_{0\perp}$ is defined via a Legendre transformation, see appendix \ref{appendix:legendre_transform_EM}, it likewise must be scaled accordingly. 

It will be convenient to perform a change of variables so that $\bm{D}_{0\perp}$ is the evolving quantity and not $\bm{E}_{0\perp}$. Consider the coordinate change
\begin{multline}
    \Phi[\bm{E}_{0\perp}, \bm{B}_{0\perp}, E_z, v_{\alpha,z}^{(2)}, n_\alpha^{(2)}] \\
    = (\bm{D}_{0\perp}, \bm{B}_{0\perp}, E_z, v_{\alpha,z}^{(2)}, n_\alpha^{(2)}) \,.
\end{multline}
Then
\begin{multline}
	D \Phi[\bm{E}_{0\perp}, \bm{B}_{0\perp}, E_z, v_{\alpha,z}^{(2)}, n_\alpha^{(2)}] \\
    = 
	\begin{pmatrix}
		\mathcal{E} & 0 & 0 & 0 & \frac{4 \pi q_\alpha^2 }{m_\alpha \omega_0^2} ( \bm{E}_{0\perp} \cdot ) \\
		& I & & \\
		& & I & \\
		& & & I & \\
		& & & & I
	\end{pmatrix} \,,
\end{multline}
where $\mathcal{E} = I - 8 \pi \frac{\delta^2 K}{\delta \bm{E}_{0\perp} \delta \bm{E}_{0\perp} }$ and $(\bm{E}_{0\perp} \cdot)$ means the operator which applies the dot product of $\bm{E}_{0\perp}$ to a given vector. The derivative of the inverse transformation is then
\begin{multline}
    D \Phi^{-1} [\bm{E}_{0\perp}, \bm{B}_{0\perp}, E_z, v_{\alpha,z}^{(2)}, n_\alpha^{(2)}] \\
    = 
    \begin{pmatrix}
    \mathcal{E}^{-1} & 0 & 0 & 0 & -\frac{4 \pi q_\alpha^2 }{m_\alpha \omega_0^2} \mathcal{E}^{-1} ( \bm{E}_{0\perp} \cdot ) \\
    & I & & \\
    & & I & \\
    & & & I & \\
    & & & & I
    \end{pmatrix}.
\end{multline}
Therefore, if we define 
\begin{multline}
    \overline{H}[\bm{D}_{0\perp}, \bm{B}_{0\perp}, E_z, v_{\alpha,z}^{(2)}, n_\alpha^{(2)}] \\
    = 
    H[\bm{E}_{0\perp}, \bm{B}_{0\perp}, E_z, v_{\alpha,z}^{(2)}, n_\alpha^{(2)}],
\end{multline}
then (due to the self-adjointness of $\mathcal{E}$)
\begin{multline}
	\frac{\delta \overline{H}}{\delta n_\alpha^{(2)}} 
		= \frac{\delta H}{\delta n_\alpha^{(2)}} + \left( \frac{\delta \bm{E}_{0\perp} }{\delta n_\alpha^{(2)} } \right)^* \frac{\delta H}{\delta \bm{E}_{0\perp} } \\
		= \frac{q_\alpha^2 }{4 m_\alpha \omega_0^2} | \bm{E}_{0\perp} |^2 + ( n_\alpha^{(0)} \overline{U}_\alpha(n_\alpha^{(2)}) )_{n_\alpha^{(2)}} 
			+ \left( \frac{\delta \bm{E}_{0\perp} }{\delta n_\alpha^{(2)} } \right)^* \mathcal{E} \frac{\bm{E}_{0\perp}}{8 \pi} \\
		= - \frac{q_\alpha^2 }{4 m_\alpha \omega_0^2} | \bm{E}_{0\perp} |^2 + ( n_\alpha^{(0)} \overline{U}_\alpha(n_\alpha^{(2)}) )_{n_\alpha^{(2)}} \,,
\end{multline}
where the asterisks denote the $L^2$ adjoint of an operator. Similarly, one finds that
\begin{equation}
	\frac{\delta \overline{H}}{\delta \bm{D}_{0\perp} } = \left( \frac{\delta \bm{E}_{0\perp} }{\delta \bm{D}_{0\perp} } \right)^* \frac{\delta H}{\delta \bm{E}_{0\perp} } 
		= \frac{\bm{E}_{0\perp}}{8 \pi}.
\end{equation}
The functional derivatives of the remaining variables do not change due to the coordinate transformation. Hence, we find that
\begin{multline}
	\frac{\delta \overline{H}}{\delta \bm{B}_{0\perp} } = \frac{\bm{B}_{0\perp}}{8 \pi},
	\quad 
	\frac{\delta \overline{H}}{\delta E_z} = \frac{E_z}{4 \pi}, \\
	\quad \text{and} \quad
	\frac{\delta \overline{H}}{\delta v_{\alpha,z}^{(2)} } = m_\alpha n_\alpha^{(0)} v_{\alpha,z}^{(2)}.
\end{multline}
These along with the Poisson bracket generate the flow for the 1D ponderomotive Maxwell system. 

\subsection{Deriving the Poisson bracket}
The derivation of the Poisson bracket for the reduced model is heuristic and lacks a complete justification. This is because it is awkward to handle the inclusion of the WKB ansatz for the rapidly oscillating fields directly in the Poisson bracket. Instead, one should consider the WKB ansatz beginning from the phase-space Lagrangian for the two-fluid Maxwell-Euler and then derive the Poisson bracket after performing the asymptotic reductions in the phase-space Lagrangian \cite{10.1063/1.5099383}. 

The two-fluid Maxwell-Euler system was shown to be Hamiltonian \cite{PhysRevA.25.2437}. The bracket given therein uses momentum coordinates, $\bm{M}_\alpha = \rho_\alpha \bm{v}_\alpha$. One may change variables to formulate the model in terms of velocity \cite{pjm82} to obtain the Poisson bracket
{\allowdisplaybreaks
\begin{align*}
		\{F, G\} &= \sum_\alpha \bigg[ \frac{1}{m_\alpha} \int \left( \frac{\nabla \times \bm{v}_\alpha}{n_\alpha} \cdot \frac{\delta F}{\delta \bm{v}_\alpha} \times \frac{\delta G}{\delta \bm{v}_\alpha} \right) \mathsf{d}^3 \bm{x} \\
            &-\frac{1}{m_\alpha} \int \left( \frac{\delta F}{\delta \bm{v}_\alpha} \cdot \nabla \frac{\delta G}{\delta n_\alpha} - \frac{\delta G}{\delta \bm{v}_\alpha} \cdot \nabla \frac{\delta F}{\delta n_\alpha} \right) \mathsf{d}^3 \bm{x} \\
		&+ \frac{q_\alpha}{c m_\alpha^2} \int \left( \frac{\bm{B}}{n_\alpha} \cdot \frac{\delta F}{\delta \bm{v}_\alpha} \times \frac{\delta G}{\delta \bm{v}_\alpha} \right) \mathsf{d}^3 \bm{x} 
		\tag{\stepcounter{equation}\theequation}\\ 
		&- \frac{4 \pi q_\alpha}{m_\alpha} \int \left( \frac{\delta F}{\delta \bm{D}} \cdot \frac{\delta G}{\delta \bm{v}_\alpha} - \frac{\delta G}{\delta \bm{D}} \cdot \frac{\delta F}{\delta \bm{v}_\alpha} \right) \mathsf{d}^3 \bm{x} \bigg] \\
		&+ 4 \pi c \int \left( \frac{\delta F}{\delta \bm{D}} \cdot \nabla \times \frac{\delta G}{\delta \bm{B}} - \frac{\delta G}{\delta \bm{D}} \cdot \nabla \times \frac{\delta F}{\delta \bm{B}} \right) \mathsf{d}^3 \bm{x} \,,
\end{align*}
}%
where we have made a further alteration to the model so that $\bm{D} = \bm{E} - 4 \pi \delta K/\delta \bm{E}$. This notational change is made to better accommodate a self-consistent treatment of nonlinear polarization \cite{morrison_gauge_free_lifting} which arises in the reduced model. Because in this parent two-fluid model, $K = K[n_\alpha, \bm{v}_\alpha]$, it follows that $\bm{D} = \bm{E}$. In passing to the reduced model, this will no longer be the case. 

The Poisson bracket for the full longitudinal/transverse split Maxwell-Euler system may be written
\begin{widetext}
{\allowdisplaybreaks
\begin{align*}
		\{F, G\} &= \sum_\alpha \bigg[ 
				- \frac{1}{m_\alpha} 
				\int \left( \frac{\delta F}{\delta v_{\alpha,z}} \partial_z \frac{\delta G}{\delta n_\alpha} 
					- \frac{\delta G}{\delta v_{\alpha,z}} \partial_z \frac{\delta F}{\delta n_\alpha} \right) \mathsf{d}^3 \bm{x} \\
			&\hspace{10em}
				+ \int \left( \frac{\delta F}{\delta \bm{v}_{\alpha,\perp}} \cdot \nabla_\perp \frac{\delta G}{\delta n_\alpha} 
					- \frac{\delta G}{\delta \bm{v}_{\alpha,\perp}} \cdot \nabla_\perp \frac{\delta F}{\delta n_\alpha} \right) \mathsf{d}^3 \bm{x}
			\bigg] \\
			&+ \sum_\alpha \frac{1}{m_\alpha}
			\bigg[
				\int \frac{ \nabla_\perp v_{\alpha,z} - \partial_z \bm{v}_{\alpha, \perp} }{n_\alpha} \cdot
					\left( \frac{\delta F}{\delta v_{\alpha,z}} \frac{\delta G}{\delta \bm{v}_{\perp,\alpha} } 
					- \frac{\delta G}{\delta v_{\alpha,z}} \frac{\delta F}{\delta \bm{v}_{\perp,\alpha} } \right) \mathsf{d}^3 \bm{x} \\
			&\hspace{10em}
				+ \int \frac{\text{curl}_\perp( \bm{v}_{\alpha, \perp} )}{n_\alpha} 
					\left( \frac{\delta F}{\delta \bm{v}_{\alpha, \perp}} \cdot \hat{z} \times \frac{\delta G}{\delta \bm{v}_{\alpha, \perp} } \right)
					\mathsf{d}^3 \bm{x}
			\bigg] \\
			&+ \sum_\alpha 
			\frac{q_\alpha}{m_\alpha^2 c} \bigg[
				\int \frac{ \hat{z} \times \bm{B}_\perp}{n_\alpha} \cdot
					\left( \frac{\delta F}{\delta v_{\alpha,z}} \frac{\delta G}{\delta \bm{v}_{\perp,\alpha} } 
					- \frac{\delta G}{\delta v_{\alpha,z}} \frac{\delta F}{\delta \bm{v}_{\perp,\alpha} } \right) \mathsf{d}^3 \bm{x} 
				- \int \frac{B_z}{n_\alpha} 
					\left( \frac{\delta F}{\delta \bm{v}_{\alpha, \perp}} \cdot \hat{z} \times \frac{\delta G}{\delta \bm{v}_{\alpha, \perp} } \right) 
					\mathsf{d}^3 \bm{x} 
			\bigg] 
			\tag{\stepcounter{equation}\theequation}\\ 
			&- \sum_\alpha \frac{4 \pi q_\alpha}{m_\alpha}
			\bigg[ 
				\int \left( \frac{\delta F}{\delta D_{z}} \frac{\delta G}{\delta v_{\alpha,z}} 
					- \frac{\delta G}{\delta D_{z}}\frac{\delta F}{\delta v_{\alpha,z}}  \right) \mathsf{d}^3 \bm{x} 
				+ \int \left( \frac{\delta F}{\delta \bm{D}_{\perp}} \cdot \frac{\delta G}{\delta \bm{v}_{\alpha,\perp}} 
					- \frac{\delta G}{\delta \bm{D}_{\perp}} \cdot \frac{\delta F}{\delta \bm{v}_{\alpha,\perp}} \right) 
					\mathsf{d}^3 \bm{x}
			\bigg] \\
			&+ 4 \pi c \int \bigg[ \frac{\delta F}{\delta \bm{D}_\perp} \times \nabla_\perp \frac{\delta G}{\delta B_z} - 
				 \frac{\delta G}{\delta \bm{D}_\perp} \times \nabla_\perp \frac{\delta F}{\delta B_z} 
			+ \frac{\delta F}{\delta \bm{B}_\perp} \times \nabla_\perp \frac{\delta G}{\delta D_z} - 
				 \frac{\delta G}{\delta \bm{B}_\perp} \times \nabla_\perp \frac{\delta F}{\delta D_z} \\
			&\hspace{10em} 
			- \hat{z} \cdot \left( \frac{\delta F}{\delta \bm{D}_\perp} \times \partial_z \frac{\delta G}{\delta \bm{B}_\perp} 
				- \frac{\delta G}{\delta \bm{D}_\perp} \times \partial_z \frac{\delta F}{\delta \bm{B}_\perp} \right) \bigg] \mathsf{d}^3 \bm{x}.
\end{align*}
}%
\end{widetext}
Using the asymptotic scaling assumptions and the results from the previous section, we find that, to leading order, all transverse gradients vanish, the bracket generating the Lorentz force (the fifth line of the bracket) cancels with the vorticity bracket (the third and fourth lines of the bracket), the longitudinal magnetic field vanishes, and the current in the transverse plane vanishes (having been instead modeled as a nonlinear dielectric function rather than an independent field). Hence, we obtain the following Poisson bracket for the evolution of the fields $(\bm{D}_{0\perp}, \bm{B}_{0\perp}, E_z, v_{\alpha,z}^{(2)}, n_{\alpha}^{(2)})$: 
\begin{widetext}
{\allowdisplaybreaks
\begin{align*}
    \{F, G\} &= -\sum_\alpha \frac{1}{m_\alpha} 
            \int \bigg( \frac{\delta F}{\delta v_{\alpha,z}^{(2)}} \partial_z \frac{\delta G}{\delta n_\alpha^{(2)}} 
            - \frac{\delta G}{\delta v_{\alpha,z}^{(2)}} \partial_z \frac{\delta F}{\delta n_\alpha^{(2)}} \bigg) \mathsf{d} z 
            - \sum_\alpha \frac{4 \pi q_\alpha}{m_\alpha}
            \int \left( \frac{\delta F}{\delta E_{z}} \frac{\delta G}{\delta v_{\alpha,z}^{(2)}} 
                - \frac{\delta G}{\delta E_{z}}\frac{\delta F}{\delta v_{\alpha,z}^{(2)}}  \right) \mathsf{d} z
        \tag{\stepcounter{equation}\theequation}\\ 
            &\quad
            + 8 \pi c \int \hat{z} \cdot \left( \frac{\delta F}{\delta \bm{D}_{0\perp} } \times \partial_z \frac{\delta G}{\delta \bm{B}_{0\perp} } 
            - \frac{\delta G}{\delta \bm{D}_{0\perp} } \times \partial_z \frac{\delta F}{\delta \bm{B}_{0\perp} } \right) \mathsf{d} z.
\end{align*}
}%
\end{widetext}
As elucidated in equation \eqref{eq:phase_space_lagrangian}, the averaging procedure causes the electromagnetic symplectic form to be halved. Hence, as the (canonical) Poisson bracket is the inverse of the symplectic form, the electromagnetic Poisson bracket is multiplied by two. 

It is generally a delicate procedure to freely eliminate terms from the Poisson bracket based on asymptotic arguments as the resulting Poisson bracket may fail to satisfy the Jacobi identity. However, in this case, there is no cause for concern since the resulting bracket is field-free and anti-symmetric thus automatically satisfying the Jacobi identity \cite{pjm82}. 

\subsection{1D ponderomotive Maxwell equations of motion}
We may now obtain the equations of motion for the 1D ponderomotive Maxwell system via the usual means, i.e. for any observable $F$, $\dot{F} = \{F, H\}$. We find that
\begin{widetext}
{\allowdisplaybreaks
\begin{align*}
    \dot{F} 
        &= \{F, H\} 
        = - \sum_\alpha \frac{1}{m_\alpha} 
        \int \bigg( \frac{\delta F}{\delta v_{\alpha,z}^{(2)}} \partial_z \left( - \frac{q_\alpha^2 }{4 m_\alpha \omega_0^2} | \bm{E}_{0\perp} |^2 
            + ( n_\alpha^{(0)} \overline{U}_\alpha(n_\alpha^{(2)}) )_{n_\alpha^{(2)}} \right)
        - \left( m_\alpha n_\alpha^{(0)} v_{\alpha,z}^{(2)} \right) \partial_z \frac{\delta F}{\delta n_\alpha^{(2)}} \bigg) \mathsf{d}z \\
        &\hspace{4em} 
        - \sum_\alpha \frac{4 \pi q_\alpha}{m_\alpha}
        \int \left( \frac{\delta F}{\delta E_{z}} \left( m_\alpha n_\alpha^{(0)} v_{\alpha,z}^{(2)} \right)
            - \frac{E_z}{4 \pi} \frac{\delta F}{\delta v_{\alpha,z}^{(2)}}  \right) \mathsf{d}z 
        + c \int \hat{z} \cdot \left( \frac{\delta F}{\delta \bm{D}_{0\perp} } \times \partial_z \bm{B}_{0\perp}  
        - \bm{E}_{0\perp} \times \partial_z \frac{\delta F}{\delta \bm{B}_{0\perp} } \right) \mathsf{d}z.
\end{align*}
}%
\end{widetext}

Hence, we find that
\begin{equation}
	\begin{aligned}
		\partial_t \bm{D}_{0\perp} - c \hat{z} \times \partial_z \bm{B}_{0\perp} &= 0, \\
		\partial_t \bm{B}_{0\perp} + c \hat{z} \times \partial_z \bm{E}_{0\perp} &= 0,
	\end{aligned}
	\hspace{1em}
	\partial_t E_z + 4 \pi q_\alpha n_\alpha^{(0)} v_{\alpha,z}^{(2)} = 0 
\end{equation}
for Maxwell's equations, and
\begin{equation}
    \begin{aligned}
        m_\alpha \partial_t v_{\alpha,z}^{(2)} - q_\alpha E_z \\
            &\hspace{-3em} 
            = \partial_z \left( - \frac{q_\alpha^2 | \bm{E}_{0\perp} |^2 }{4 m_\alpha \omega_0^2} + ( n_\alpha^{(0)} \overline{U}_\alpha(n_\alpha^{(2)}) )_{n_\alpha^{(2)}} \right) \\
        \partial_t n_\alpha^{(2)} + n_\alpha^{(0)} \partial_z v_{\alpha,z}^{(2)} 
            &=0
    \end{aligned}
\end{equation}
for the fluid equations. 

If we align the $x$-axis with the electric field and the $y$-axis with the magnetic field, we obtain
\begin{equation} \label{eq:1d_maxwell_eqns_dimensional}
	\begin{aligned}
		\partial_t D_x + c \partial_z B_y &= 0, \\
		\partial_t B_y + c \partial_z E_x &= 0,
	\end{aligned}
	\hspace{1em}
	\partial_t E_z + 4 \pi q_\alpha n_\alpha^{(0)} v_{\alpha,z}^{(2)} = 0 
\end{equation}
for Maxwell's equations, and
\begin{equation} \label{eq:1d_fluid_eqns_dimensional}
	\begin{aligned}
		m_\alpha \partial_t v_{\alpha,z}^{(2)} - q_\alpha E_z \\
			&\hspace{-3em} 
                = 
                \partial_z \left( - \frac{q_\alpha^2 }{4 m_\alpha \omega_0^2} E_x^2 + ( n_\alpha^{(0)} \overline{U}_\alpha(n_\alpha^{(2)}) )_{n_\alpha^{(2)}} \right)\\
		\partial_t n_\alpha^{(2)} + n_\alpha^{(0)} \partial_z v_{\alpha,z}^{(2)} &= 0
	\end{aligned}
\end{equation}
for the fluid equations. These evolution equations are completed by the constitutive relation
\begin{equation}
	D_x = \left[ 1 + \sum_\alpha \frac{4 \pi q_\alpha^2 n_\alpha^{(0)} }{m_\alpha \omega_0^2} \left( 1 + \frac{n_\alpha^{(2)}}{n_\alpha^{(0)}} 
					- \frac{q_\alpha^2}{8 c^2 m_\alpha^2 \omega_0^2} E_x^2 \right) \right] E_x.
\end{equation}

The system of first-order equations given by \eqref{eq:1d_maxwell_eqns_dimensional} and \eqref{eq:1d_fluid_eqns_dimensional} may be rewritten as the following system of second-order nonlinear wave equations:
\begin{multline}
    \partial_t^2 D_x = c^2 \partial_z^2 E_x 
    \quad \text{and} \quad \\
        \partial_t^2 n_\alpha^{(2)} + \sum_\beta \frac{4 \pi n_\alpha^{(0)} q_\alpha q_\beta}{m_\alpha}(n_\beta^{(0)} + n_\beta^{(2)}) \\
    = \frac{1}{m_\alpha} \partial_z^2 \left(        \frac{q_\alpha^2 }{4 m_\alpha \omega_0^2} E_x^2 
				- ( n_\alpha^{(0)} \overline{U}_\alpha(n_\alpha^{(2)}) )_{n_\alpha^{(2)}} \right) \,,
\end{multline}
where we have used the fact that
\begin{equation}
	\partial_z E_z = \sum_\beta 4 \pi q_\beta (n_\beta^{(0)} + n_\beta^{(2)})
\end{equation}
which is conserved as a Casimir invariant of the bracket.

Note, if we model only the electron dynamics, i.e.\ let $n_i^{(2)} = 0$ with a neutralizing background, $n_e^{(0)} = Z n_i^{(0)}$, we find
\begin{multline}
    \partial_t^2 D_x = c^2 \partial_z^2 E_x 
    \quad \text{and} \quad \\
    \partial_t^2 n_e^{(2)} + \frac{4 \pi n_e^{(0)} q_e^2}{m_e} n_e^{(2)} \\
	= \frac{1}{m_e} \partial_z^2 \left( \frac{q_e^2 }{4 m_e \omega_0^2} E_x^2 
		- ( n_\alpha^{(0)} \overline{U}_\alpha(n_e^{(2)}) )_{n_e^{(2)}} \right) \,,
\end{multline}
where
\begin{equation}
	D_x = \left[ 1 + \frac{4 \pi q_e^2 n_e^{(0)} }{m_e \omega_0^2} \left( 1 + \frac{n_e^{(2)}}{n_e^{(0)}} 
					- \frac{q_e^2}{8 c^2 m_e^2 \omega_0^2} E_x^2 \right) \right] E_x.
\end{equation}
which closely resembles previously studied models \cite{10.1063/1.860884} except that the transverse electromagnetic field is described by a wave equation for the electric field rather than one for the vector potential. 

\subsection{Dimensionless 1D ponderomotive Maxwell system} \label{sec:dimensionless_ponder_maxwell}
Suppose we consider the electron dynamics in which $n_i^{(2)} = 0$ and $n_e^{(0)} = Z n_i^{(0)}$ neutralizes the background ion density. For notational simplicity, we denote the dynamical fields by $(E_x, B_y, E_z, v_z, n)$ since there is no longer possibility for confusion of the plasma species. The background electron density is assumed constant and denoted $n_0$. Suppose we nondimensionalize the fields such that
\begin{multline}
	v_z = c \tilde{v}_z,
	\quad
	n = n_0 \tilde{n},
	\quad 
	B_y = B_0 \tilde{B}_y, \\
	E_x = B_0 \tilde{E}_x,
	\quad \text{and} \quad
	E_z = B_0 \tilde{E}_z.
\end{multline}
Recall, the (electron) plasma and cyclotron frequencies are given by
\begin{equation}
	\omega_p = \sqrt{ \frac{4 \pi n_0 q_e^2}{m_e} },
	\quad \text{and} \quad
	\omega_c = \frac{q_e B_0}{m_e c}.
\end{equation}
We nondimensionalize space and time via $t = \omega_p^{-1} \tilde{t}$ and $z = c/\omega_p \tilde{z}$. 

Dropping the tildes for notational convenience, the dimensionless equations of motion may be written
\begin{equation} \label{eq:ponder_maxwell_1}
	\begin{aligned}
		\partial_t D_x + \partial_z B_y &= 0 \\
		\partial_t B_y + \partial_z E_x &= 0
	\end{aligned}
	\hspace{15mm}
	\partial_t E_z + \frac{\omega_p}{\omega_c} v_z = 0 
\end{equation}
for Maxwell's equations, and
\begin{equation} \label{eq:ponder_maxwell_2}
	\begin{aligned}
		\partial_t v_z - \frac{\omega_c}{\omega_p} E_z 
			+ \partial_z \left( \frac{ \omega_c^2 }{4 \omega_0^2} E_x^2 - ( \overline{U}(n) )_n \right)
				&= 0 \\
		\partial_t n + \partial_z v_z &= 0
	\end{aligned}
\end{equation}
for the fluid equations where
\begin{equation} \label{eq:ponder_maxwell_3}
	D_x = \left[ 1 + \frac{\omega_p^2}{\omega_0^2} \left( 1 + n 
					- \frac{\omega_c^2}{8 \omega_0^2} E_x^2 \right) \right] E_x.
\end{equation}
We have implicitly rescaled the internal energy so as to make it dimensionless. There are three dimensionless parameters of interest in this system: $\omega_p/\omega_c$, $\omega_p/\omega_0$, and $\omega_c/\omega_0$. As a nonlinear wave equation, this system may be written
\begin{equation} \label{eq:nondim_wave_equations}
	\begin{aligned}
		\partial_t^2 D_x &= \partial_z^2 E_x \\
		\partial_t^2 n + n
			&= \partial_z^2 \left( \frac{ \omega_c^2 }{4 \omega_0^2} E_x^2 - ( \overline{U}(n) )_n \right).
	\end{aligned}
\end{equation}
In this form, it is clear that this model is simply the Maxwell wave equation (for a plane wave) nonlinearly coupled to an acoustic wave equation. 

The dimensionless formulation of the model in equations \eqref{eq:ponder_maxwell_3} and \eqref{eq:nondim_wave_equations} as a second-order nonlinear wave equation, shows that there are only two independent parameters: $(\omega_p/\omega_0, \omega_c/\omega_0)$. The magnitude of the cyclotron frequency controls the strength of the ponderomotive force (which is intuitive since the ponderomotive force arises from the fast periodic cyclotron motion in the transverse plane), while the overall strength of the nonlinear index of refraction is controlled by the plasma frequency. There is the additional effect of the relativistic correction whose strength is proportional to the product of the squares of the two frequencies, but this term is smaller (assuming the dimensionless parameters are small). Hence, for the most part, these two parameters independently control the strength of the two nonlinear features of the model.

This system may be written as a Hamiltonian system by letting
\begin{multline}
    H = \frac{1}{2} \int \bigg[ \frac{\omega_p^2}{\omega_c^2} \left( v_z^2 + \overline{U}(n) \right) + E_z^2 + \frac{1}{2} \bigg( B_y^2 \\
        + \left(1 + \frac{\omega_p^2}{\omega_0^2} \left( 1 + n 
	- \frac{3 \omega_c^2}{16 \omega_0^2} E_x^2 \right) \right) E_x^2 \bigg) \bigg] \mathsf{d} z. 
\end{multline}
One can show that
\begin{multline}
	\frac{\delta H}{\delta v_z} = \frac{\omega_p^2}{\omega_c^2} v_z,
	\quad
	\frac{\delta H}{\delta n} = - \frac{\omega_p^2}{4 \omega_0^2} E_x^2 + \frac{\omega_p^2}{\omega_c^2} (\overline{U}(n))_n, \\
	\frac{\delta H}{\delta E_z} = E_z, 
 	\quad
	\frac{\delta H}{\delta D_x } = \frac{E_x}{2},
	\quad \text{and} \quad
	\frac{\delta H}{\delta B_y } = \frac{B_y}{2}.
\end{multline}

The Poisson bracket is given by
\begin{equation}
	\begin{aligned}
		\{F, G\} &= - \frac{\omega_c^2}{\omega_p^2} 
				\int \left( \frac{\delta F}{\delta v_z} \partial_z \frac{\delta G}{\delta n} 
					- \frac{\delta G}{\delta v_z} \partial_z \frac{\delta F}{\delta n} \right) \mathsf{d} z \\
			    &- \frac{\omega_c}{\omega_p} \int \left( \frac{\delta F}{\delta E_z} \frac{\delta G}{\delta v_z} 
					- \frac{\delta G}{\delta E_z}\frac{\delta F}{\delta v_z}  \right) \mathsf{d} z \\
			    &\quad
			    - 2 \int \left( \frac{\delta F}{\delta D_x } \partial_z \frac{\delta G}{\delta B_y } 
				- \frac{\delta G}{\delta D_x } \partial_z \frac{\delta F}{\delta B_y } \right) \mathsf{d} z.
	\end{aligned}
\end{equation}
One can show that this bracket and Hamiltonian generate the desired equations of motion. This Poisson bracket possesses several Casimir invariants (assuming periodic or homogeneous boundary conditions). Three are the consequence of the fact that the partial derivative of a constant is zero:
\begin{equation}
	C_{D_x} = \int D_x \mathsf{d} z,
	\quad
	C_{B_y} = \int B_y \mathsf{d} z,
	\quad \text{and} \quad
	C_n = \int n \mathsf{d} z.
\end{equation}
The final of these represents conservation of the total number of particles whereas the other two conservation laws arise from the symmetric configuration of plane electromagnetic waves (analogous conservation laws do not exist in the full 3D Maxwell's equations). Finally, we also see that
\begin{equation}
	C_{E_z} = \int \mathcal{C} \left( \partial_z E_z + \frac{\omega_p}{\omega_c} n \right) \mathsf{d} z 
	\quad \forall \mathcal{C} \in C^\infty
\end{equation}
is a Casimir invariant. This invariant corresponds to conservation of charge. 

\section{A structure preserving discretization of the 1D ponderomotive Maxwell system} \label{sec:ponder_maxwell_discretization}
We use a finite element exterior calculus (FEEC) spectral element method to discretize the 1D ponderomotive Maxwell system. Moreover, we consider both conforming and broken FEEC methods (broken FEEC methods are discontinuous Galerkin methods which nonetheless preserve the conforming de Rham complex structure \cite{conga_pic_1, conga_pic_2, campos-pinto2016:CONGA, campos-pinto_and_guclu_2021, güçlü2022broken}). Here, we will only provide the briefest possible exposition of notation used in this section. See appendix \ref{appendix:1d_spectral_element_feec} for a more detailed explanation of the spectral element FEEC method and the definitions of the notation used herein. 

Briefly, in 1D, the FEEC method models fields as expansions in one of two different bases:
\begin{equation}
    \phi_h = \sum_{j=1}^{N_0} \upphi_j \Lambda_{0,j}(z)
    \quad \text{and} \quad
    \rho_h = \sum_{j=1}^{N_1} \uprho_j \Lambda_{1,j}(z),
\end{equation}
where $\{ \Lambda_{\ell, j} \}_{j=1}^{N_\ell}$ are the finite element basis functions and $\bm{\upphi}$ and $\bm{\uprho}$ are the Galerkin coefficients (the degrees of freedom). The reason for using two different bases of finite element expansions is that this allows us to capture physically meaningful qualities of our fields by specializing the bases used to represent them. The basis functions, $\{ \Lambda_{\ell, j} \}_{j=1}^{N_\ell}$, provide bases for the finite element spaces $V^0_h$ and $V^1_h$ which in turn approximate the infinite dimensional function spaces $V^0 \coloneqq H^1 \Lambda^0 (\Omega)$ and $V^1 \coloneqq L^2 \Lambda^1 (\Omega)$. The former space models fields which transform as scalar $0$-forms while the latter models fields which transform as scalar $1$-forms (see appendix \ref{appendix:1d_spectral_element_feec}). 

The interpolation operators $\mathcal{I}^\ell \colon \mathcal{C}_\ell \to V^\ell_h$ are simply those operators which take a given vector of Galerkin coefficients and returns the interpolated field: e.g. $\mathcal{I}^0 \bm{\upphi} = \phi_h$. We also define degrees of freedom operators $\bm{\sigma}^\ell \colon V^\ell \to \mathcal{C}_\ell$ which simply take a field and return the appropriate degrees of freedom associated with that field, e.g. $\bm{\sigma}^0(\phi_h) = \bm{\upphi}$. 

Finally, without going too much into the technical details, there are also dual spaces, $(V_h^\ell)^*$, dual interpolation, $\mathcal{I}_\ell^*$, and dual degrees of freedom operators, $\bm{\sigma}_\ell^*$, which are introduced to discretize the adjoint of the derivative. We represent some variables as coming from these dual spaces and thus have coefficients in spaces which we denote $\mathcal{C}_\ell^*$. Roughly speaking $\bm{\sigma}_\ell^* \colon (V^\ell)^* \to \mathcal{C}_\ell^*$ are $L^2$-projections. Any equation which contains an adjoint derivative must be discretized by applying such an $L^2$-projection in order to integrate by parts and discretize the adjoint derivative weakly. 

The finite element mass matrices, $\mathbb{M}_\ell \colon \mathcal{C}_\ell \to \mathcal{C}_\ell^*$, are defined $(\mathbb{M}_\ell)_{ij} = (\Lambda_{\ell,i}, \Lambda_{\ell,j})_{L^2(\Omega)}$. By proposition \ref{prop:change_of_basis} in appendix \ref{appendix:1d_spectral_element_feec}, a variable on the dual complex is related to its primal complex counterpart by the mass matrix: 
\begin{equation}
	\tilde{\bmsf{A}} \in \mathcal{C}^*_\ell \iff \bmsf{A} \in \mathcal{C}^\ell 
	\quad \text{and} \quad
	\tilde{\bmsf{A}} = \mathbb{M}_\ell \bmsf{A}.
\end{equation}

\subsection{Direct projection of the equations of motion}
In this section, we discretize the dimensionless 1D ponderomotive Maxwell equations derived in the previous section. For simplicity, we have taken the internal energy to be $\overline{U}(n) = 0$ so that there is no thermodynamic pressure. Before considering the Hamiltonian structure preserving discretization in detail, we first look at a discretization achieved via a direct application of the degrees of freedom operators to the equations of motion.

The first step to discretize the model is to decide which variables belong in which vector space. This is achieved by inspecting the equations and the Poisson bracket. We can see from the Poisson bracket that: $D_x$ and $B_y$ are dual to each other, $n$ and $v_z$ are dual to each other, and $E_z$ and $v_z$ are dual to each other. One option (there are others) is to let $(E_z, B_y, n)$ be modeled on the primal sequence while $(D_x, v_z)$ are on the dual sequence. We chose to model $v_z$ on the dual sequence so that the derivative on the electric field squared might be weakened via integration by parts. Hence,
\begin{equation}
	E_z \in V^0,
	\quad
	B_y, n \in V^1,
	\quad \text{and} \quad
	\tilde{D}_x, \tilde{v}_z \in V_0^* \,,
\end{equation}
where the tildes indicate that a variable is an element of the dual space. We shall occasionally need to make use of representations of a same field in both the primal and dual space. The presence or absence of a tilde indicates which space is meant: e.g. $B_y \in V^1$ while $\tilde{B}_y \in (V^1)^*$. The two representations are related through the $L^2$ duality map. A subscript ``$h$'' denotes a field's FEM representation: e.g. $A \in V^\ell \implies A_h = \Pi^\ell A$. 

We may directly discretize the equations by applying the appropriate degrees of freedom operators to each equation:
\begin{equation}
	\begin{aligned}
		\partial_t \bm{\sigma}_0^* (\tilde{D}_x) + \mathbbm{d}_0^*  \bm{\sigma}_1^* ( \tilde{B}_y ) &= 0 \\
		\partial_t \bm{\sigma}^1 (B_y) + \mathbbm{d}_0 \bm{\sigma}^0( E_x ) &= 0 \\
		\partial_t \bm{\sigma}^0 (E_z) + \frac{\omega_p}{\omega_c} \bm{\sigma}^0 ( v_z ) &= 0 
	\end{aligned}
	\implies
	\begin{aligned}
		\partial_t \tilde{\bmsf{D}}_x + \mathbbm{d}_0^* \tilde{\bmsf{B}}_y &= 0 \\
		\partial_t \bmsf{B}_y + \mathbbm{d}_0 \bmsf{E}_x &= 0 \\
		\partial_t \bmsf{E}_z + \frac{\omega_p}{\omega_c} \bmsf{v}_z &= 0 
	\end{aligned}
\end{equation}
for Maxwell's equations, and the fluid equations become
\begin{multline}
    \begin{aligned}
        \partial_t \bm{\sigma}_0^* (\tilde{v}_z) - \frac{\omega_c}{\omega_p} \bm{\sigma}_0^* ( \tilde{E}_z) 
            + \frac{ \omega_c^2 }{4 \omega_0^2} \mathbbm{d}_0^* \bm{\sigma}_1^* \left( \widetilde{E_x^2} \right)
                &= 0 \\
        \partial_t \bm{\sigma}^1( n ) + \mathbbm{d}_0 \bm{\sigma}^1( v_z ) &= 0
    \end{aligned}
        \\
    \implies
    \begin{aligned}
        \partial_t \tilde{\bmsf{v}}_z - \frac{\omega_c}{\omega_p} \tilde{\bmsf{E}}_z 
            + \frac{ \omega_c^2 }{4 \omega_0^2} \mathbbm{d}_0^* \tilde{\bmsf{I}} 
                &= 0 \\
        \partial_t \bmsf{n} + \mathbbm{d}_0 \bmsf{v}_z &= 0 \,,
    \end{aligned}
\end{multline}
where $\tilde{\bmsf{I}} = \bm{\sigma}_1^* \left( E_{x,h}^2 \right)$, $E_{x,h} = \Pi_0 E_x$, and the adjoint discrete exterior derivative is defined $\mathbbm{d}_0^* = \mathbb{M}_0^{-1} \mathbbm{d}_0^{T} \mathbb{M}_1$. The constitutive relation is given by
\begin{equation}
    \begin{aligned}
        (\tilde{\bmsf{D}}_x)_i 
            &= \sigma_{0,i}^* \left( \tilde{D}_{x,h} \right) \\
            &= \sigma_{0,i}^* \left(  \left[ 1 + \frac{\omega_p^2}{\omega_0^2} \left( 1 
                + n_h - \frac{\omega_c^2}{8 \omega_0^2} E_{x,h}^2 \right) \right] E_{x,h} \right) \\
            &\hspace{-2em}
            = \left( \Lambda_{0,i}, \left[ 1 + \frac{\omega_p^2}{\omega_0^2} \left( 1 
                + n_h - \frac{\omega_c^2}{8 \omega_0^2} E_{x,h}^2 \right) \right] E_{x,h} \right)_{L^2(\Omega)}.
    \end{aligned}
\end{equation}
We shall find that precisely the same semi-discrete equations are obtained by directly discretizing the Hamiltonian structure. Hence, this system of ODEs is Hamiltonian. 

\subsection{A Hamiltonian structure preserving discretization of the 1D ponderomotive Maxwell system}
In order to derive the spatially discrete Hamiltonian structure, we simply make the ansatz:
\begin{equation}
	E_{z,h} \in V^0_h,
	\quad
	B_{y,h}, n_h \in V^1_h,
	\quad \text{and} \quad
	\tilde{D}_{x,h}, \tilde{v}_{z,h} \in V_{0,h}^*.
\end{equation}
We then restrict the Hamiltonian and Poisson bracket to act on functionals of the form
\begin{equation}
	F[\tilde{D}_{x,h}, B_{y,h}, \tilde{v}_{z,h}, E_{z,h}, n_h] 
		= \msf{F}[ \tilde{\bmsf{D}}_x, \bmsf{B}_y, \tilde{\bmsf{v}}_z, \bmsf{E}_z, \bmsf{n}].
\end{equation}
To compute the discretized bracket, we use the functional chain rule, the fact that $( \left. \bm{\sigma}^\ell \right|_{V_h^\ell} )^\dagger = \mathcal{I}_\ell^*$ and $( \left. \bm{\sigma}_\ell^* \right|_{(V_h^\ell)^*})^\dagger = \mathcal{I}^\ell$ where $\dagger$ denotes the adjoint operator (see appendix \ref{appendix:disc_functional_derivatives}), and $\int \Lambda_{\ell,i} \tilde{\Lambda}_j^\ell \mathsf{d} z = \delta_{ij}$. We then find that
\begin{widetext}
\begin{equation}
	\begin{aligned}
		\{F, G\} &= [ \msf{F}, \msf{G}] \\
		&= - \frac{\omega_c^2}{\omega_p^2} 
				\int \left( 
					(\bm{\sigma}^1)^\dagger \left(\frac{\partial \msf{F}}{\partial \bmsf{n}} \right)
					      \partial_z
					     (\bm{\sigma}_0^*)^\dagger \left(\frac{\partial \msf{G}}{\partial \tilde{\bmsf{v}}_z} \right)
					-
					(\bm{\sigma}^1)^\dagger \left(\frac{\partial \msf{G}}{\partial \bmsf{n}} \right)
					      \partial_z
					     (\bm{\sigma}_0^*)^\dagger \left(\frac{\partial \msf{F}}{\partial \tilde{\bmsf{v}}_z} \right)
					 \right) \mathsf{d} z \\
			    &\quad
			    - \frac{\omega_c}{\omega_p} \int \left( 
			    		(\bm{\sigma}^0)^\dagger \left( \frac{\partial \msf{F}}{\partial \bmsf{E}_z} \right)
					(\bm{\sigma}_0^*)^\dagger \left( \frac{\partial \msf{G}}{\partial \tilde{\bmsf{v}}_z} \right)
					- 
			    		(\bm{\sigma}^0)^\dagger \left( \frac{\partial \msf{G}}{\partial \bmsf{E}_z} \right)
					(\bm{\sigma}_0^*)^\dagger \left( \frac{\partial \msf{F}}{\partial \tilde{\bmsf{v}}_z} \right)
					\right) \mathsf{d} z \\
			    &\quad
			    - 2 \int \left( 
					(\bm{\sigma}^1)^\dagger \left(\frac{\partial \msf{F}}{\partial \bmsf{B}_y} \right)
					      \partial_z
					     (\bm{\sigma}_0^*)^\dagger \left(\frac{\partial \msf{G}}{\partial \tilde{\bmsf{D}}_x} \right)
					-
					(\bm{\sigma}^1)^\dagger \left(\frac{\partial \msf{G}}{\partial \bmsf{B}_y} \right)
					      \partial_z
					     (\bm{\sigma}_0^*)^\dagger \left(\frac{\partial \msf{F}}{\partial \tilde{\bmsf{D}}_x} \right)
				\right) \mathsf{d} z \,,
	\end{aligned}
\end{equation}
which in turn simplifies to become
\begin{equation}
	\begin{aligned}
		[ \msf{F}, \msf{G}] &= - \frac{\omega_c^2}{\omega_p^2} \bigg[
			\left(\frac{\partial \msf{F}}{\partial \bmsf{n}} \right)_i
				(\mathbbm{d}_0)_{ij}
				\left(\frac{\partial \msf{G}}{\partial \tilde{\bmsf{v}}_z} \right)_j
				-
				\left(\frac{\partial \msf{G}}{\partial \bmsf{n}} \right)_i
				(\mathbbm{d}_0)_{ij}
				\left(\frac{\partial \msf{F}}{\partial \tilde{\bmsf{v}}_z} \right)_j
				\bigg] 
                    - \frac{\omega_c}{\omega_p} \bigg[
				\left( \frac{\partial \msf{F}}{\partial \bmsf{E}_z} \right)_i
				\left( \frac{\partial \msf{G}}{\partial \tilde{\bmsf{v}}_z} \right)_i
				- 
			   	\left( \frac{\partial \msf{G}}{\partial \bmsf{E}_z} \right)_i
				\left( \frac{\partial \msf{F}}{\partial \tilde{\bmsf{v}}_z} \right)_i
				\bigg] \\
			    &\quad
			- 2 \left[ \left(\frac{\partial \msf{F}}{\partial \bmsf{B}_y} \right)_i
				(\mathbbm{d}_0)_{ij}
				\left(\frac{\partial \msf{G}}{\partial \tilde{\bmsf{D}}_x} \right)_j
				-
				\left(\frac{\partial \msf{G}}{\partial \bmsf{B}_y} \right)_i
				(\mathbbm{d}_0)_{ij}
				\left(\frac{\partial \msf{F}}{\partial \tilde{\bmsf{D}}_x} \right)_j \right].
	\end{aligned}
\end{equation}
\end{widetext}
Hence, simply by restricting the functional derivatives to act on fields which are Galerkin expansions in the appropriate finite element spaces, we obtain a Poisson bracket for the finite dimensional dynamical system with the coefficients as the dynamical quantities. Because the Poisson bivector is field independent and antisymmetric, this bracket satisfies the Jacobi identity \cite{pjm82}. 

The discrete Hamiltonian is likewise obtained by restricting the Hamiltonian functional to act only on fields in the finite element space. Let $E_{x,h} \in V^0$ and the other fields are as before:
\begin{widetext}
\begin{equation}
	\begin{aligned}
	\msf{H}[\bmsf{E}_x, \bmsf{B}_y, \tilde{\bmsf{v}}_z, \bmsf{E}_z, \bmsf{n}] 
		&= H[E_{x,h}, B_{y,h}, \tilde{v}_{z,h}, E_{z,h}, n_h] \\
		&\hspace{-15mm}= \frac{1}{2} \int \left[ \frac{\omega_p^2}{\omega_c^2} \tilde{v}_{z,h}^2 + E_{z,h}^2 
			+ \frac{1}{2} \Bigg( \left(1 + \frac{\omega_p^2}{\omega_0^2} \left( 1 + n_h 
				- \frac{3 \omega_c^2}{16 \omega_0^2} E_{x,h}^2 \right) \right) E_{x,h}^2 + B_{y,h}^2 \Bigg) \right] \mathsf{d} z \\
		&= \frac{1}{2} \left[ 
			\frac{\omega_p^2}{\omega_c^2} \tilde{\bmsf{v}}_z^T \mathbb{M}_0^{-1} \tilde{\bmsf{v}}_z 
			+ \bmsf{E}_z^T \mathbb{M}_0 \bmsf{E}_z + 
            \frac{1}{2} \left( \tilde{\bmsf{E}}_*^T \bmsf{E}_x
			+ \bmsf{B}_y^T \mathbb{M}_1 \bmsf{B}_y \right)
		\right] \,,
	\end{aligned}
\end{equation}
where we have the Galerkin projection
\begin{equation}
	(\tilde{\bmsf{E}}_*)_i = \left(\Lambda_{0,i}, \left[ 1 + \frac{\omega_p^2}{\omega_0^2} \left( 1 + n_h 
				- \frac{3 \omega_c^2}{16 \omega_0^2} E_{x,h}^2 \right) \right] E_{x,h} \right)_{L^2(\Omega)}.
\end{equation}
\end{widetext}
As in the continuous setting, we must define a constitutive relation: $\tilde{\bmsf{D}}_x = \tilde{\bmsf{D}}_x( \bmsf{E}_x, \bmsf{n})$. This is accomplished through a direct Galerkin projection of the continuous constitutive relation:
\begin{widetext}
\begin{equation} \label{eq:disc_D_cons_law}
	(\tilde{\bmsf{D}}_x)_i = \left( \Lambda_{0,i}, \left[ 1 + \frac{\omega_p^2}{\omega_0^2} \left( 1 + n_h 
					- \frac{\omega_c^2}{8 \omega_0^2} E_{x,h}^2 \right) \right] E_{x,h} \right)_{L^2(\Omega)}.
\end{equation}
\end{widetext}
Notice the difference between the coefficients on the cubic nonlinearity in the definition of $\tilde{\bmsf{E}}_*$ and $\tilde{\bmsf{D}}_x$. Recall, we reconstruct the finite element fields from the coefficients via a Galerkin expansion:
\begin{equation}
	E_{x,h} = \sum_i (\msf{E}_x)_i \Lambda_{0,i}(z)
	\quad \text{and} \quad
	n_h = \sum_i \msf{n}_i \Lambda_{1,i}(z).
\end{equation}
As in the continuous setting, the Hamiltonian and Poisson bracket are defined in terms of different variables. Hence, we perform a change of variables in order to define the Hamiltonian in terms of the correct fields:
\begin{equation}
	\overline{\msf{H}}[\tilde{\bmsf{D}}_x, \bmsf{B}_y, \tilde{\bmsf{v}}_z, \bmsf{E}_z, \bmsf{n}] 
		= \msf{H}[\bmsf{E}_x, \bmsf{B}_y, \tilde{\bmsf{v}}_z, \bmsf{E}_z, \bmsf{n}] \,,
\end{equation}
where we use the implicit function theorem to find the coordinates $(\bmsf{E}_x, \bmsf{B}_y, \tilde{\bmsf{v}}_z, \bmsf{E}_z, \bmsf{n})$ in terms of $(\tilde{\bmsf{D}}_x, \bmsf{B}_y, \tilde{\bmsf{v}}_z, \bmsf{E}_z, \bmsf{n})$. 

\subsection{Derivatives of the discrete Hamiltonian}
Just as in the continuous setting, we take derivatives of the Hamiltonian using the chain rule. This directly mirrors the process in the continuous setting. Define  a functional $\msf{K}[\bmsf{E}_x, \tilde{\bmsf{v}}_z, \bmsf{n}]$ such that
\begin{equation}
	\msf{H} = \msf{K} - \left(\frac{\partial K}{\partial \bmsf{E}_x } \right)^T \bmsf{E}_x
		+ \frac{1}{4} \left( \bmsf{E}_x^T \mathbb{M}_0 \bmsf{E}_x + \bmsf{B}_y^T \mathbb{M}_1 \bmsf{B}_y \right).
\end{equation}
Evidently, this is accomplished by letting
\begin{equation}
	\msf{K} = \frac{1}{2} \left[ 
			\frac{\omega_p^2}{\omega_c^2} \tilde{\bmsf{v}}_z^T \mathbb{M}_0^{-1} \tilde{\bmsf{v}}_z 
			+ \bmsf{E}_z^T \mathbb{M}_0 \bmsf{E}_z + \mathcal{N}[\bmsf{E}_x, \bmsf{n}]
			\right] \,,
\end{equation}
where
\begin{equation}
	\mathcal{N}[\bmsf{E}_x, \bmsf{n}] = - \frac{\omega_p^2}{\omega_0^2} \int \left( 1 + n_h 
					- \frac{\omega_c^2}{16 \omega_0^2} E_{x,h}^2 \right) E_{x,h}^2 \mathsf{d} z.
\end{equation}
It then follows that
\begin{equation}
    \tilde{\bmsf{D}}_x 
        = \mathbb{M}_0 \bmsf{E}_x - \frac{\partial \msf{K}}{\partial \bmsf{E}_x}
\end{equation}
corresponds with the expression in equation \ref{eq:disc_D_cons_law}. Hence, the polarization is characterized by this $\msf{K}$ functional. This will prove a convenient tool in simplifying the chain rule calculation. 

We begin by taking partial derivatives of $\msf{H}$:
\begin{equation}
	\frac{\partial \msf{H}}{\partial \tilde{\bmsf{v}}_z} = \frac{\omega_p^2}{\omega_c^2} \mathbb{M}_0^{-1} \tilde{\bmsf{v}}_z,
	\quad
	\frac{\partial \msf{H}}{\partial \bmsf{E}_z} = \mathbb{M}_0 \bmsf{E}_z,
	\quad \text{and} \quad 
	\frac{\partial \msf{H}}{\partial \bmsf{B}_y} = \frac{\mathbb{M}_1 \bmsf{B}_y}{2}.
\end{equation}
If we define the following matrix
\begin{equation}
	\mathcal{E}[\bmsf{E}_x, \bmsf{n}] = \mathbb{M}_0 - \frac{\partial^2 \msf{K}}{\partial \bmsf{E}_x \partial \bmsf{E}_x},
\end{equation}
then we may write
\begin{equation}
	\frac{\partial \msf{H}}{\partial \bmsf{E}_x} = \mathcal{E}[\bmsf{E}_x, \bmsf{n}] \bmsf{E}_x.
\end{equation}
Finally, the derivative with respect to the density is
\begin{equation}
	\left( \frac{\partial \msf{H}}{\partial \bmsf{n}} \right)_i = \frac{\omega_p^2}{4 \omega_0^2} \left( \Lambda_{1,i}, E_{x,h}^2 \right)_{L^2(\Omega)}.
\end{equation}

We find that if we define the coordinate change
\begin{equation}
	\Phi[\bmsf{E}_x, \bmsf{B}_y, \bmsf{E}_z, \tilde{\bmsf{v}}_z, \bmsf{n}] = (\tilde{\bmsf{D}}_x, \bmsf{B}_y, \bmsf{E}_z, \tilde{\bmsf{v}}_z, \bmsf{n})
\end{equation}
then
\begin{multline}
    D \Phi[\bmsf{E}_x, \bmsf{B}_y, \bmsf{E}_z, \tilde{\bmsf{v}}_z, \bmsf{n}](\delta \bmsf{E}_x, \delta \bmsf{B}_y, \delta \bmsf{E}_z, \delta \tilde{\bmsf{v}}_z, \delta \bmsf{n}) \\
    = 
    (\mathcal{E}[\bmsf{E}_x, \bmsf{n}] \delta \bmsf{E}_x,
        \delta \bmsf{B}_y,
        \delta \bmsf{E}_z, 
        \delta \tilde{\bmsf{v}}_z, 
        \mathcal{O}[\bmsf{E}_x] \delta \bmsf{n}) \,,
\end{multline}
where
\begin{equation}
	(\mathcal{O}[\bmsf{E}_x] \delta \bmsf{n})_i = \frac{\omega_p^2}{\omega_0^2} \sum_j \int \Lambda_{0,i}(z) \Lambda_j^1(z) E_{x,h}(z) \delta \msf{n}_j \mathsf{d} z.
\end{equation}
Hence, we find that
\begin{equation}
	D \Phi^{-1} [\bmsf{E}_x, \bmsf{B}_y, \bmsf{E}_z, \tilde{\bmsf{v}}_z, \bmsf{n}] = 
	\begin{pmatrix}
		\mathcal{E}^{-1} & 0 & 0 & 0 & - \mathcal{E}^{-1} \mathcal{O}[\bmsf{E}_x] \\
		& \mathbb{I} & & \\
		& & \mathbb{I} & \\
		& & & \mathbb{I} & \\
		& & & & \mathbb{I}
	\end{pmatrix} \,,
\end{equation}
where $\mathbb{I}$ is the identity matrix. 

We define $\overline{\msf{H}}[\tilde{\bmsf{D}}_x, \bmsf{B}_y, \bmsf{E}_z, \tilde{\bmsf{v}}_z, \bmsf{n}]= \msf{H}[\bmsf{E}_x, \bmsf{B}_y, \bmsf{E}_z, \tilde{\bmsf{v}}_z, \bmsf{n}]$. Due to the self-adjointness of $\mathcal{E}$, we find
\begin{equation}
	\begin{aligned}
	\frac{\partial \overline{\msf{H}}}{\partial \bmsf{n}} 
		&= \frac{\partial \msf{H}}{\partial \bmsf{n}} 
			+ \left( \frac{\partial \bmsf{E}_x }{\partial \bmsf{n} } \right)^\dagger \frac{\partial \msf{H}}{\partial \bmsf{E}_x } \\
		&= \frac{\omega_p^2}{4 \omega_0^2} \left( \bm{\Lambda}^1, E_{x,h}^2 \right)_{L^2(\Omega)} - \frac{1}{2} \mathcal{O}^\dagger[\bmsf{e}_x] \bmsf{E}_x \\
		&= - \frac{\omega_p^2}{4 \omega_0^2} \left( \bm{\Lambda}^1, E_{x,h}^2 \right)_{L^2(\Omega)}
	\end{aligned}
\end{equation}
since 
\begin{multline}
    (\mathcal{O}^\dagger[\bmsf{E}_x] \bmsf{E}_x)_i = \frac{\omega_p^2}{\omega_0^2} \sum_j \int \Lambda_j^0(z) \Lambda_{1,i}(z) E_{x,h}(z)(\msf{E}_x)_j \mathsf{d} z \\
    = \frac{\omega_p^2}{\omega_0^2} \left( \Lambda_{1,i}, E_{x,h}^2 \right)_{L^2(\Omega)}.
\end{multline}
Similarly, one finds that
\begin{equation}
	\frac{\delta \overline{\msf{H}}}{\delta \tilde{\bmsf{D}}_x } = \left( \frac{\partial \bmsf{E}_x }{\partial \tilde{\bmsf{D}}_x } \right)^\dagger \frac{\partial \msf{H}}{\partial \bmsf{E}_x } 
		= \frac{\bmsf{E}_x}{2}.
\end{equation}
The derivatives with respect to the remaining variables are unchanged by the coordinate transformation:
\begin{multline}
    \frac{\partial \overline{\msf{H}}}{\partial \tilde{\bmsf{v}}_z} 
	= \frac{\partial \msf{H}}{\partial \tilde{\bmsf{v}}_z} = \frac{\omega_p^2}{\omega_c^2} \mathbb{M}_0^{-1} \tilde{\bmsf{v}}_z,
    \quad
    \frac{\partial \overline{\msf{H}}}{\partial \bmsf{E}_z} 
        = \frac{\partial \msf{H}}{\partial \bmsf{E}_z} = \mathbb{M}_0 \bmsf{E}_z,
    \\
    \text{and} \quad 
    \frac{\partial \overline{\msf{H}}}{\partial \bmsf{B}_y} 
        = \frac{\partial \msf{H}}{\partial \bmsf{B}_y} = \frac{\mathbb{M}_1 \bmsf{B}_y}{2}.
\end{multline}
These along with the spatially discrete Poisson bracket generate the flow for the discrete 1D ponderomotive Maxwell system.

\subsection{Spatially discrete equations of motion}
The spatially discrete 1D ponderomotive Maxwell system may be obtained via the usual means for a Hamiltonian system $\dot{\msf{F}} = [ \msf{F}, \msf{H} ]$. Doing so, we find the discrete equations of motion
\begin{equation}
	\begin{aligned}
		\dot{\bmsf{B}}_y &= - \mathbbm{d}_0 \bmsf{E}_x, \\
		\dot{\tilde{\bmsf{D}}}_x &= \mathbbm{d}_0^T \mathbb{M}_1 \bmsf{B}_y, \\
		\dot{\bmsf{E}}_z &= - \frac{\omega_p}{\omega_c} \mathbb{M}_0^{-1} \tilde{\bmsf{v}}_z,
	\end{aligned}
	\quad \text{and} \quad
	\begin{aligned}
		\dot{\tilde{\bmsf{v}}}_z &= - \frac{\omega_c^2}{4 \omega_0^2} \mathbbm{d}_0^T \tilde{\bmsf{I}} 
			+ \frac{\omega_c}{\omega_p} \mathbb{M}_0 \bmsf{E}_z, \\
		\dot{\bmsf{n}} &= - \mathbbm{d}_0 \mathbb{M}_0^{-1} \tilde{\bmsf{v}}_z
	\end{aligned}
\end{equation}
where $\tilde{\bmsf{D}}_x$ is given by equation \ref{eq:disc_D_cons_law} and 
\begin{equation} \label{eq:disc_intensity_eqn}
	\tilde{\bmsf{I}} = \left( \bm{\Lambda}^1, E_{x,h}^2 \right)_{L^2(\Omega)}.
\end{equation}
Notice, this is identical to the system obtained via the direct application of the appropriate degrees of freedom operators. It is possible to write a discrete analog of the nonlinear wave equations:
\begin{equation}
	\frac{\mathsf{d}^2 \tilde{\bmsf{D}}_x }{\mathsf{d} t^2} = \mathbbm{d}_0^T \mathbb{M}_1 \mathbbm{d}_0 \bmsf{E}_x
	\quad \text{and} \quad
	\frac{\mathsf{d}^2 \bmsf{n} }{\mathsf{d} t^2} + \bmsf{n} 
		= \frac{\omega_c^2}{4 \omega_0^2} \mathbbm{d}_0 \mathbb{M}_0^{-1} \mathbbm{d}_0^T \tilde{\bmsf{I}}.
\end{equation}
What is gained from this discretization approach based on the Hamiltonian and Poisson bracket is a clear motivation for the choice of finite dimensional spaces used for each variable and elucidation of the discrete Hamiltonian structure. 

\subsection{Discrete conservation laws} \label{sec:disc_cons_laws}
From the discrete Hamiltonian structure, we immediately know the conserved energy of the system as well as several discrete conservation laws associated with Casimir invariants of the bracket. Conservation of the Hamiltonian comes from the antisymmetry of the Poisson bracket:
\begin{equation}
	\dot{\msf{H}} = [\msf{H}, \msf{H}] = 0.
\end{equation}
From appendix \ref{appendix:casimir_lemmas}, we know that the discrete Casimir invariants are given by
\begin{multline}
	C_{D_x} = \int \tilde{D}_{x,h} \mathsf{d} z = \mathbbm{1}^T \mathbb{M}_0 \bmsf{D}_x = \mathbbm{1}^T \tilde{\bmsf{D}}_x, \\
	C_{B_y} = \int B_{y,h} \mathsf{d} z =  \mathbbm{1}^T \bmsf{B}_y,
	\quad \text{and} \quad
	C_n = \int n_h \mathsf{d} z =  \mathbbm{1}^T \bmsf{n} \,,
\end{multline}
where $\mathbbm{1} = (1, 1, \cdots, 1)^T$. Hence,
\begin{equation}
	\frac{\partial C_D}{\partial \tilde{\bmsf{D}}_x} = \mathbbm{1},
	\quad
	\frac{\partial C_B}{\partial \bmsf{B}_y} = \mathbbm{1},
	\quad \text{and} \quad
	\frac{\partial C_n}{\partial \bmsf{n}} = \mathbbm{1}.
\end{equation}
From this, and the fact that $\mathbbm{1} \in \text{null}(\mathbbm{d}_0)$ and $\mathbbm{1} \in \text{null}(\mathbbm{d}_0^T)$, we conclude that these are Casimir invariants of the discrete Poisson bracket. 

A discrete charge conservation law arises as another Casimir invariant of the discrete bracket:
\begin{equation}
	\frac{\mathsf{d}}{\mathsf{d} t} \left( \mathbbm{d}_0 \bmsf{E}_z - \frac{\omega_p}{\omega_c} \bmsf{n} \right) = 0.
\end{equation}
Hence, if we set $\mathbbm{d}_0 \bmsf{E}_z = \frac{\omega_p}{\omega_c} \bmsf{n}$ as an initial condition, which is the discrete dimensionless Gauss's law, it remains conserved for all of time. For the purposes of numerically studying the charge conservation law, we define the scalar quantity:
\begin{multline}
    C_{E_z} = \left\| \partial_z E_{z,h} - \frac{\omega_p}{\omega_c} n_h \right\|_{L^2} 
    \\
    = \sqrt{ \left( \mathbbm{d}_0 \bmsf{E}_z - \frac{\omega_p}{\omega_c} \bmsf{n} \right)^T \mathbb{M}_1 \left( \mathbbm{d}_0 \bmsf{E}_z - \frac{\omega_p}{\omega_c} \bmsf{n} \right)}.
\end{multline}

\subsection{Temporal discretization via Hamiltonian splitting} \label{section:temporal_discretization}
The Hamiltonian splits into two pieces giving us two exactly integrable subsystems. Define
\begin{multline}
    \msf{H}_E = \frac{1}{2} \left[  
            \bmsf{E}_z^T \mathbb{M}_0 \bmsf{E}_z + \frac{1}{2} \tilde{\bmsf{E}}_*^T \bmsf{E}_x
        \right]
    \\
        \text{and} \quad
    \msf{H}_{Bv} = \frac{1}{2} \left[ 
            \frac{\omega_p^2}{\omega_c^2} \tilde{\bmsf{v}}_z^T \mathbb{M}_0^{-1} \tilde{\bmsf{v}}_z 
            + \frac{1}{2} \bmsf{B}_y^T \mathbb{M}_1 \bmsf{B}_y
        \right].
\end{multline}
It is straightforward to show that these give rise to two subsystems:
\begin{equation}
	\begin{aligned}
		\dot{\bmsf{B}}_y &= - \mathbbm{d}_0 \bmsf{E}_x, \\
		\dot{\tilde{\bmsf{D}}}_x &= 0, \\
		\dot{\bmsf{E}}_z &= 0,
	\end{aligned}
	\quad \text{and} \quad
	\begin{aligned}
		\dot{\tilde{\bmsf{v}}}_z &= - \frac{\omega_c^2}{4 \omega_0^2} \mathbbm{d}_0^T \tilde{\bmsf{I}} 
			+ \frac{\omega_c}{\omega_p} \mathbb{M}_0 \bmsf{E}_z, \\
		\dot{\bmsf{n}} &= 0
	\end{aligned}
\end{equation}
for $\msf{H}_E$ and
\begin{equation}
	\begin{aligned}
		\dot{\bmsf{B}}_y &= 0, \\
		\dot{\tilde{\bmsf{D}}}_x &= \mathbbm{d}_0^T \mathbb{M}_1 \bmsf{B}_y, \\
		\dot{\bmsf{E}}_z &= - \frac{\omega_p}{\omega_c} \mathbb{M}_0^{-1} \tilde{\bmsf{v}}_z,
	\end{aligned}
	\quad \text{and} \quad
	\begin{aligned}
		\dot{\tilde{\bmsf{v}}}_z &= 0, \\
		\dot{\bmsf{n}} &= - \mathbbm{d}_0 \mathbb{M}_0^{-1} \tilde{\bmsf{v}}_z
	\end{aligned}
\end{equation}
for $\msf{H}_{Bv}$. It is clear that these two systems are exactly integrable with update rules given by
\begin{multline} \label{eq:updateBV}
	\tilde{\bmsf{v}}_z(t + \Delta t) = \tilde{\bmsf{v}}_z(t) + \Delta t \left( - \frac{\omega_c^2}{4 \omega_0^2} \mathbbm{d}_0^T \tilde{\bmsf{I}}(t) 
			+ \frac{\omega_c}{\omega_p} \mathbb{M}_0 \bmsf{E}_z(t) \right), \\
	\quad \text{and} \quad
   	\bmsf{B}_y(t + \Delta t) = \bmsf{B}_y(t) - \Delta t \mathbbm{d}_0 \bmsf{E}_x(t)
\end{multline}
for the $\msf{H}_E$ subsystem, and
\begin{multline} \label{eq:updateEN}
	\bmsf{n}(t + \Delta t) = \bmsf{n}(t) - \Delta t \mathbbm{d}_0 \mathbb{M}_0^{-1} \tilde{\bmsf{v}}_z(t), \\
	\bmsf{E}_z(t + \Delta t) = \bmsf{E}_z(t) - \frac{\omega_p}{\omega_c} \Delta t \mathbb{M}_0^{-1} \tilde{\bmsf{v}}_z(t), \\
	\text{and} \quad
    	\tilde{\bmsf{D}}_x(t + \Delta t) = \tilde{\bmsf{D}}_x(t) + \Delta t \mathbbm{d}_0^T \mathbb{M}_1 \bmsf{B}_y(t)
\end{multline}
for the $\msf{H}_{Bv}$ subsystem. Note, each time we solve the $\msf{H}_{Bv}$ subsystem, we must use the constitutive relation in equation \ref{eq:disc_D_cons_law} to solve for $\bmsf{E}_x$ and perform the Galerkin projection in equation \ref{eq:disc_intensity_eqn} to get $\tilde{\bmsf{I}}$ before we can advance the $\msf{H}_E$ subsystem again. 

In order to approximate the flow for the full system we use operator splitting methods \cite{trotter1959product}. Strang splitting \cite{doi:10.1137/0705041} is a popular symmetric second order method:
\begin{equation}
	\begin{aligned}
		e^{ \Delta t \{ \cdot, H \} } 
			&= e^{ \Delta t \{ \cdot, H_E \} + \Delta t \{\cdot, H_B \} } \\
			&= e^{ \Delta t/2 \{ \cdot H_E \} } e^{ \Delta t \{ \cdot, H_B \} } e^{ \Delta t/2 \{ \cdot H_E \} } + O(\Delta t^{3}).
	\end{aligned}
\end{equation}
This is a favorable approach at second order because only one nonlinear solve is needed per time step (since the $\msf{H}_{Bv}$ flow is only computed once per time step). There are also higher order splitting methods which achieve a higher temporal order by having a larger number of substeps for each time step \cite{YOSHIDA1990262}. The derivation of symplectic integrators based on splitting methods for exponential maps is a well established theory \cite{geometric_numerical_integration}. We only consider second and sixth order symmetric splitting methods. 

The symplecticity of the algorithm is only guaranteed up to the precision of the nonlinear solver. Hence, while in the idealized world of infinite precision arithmetic this algorithm is a Poisson integrator, in reality, the flow is only approximately so to the precision of the nonlinear solver. As the nonlinear system is a perturbation of a linear symmetric positive definite system by a nonlinear operator with a small coefficient, it is sufficient to use Picard iteration with Anderson acceleration \cite{10.1145/321296.321305} to solve for $\bmsf{E}_x$ at each timestep. 

\subsection{Full algorithm to advance 1D ponderomotive Maxwell system}
In the following, we shall assume that we may treat the subroutines associated with the FEEC method as a black box. Only five principle routines are needed to accomplish each timestep: (1) a routine to perform the nonlinear solve for $\bmsf{E}_x$, (2) a routine to compute $\tilde{\bmsf{I}}$ via an $L^2$ projection, (3) a routine to advance the $\msf{H}_E$ subsystem, (4) a routine to advance the $\msf{H}_{Bv}$ subsystem, and (5) a routine which composes together the partial flows to get the flow of the entire system to some specified temporal order. 

We assume the availability of some Gaussian quadrature scheme to perform Gauss-Lobatto quadrature at a given number of quadrature points: $\tilde{\bmsf{g}} \leftarrow \texttt{gaussQuadrature}\left( g, N_Q \right)$. Moreover, for algorithmic convenience, we consider an expanded state vector: 
\begin{equation}
	\texttt{stateVec} := (\tilde{\bmsf{D}}_x, \bmsf{E}_x, \tilde{\bmsf{I}}, \bmsf{E}_z, \bmsf{B}_y, \tilde{\bmsf{v}}_z, \bmsf{n}). 
\end{equation}
While $\bmsf{E}_x$ and $\tilde{\bmsf{I}}$ are technically redundant information which might be obtained from the other fields, it is more convenient to store this information rather than recompute as needed. We also have a helper routine given in algorithm \ref{alg:assemble} which simply projects a field onto a given FEM basis. 

\setlength{\textfloatsep}{0.5em}
\begin{algorithm}[H] 
	\caption{$\texttt{assemble}_\ell(f; \ N_Q)$}
	\label{alg:assemble}
	\begin{algorithmic}
		\State $\tilde{\bmsf{f}} \leftarrow \texttt{gaussQuadrature}\left( \left( \Lambda_{\ell,i}, f \right)_{L^2}, N_Q \right)$ \Comment{\textit{Quadrature with $N_Q$ points}}
		\State \texttt{return} $\tilde{\bmsf{f}}$
	\end{algorithmic}
\end{algorithm}

We compute $\bmsf{E}_x$ via fixed point iteration. The initial guess for $\bmsf{E}_x$ is its value at the previous time step. Assuming a sufficiently small time step, the fixed point iteration converges relatively quickly. In order to further speed up the fixed point iteration, we employed Anderson acceleration, although this is not reflected in the basic algorithm outlined in algorithm \ref{alg:picardSolveE}. The algorithm to compute $\tilde{\bmsf{I}}$ is fairly trivial and given in \ref{alg:computeIntensity}.

\begin{algorithm}[H] 
	\caption{\texttt{picardSolveE}($\texttt{stateVec}; \ \texttt{ATOL}, \texttt{RTOL}, N_Q$)}
	\label{alg:picardSolveE}
	\hspace*{\algorithmicindent} \textbf{Input:} $\tilde{\bmsf{D}}_x$, $\bmsf{E}_x$, $\bmsf{n}$ \\
	\hspace*{\algorithmicindent} \textbf{Output:} $\bmsf{E}_x$
	\vspace{0.25em}
	\hrule
	\vspace{0.25em}
	\begin{algorithmic}
		\State 
            {\texttt{field}$(\bm{x}; \ \bmsf{E}_x, \bmsf{n}):= \frac{\omega_p^2}{\omega_0^2} \bigg[1 + \sum_i \mathsf{n}_i \Lambda_{1,i}(\bm{x})$ \\
            \hspace{3em} $- \frac{\omega_c^2}{8 \omega_0^2} \left| \sum_i (\mathsf{E}_x)_i \Lambda_{0,i}(\bm{x}) \right|^2 \bigg] \sum_i (\mathsf{E}_x)_i \Lambda_{0,i}(\bm{x}) $
		\State $\bmsf{D}_x \leftarrow \mathbb{M}_1^{-1} \tilde{\bmsf{D}}_x$
		\State $\texttt{res}(\bmsf{E}_x, \bmsf{P}_x) := \bmsf{E}_x - \bmsf{P}_x - \bmsf{D}_x$ 
            }
		\Do
			\State $\tilde{\bmsf{P}}_x \leftarrow \texttt{assemble}_1(\texttt{field}(\bm{x}; \ \bmsf{E}_x, \bmsf{n}), N_Q)$
			\State $\bmsf{P}_x \leftarrow \mathbb{M}_1^{-1} \tilde{\bmsf{P}}_x$
			\State $\bmsf{R} \leftarrow \texttt{res}(\bmsf{E}_x, \bmsf{P}_x)$
			\State $\bmsf{E}_x \leftarrow \bmsf{E}_x - \bmsf{R}$
		\doWhile{ ( $| \mathsf{R}_i | < \texttt{RTOL} | \mathsf{D}_i | + \texttt{ATOL}, \ \forall i$ ) } 
	\end{algorithmic}
\end{algorithm}

\begin{algorithm}[H] 
	\caption{\texttt{computeIntensity}($\texttt{stateVec}; \ N_Q$)}
	\label{alg:computeIntensity}
	\hspace*{\algorithmicindent} \textbf{Input:} $\bmsf{E}_x$ \\
	\hspace*{\algorithmicindent} \textbf{Output:} $\tilde{\bmsf{I}}$
	\vspace{0.25em}
	\hrule
	\vspace{0.25em}
	\begin{algorithmic}
		\State $\texttt{field}(\bm{x}; \ \bmsf{E}_x) := \left| \sum_i (\mathsf{E}_x)_i \Lambda_{0,i}(\bm{x}) \right|^2$
		\State $\tilde{\bmsf{I}} \leftarrow \texttt{assemble}_1(\texttt{field}(\bm{x}; \ \bmsf{E}_x), N_Q)$
	\end{algorithmic}
\end{algorithm}

The remaining three routines are straightforward. We simply use the update rules given in equations \ref{eq:updateBV} and \ref{eq:updateEN} to define the two partial flows given by $\msf{H}_{E}$ and $\msf{H}_{Bv}$, and then compose them together using Strang splitting in algorithm \ref{alg:strangSplitting}. In several tests, we also use higher order splitting methods \cite{YOSHIDA1990262}. We do not write down an algorithm for these methods as the idea is analogous to Strang splitting, but with more sub-steps. A single call of $\texttt{adv2ndOrd}$ advances the system a single time step. In order to ensure stability of the scheme, we require $\Delta t/\Delta x < 1$ where $\Delta x$ is the minimum Gauss-Lobatto grid-spacing among all of the mapped elements. In practice, we let $\Delta t/\Delta x = 1/2$. 

\begin{algorithm}[H]
	\caption{\texttt{updateBV}($\Delta t$; \ \texttt{stateVec})}
	\label{alg:faradUpdate}
	\hspace*{\algorithmicindent} \textbf{Input:} \texttt{stateVec} \\
	\hspace*{\algorithmicindent} \textbf{Output:} $\tilde{\bmsf{v}}_z, \bmsf{B}_y$
	\vspace{0.25em}
	\hrule
	\vspace{0.25em}
	\begin{algorithmic}
		\State $\tilde{\bmsf{v}}_z \leftarrow \tilde{\bmsf{v}}_z + \Delta t \left( - \frac{\omega_c^2}{4 \omega_0^2} \mathbbm{d}_0^T \tilde{\bmsf{I}} + \frac{\omega_c}{\omega_p} \mathbb{M}_0 \bmsf{E}_z \right)$
   		\State $\bmsf{B}_y \leftarrow \bmsf{B}_y - \Delta t\, \mathbbm{d}_0 \bmsf{E}_x$
	\end{algorithmic}
\end{algorithm}%
\begin{algorithm}[H]
	\caption{\texttt{updateEN}($\Delta t; \ \texttt{stateVec}, 
                \ \texttt{ATOL}, \ \texttt{RTOL}, N_Q$)}
	\label{alg:contnUpdate}
	\hspace*{\algorithmicindent} \textbf{Input:} \texttt{stateVec} \\
	\hspace*{\algorithmicindent} \textbf{Output:} $\bmsf{n}, \bmsf{E}_z, \tilde{\bmsf{D}}_x$
	\vspace{0.25em}
	\hrule
	\vspace{0.25em}
	\begin{algorithmic}
		\State $\bmsf{n} \leftarrow 
			\bmsf{n} - \Delta t\, \mathbbm{d}_0 \mathbb{M}_0^{-1} \tilde{\bmsf{v}}_z$
		\State $\bmsf{E}_z \leftarrow 
			\bmsf{E}_z - \frac{\omega_p}{\omega_c} \Delta t\, \mathbb{M}_0^{-1} \tilde{\bmsf{v}}_z$
    		\State $\tilde{\bmsf{D}}_x \leftarrow \tilde{\bmsf{D}}_x + \Delta t\, \mathbbm{d}_0^T \mathbb{M}_1 \bmsf{B}_y$
      	\State \texttt{picardSolveE}($\texttt{stateVec}; \ \texttt{ATOL}, \texttt{RTOL}, N_Q$)
		\State \texttt{computeIntensity}($\texttt{stateVec}; \ N_Q$)
	\end{algorithmic}
\end{algorithm}%
\begin{algorithm}[H]
	\caption{\texttt{adv2ndOrd}($\Delta t; \ \texttt{stateVec}; \ \texttt{ATOL}, \texttt{RTOL}, N_Q$)}
	\label{alg:strangSplitting}
	\hspace*{\algorithmicindent} \textbf{Input:} \texttt{stateVec} \\
	\hspace*{\algorithmicindent} \textbf{Output:} \texttt{stateVec}
	\vspace{0.25em}
	\hrule
	\vspace{0.25em}
	\begin{algorithmic}
		\State \texttt{updateBV}($\Delta t/2$; \ \texttt{stateVec})
		\State \texttt{updateEN}($\Delta t; \ \texttt{stateVec}, 
                \ \texttt{ATOL}, \ \texttt{RTOL}, N_Q$)
		\State \texttt{updateBV}($\Delta t/2$; \ \texttt{stateVec})
	\end{algorithmic}
\end{algorithm}

\section{Numerical results for 1D ponderomotive Maxwell system} \label{sec:numerical_results}
In this section, we examine the behavior of the algorithm derived in the previous section. All quantities in these tests are dimensionless, see section \ref{sec:dimensionless_ponder_maxwell}. We take $t \in (0,15)$, and let the spatial domain, $\Omega = (0, 40)$, be periodic. The domain is made long enough that any deflected waves propagating in the reverse direction do not pollute the forward propagating wave. All fields except the transverse electromagnetic fields are initially zero. The transverse electromagnetic fields are Gaussian:
\begin{equation}
	E_0(z) = B_0(z) = \exp \left( - (z-4)^2 \right).
\end{equation}
In a vacuum, this wave would propagate unchanged with unit speed. Hence, we will plot the solutions only on the domain $z \in (0,25)$ since the forward propagating wave remains in this portion of the domain. 

Let $\xi_{i,k}$ denote the mapped Gauss-Lobatto points in each element and define
\begin{equation}
	\Delta x = \min_{i,k} | \xi_{i,k} - \xi_{i+1,k} |.
\end{equation}
The CFL condition requires that $u \Delta t/\Delta x < C_{\text{max}}$ where $u$ is the maximum wave speed in our system and $C_{\text{max}}$ is the maximum Courant parameter allowed by the scheme. It is difficult to determine $u$ exactly, but we make the following heuristic argument. In linear media, the speed of light is $c/\sqrt{\epsilon}$ where $\epsilon$ is the dielectric constant. Hence, assuming our dielectric function remains greater than one, i.e.
\begin{equation}
    1 + n - \frac{\omega_c^2}{8 \omega_0^2} E_x^2 > 0,
\end{equation}
light waves travel slower than the speed of light in vacuum which, in our dimensionless units, has value unity. One should not blindly assume this inequality remains satisfied, but must consider whether the initial conditions and parameters of the system justify this assumption and whether it continues to remain satisfied through the whole simulation. We found that the inequality was satisfied in all of our simulations as we choose relatively small physical parameters and the density perturbation and electric field remained small. Likewise, determining $C_{\text{max}}$ exactly would require rigorous linear stability analysis which is beyond the scope of this paper. We empirically observe that the solution is stable when $\Delta t = \Delta x/2$ for simulations in which the nonlinearities of the system are relatively weak, and when $\Delta t = \Delta x/4$ for simulations in which the nonlinearity plays a more dominant role (the parameters associated with these two regimes will be made definite subsequently). The temporal resolution is therefore set by the spatial resolution of the method and also by the state of the system and the parametric regime in a manner which requires further investigation. 

\subsection{Numerical results}
In order to understand the properties of the numerical solver, we vary several physical and numerical parameters. Regarding the physical parameters, we consider a weakly nonlinear regime in which $(\omega_p/\omega_0, \omega_c/\omega_0) = (1/5,-1/5)$, and a strongly nonlinear regime in which $(\omega_p/\omega_0, \omega_c/\omega_0) = (2/5, - 9/10)$. The two dimensionless parameters independently control the strength of the two nonlinear features of the model: $\omega_c/\omega_0$ controls the strength of the ponderomotive force, and $\omega_p/\omega_0$ controls the strength of the nonlinear refractive index. Recall, it is easiest to see the effects of these parameters by looking at the nondimensional system formulated as a coupled system of second-order wave equations in equations \eqref{eq:nondim_wave_equations} and the nondimensional constitutive relation \eqref{eq:ponder_maxwell_3}. 

In the weakly nonlinear case, we let the plasma and cyclotron frequencies be the same while in the latter case, the cyclotron frequency is double the plasma frequency. We do not perform an exhaustive parameter sweep in this work to understand the physical effects of the relative strength of these two parameters, but one should expect different regimes in which the physics is dominated by a strong ponderomotive force or a strong nonlinear refractive index. However, as both of these phenomena lead to wave steepening, their effects may be qualitatively similar in this one-dimensional model. 

Regarding numerical parameters, we consider both high and low spatial and temporal order solutions for both conforming and broken FEEC discretizations. Recall, from the construction of the spectral element method in appendix \ref{appendix:1d_spectral_element_feec}, the polynomials interpolating the $0$-forms are one degree higher than those interpolating the $1$-forms. For the low order tests, we use a second order splitting method in time, $N=3$ (which yields cubic polynomials for the $0$-forms and quadratic for the $1$-forms), and $K=200$ cells. In the conforming FEEC case, the $0$-forms have $600$ degrees of freedom while in broken FEEC they have $800$. In both cases, the $1$-forms have $600$ degrees of freedom. For the high order tests, we use a sixth order splitting method in time, $N=6$ (which yields $6^{th}$-order polynomials for the $0$-forms and $5^{th}$-order polynomials for the $1$-forms), and $K=100$ cells. In the conforming FEEC case, the $0$-forms have $600$ degrees of freedom while in broken FEEC they have $700$. In both cases, the $1$-forms have $600$ degrees of freedom. The time-step is implicitly determined by the spatial resolution and was empirically found to require greater resolution in the regimes where the nonlinearities of the system are more pronounced. The low spatial order test cases used $\Delta t \approx 0.028$ for the weakly nonlinear case, and $\Delta t \approx 0.014$ for the strongly nonlinear case. The high spatial order test cases used $\Delta t \approx 0.017$ in the weakly nonlinear case, and $\Delta t \approx 0.0085$ in the strongly nonlinear case. In all cases, the nonlinear solver tolerance has been set to $10^{-13}$.

\subsubsection{Weakly nonlinear regime}
We first consider a weakly nonlinear regime in which $(\omega_p/\omega_0, \omega_c/\omega_0) = (1/5,-1/5)$. This choice of parameters is within the valid bounds for which the asymptotic assumptions in the derivation of the 1D ponderomotive Maxwell model hold. 

\begin{figure*}
	\centering
	
	\begin{subfigure}{\textwidth}
    		\begin{minipage}{0.47\textwidth}
   			\centering
			\includegraphics[width=\linewidth]{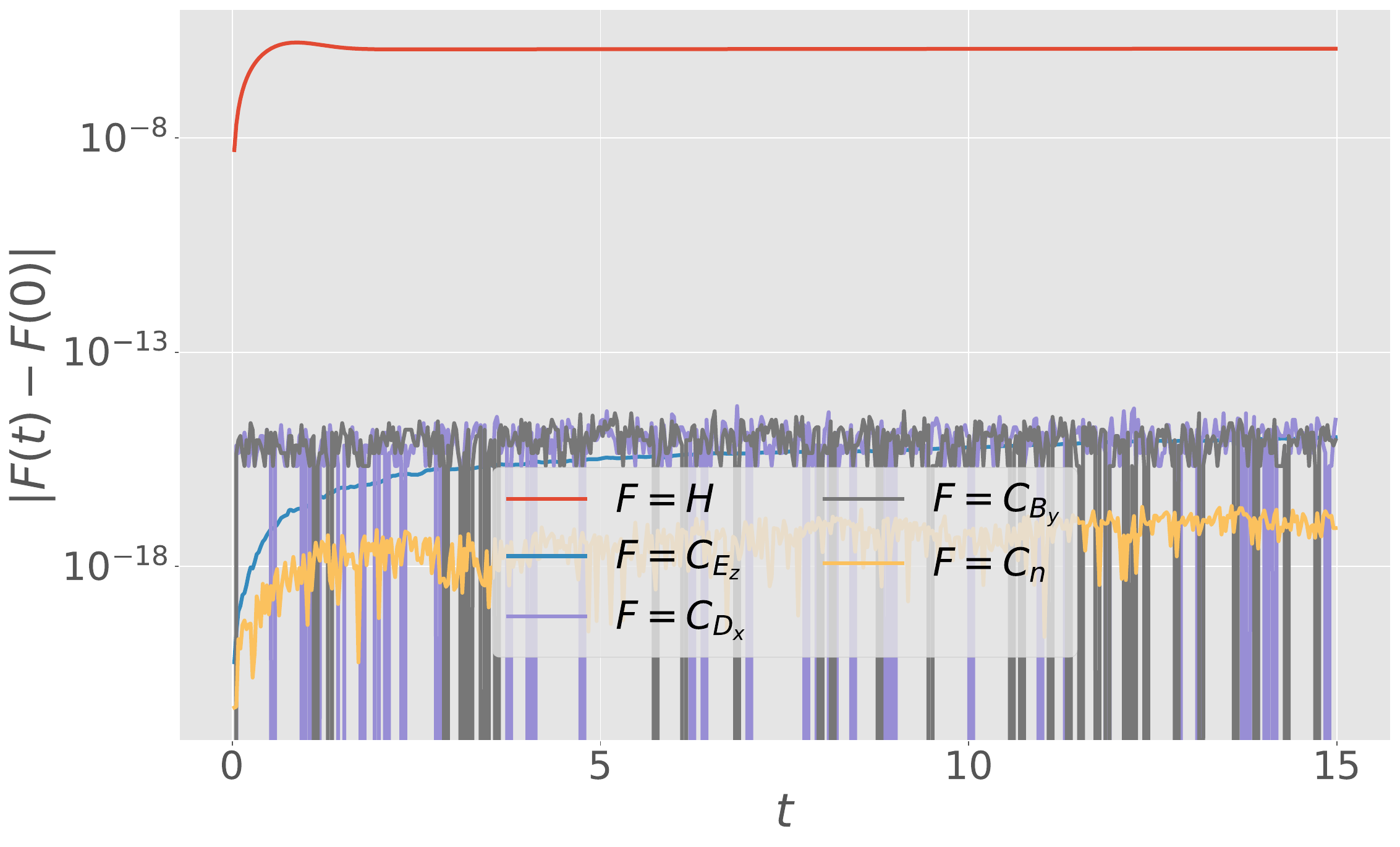}  	
		\end{minipage}\hfill
  		\begin{minipage}{0.47\textwidth}
   			\centering
        			\includegraphics[width=\linewidth]{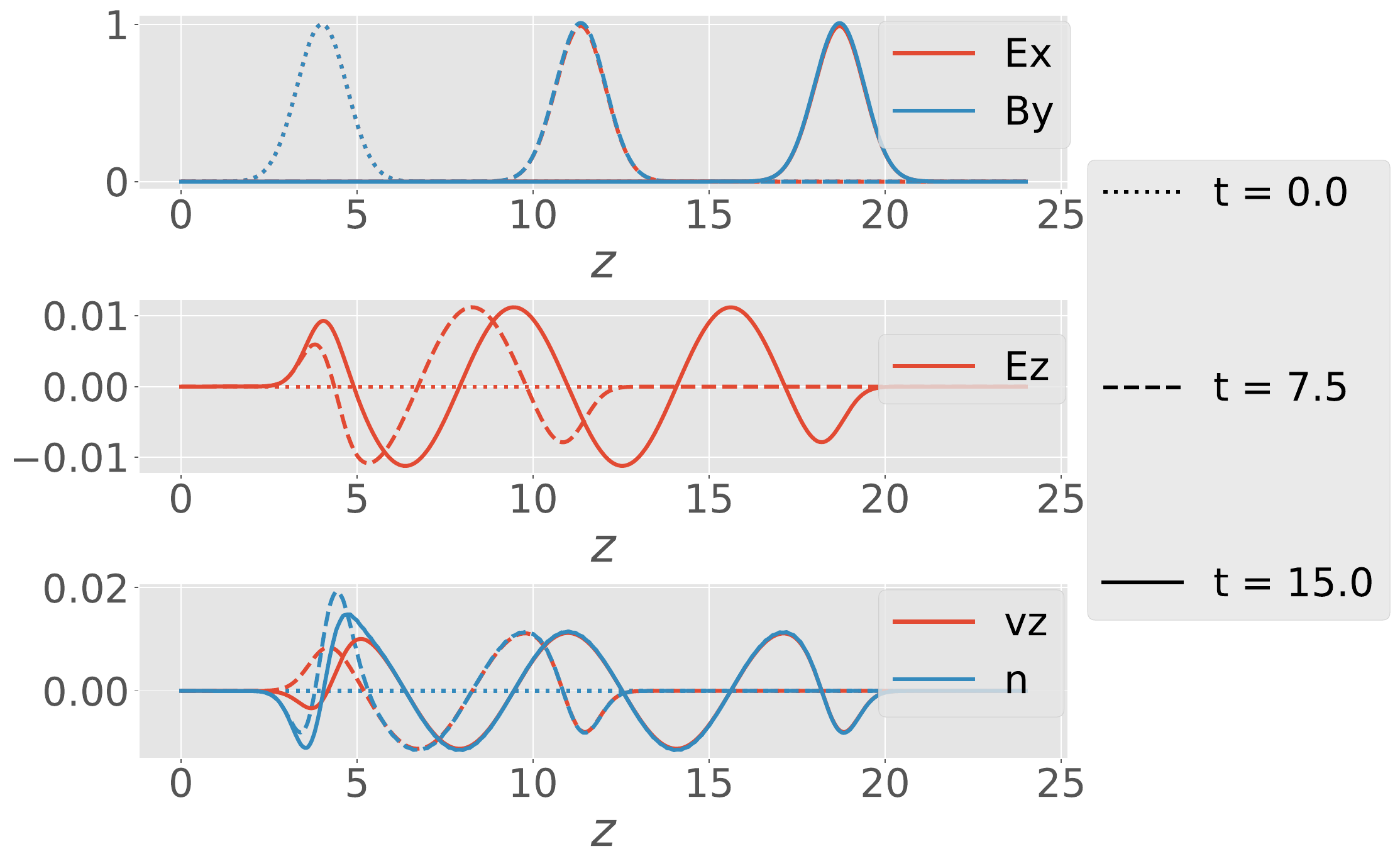}
  		\end{minipage}
		\caption{Conforming-FEEC; Temporal order: $O(\Delta t^2)$; Spatial order: $3^{rd}$ for $0$-forms and $2^{nd}$ for $1$-forms}
	\end{subfigure}
	
	\begin{subfigure}{\textwidth}
    		\begin{minipage}{0.47\textwidth}
   			\centering
			\includegraphics[width=\linewidth]{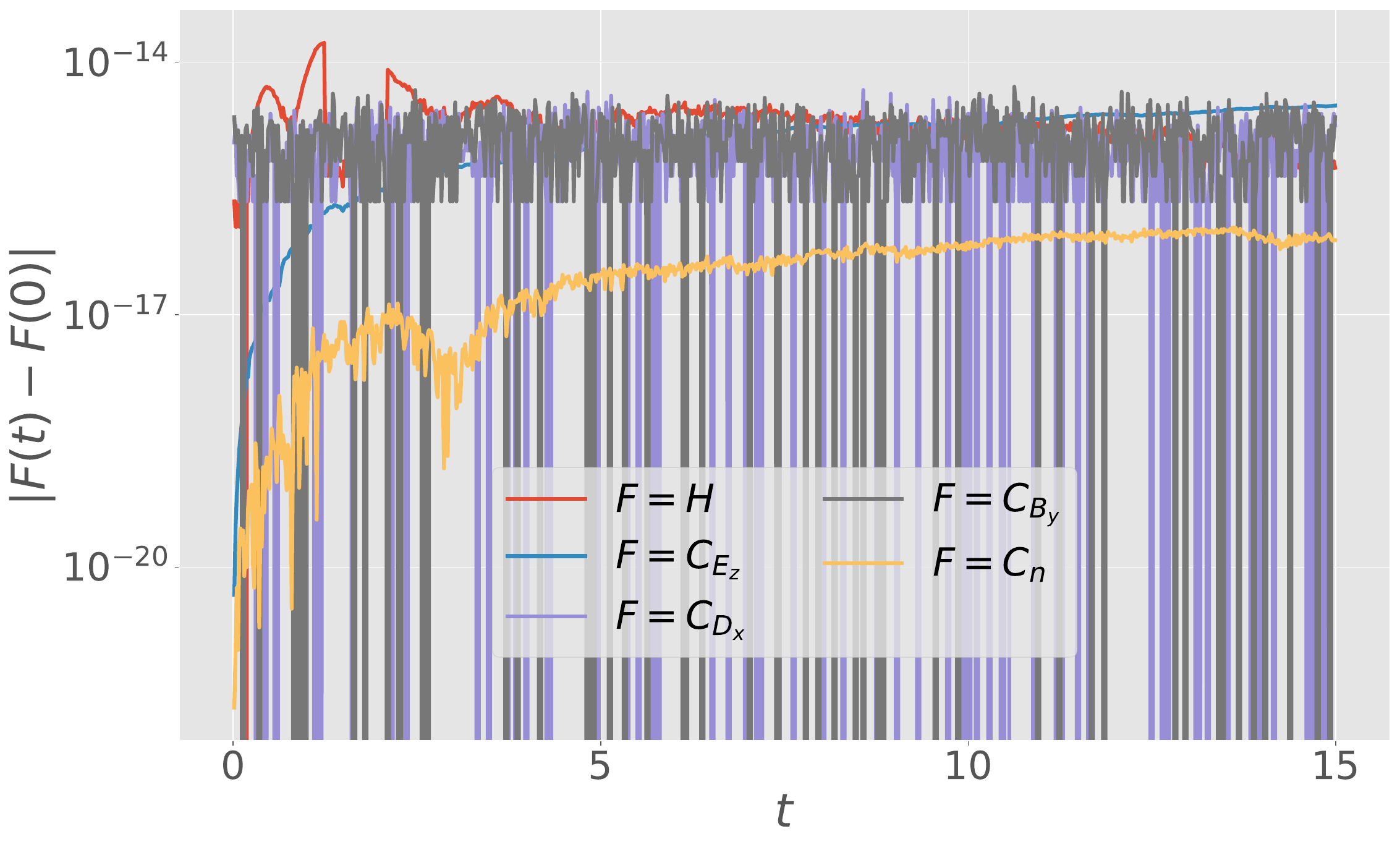}  	
		\end{minipage}\hfill
  		\begin{minipage}{0.47\textwidth}
   			\centering
        			\includegraphics[width=\linewidth]{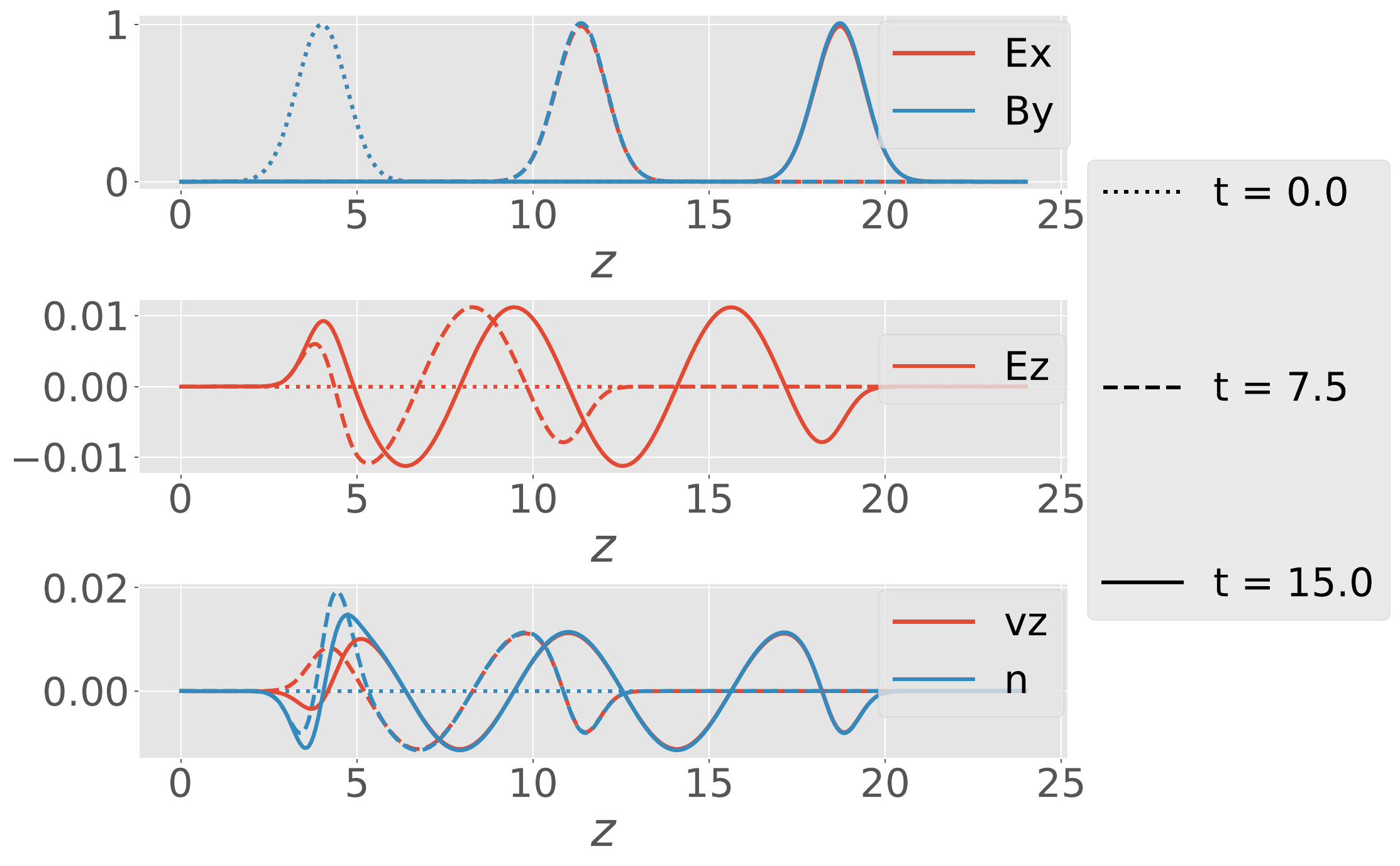}
  		\end{minipage}
		\caption{Conforming-FEEC; Temporal order: $O(\Delta t^6)$; Spatial order: $6^{th}$ for $0$-forms and $5^{th}$ for $1$-forms}
	\end{subfigure}

	\begin{subfigure}{\textwidth}
    		\begin{minipage}{0.47\textwidth}
   			\centering
			\includegraphics[width=\linewidth]{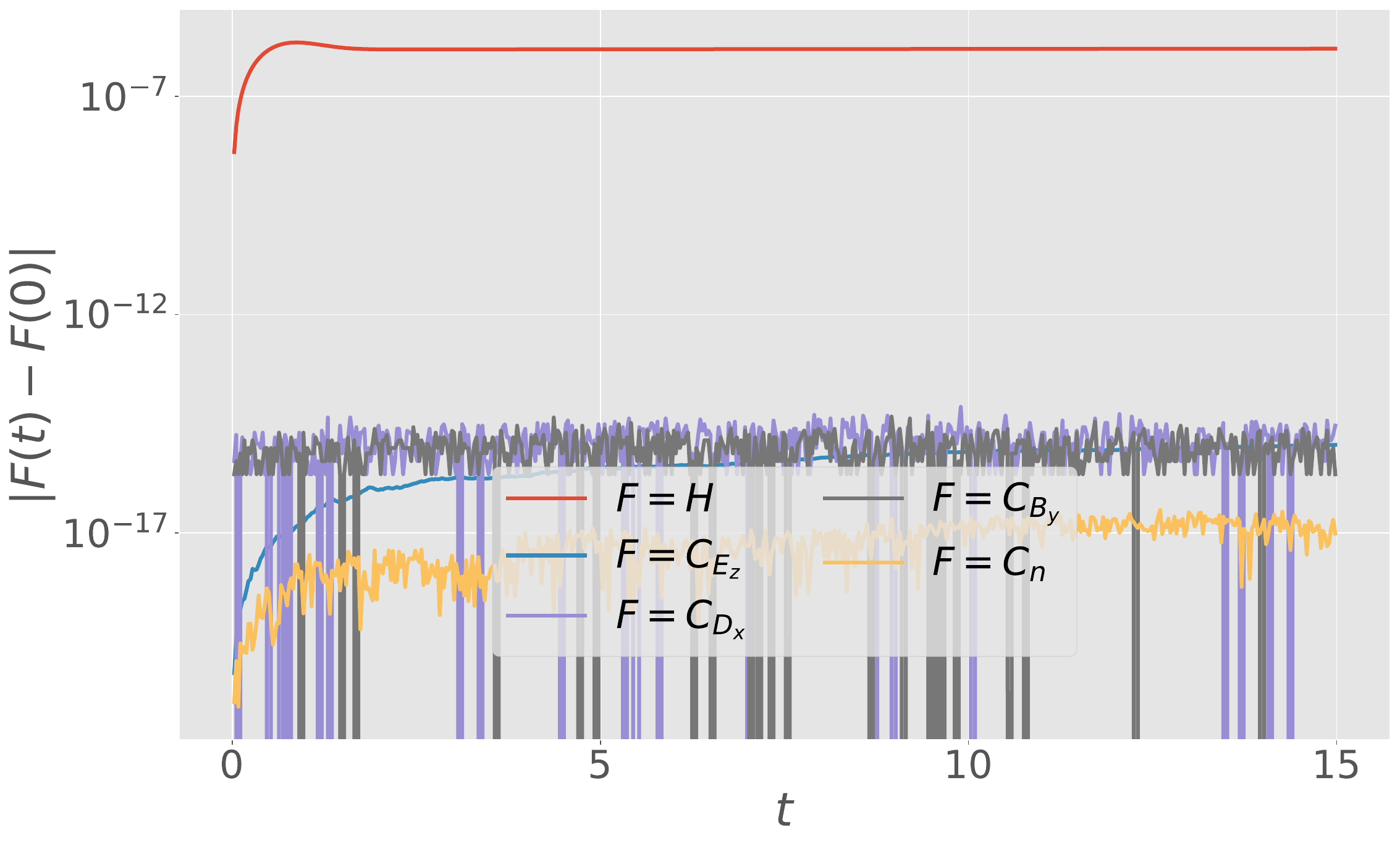}  	
		\end{minipage}\hfill
  		\begin{minipage}{0.47\textwidth}
   			\centering
        			\includegraphics[width=\linewidth]{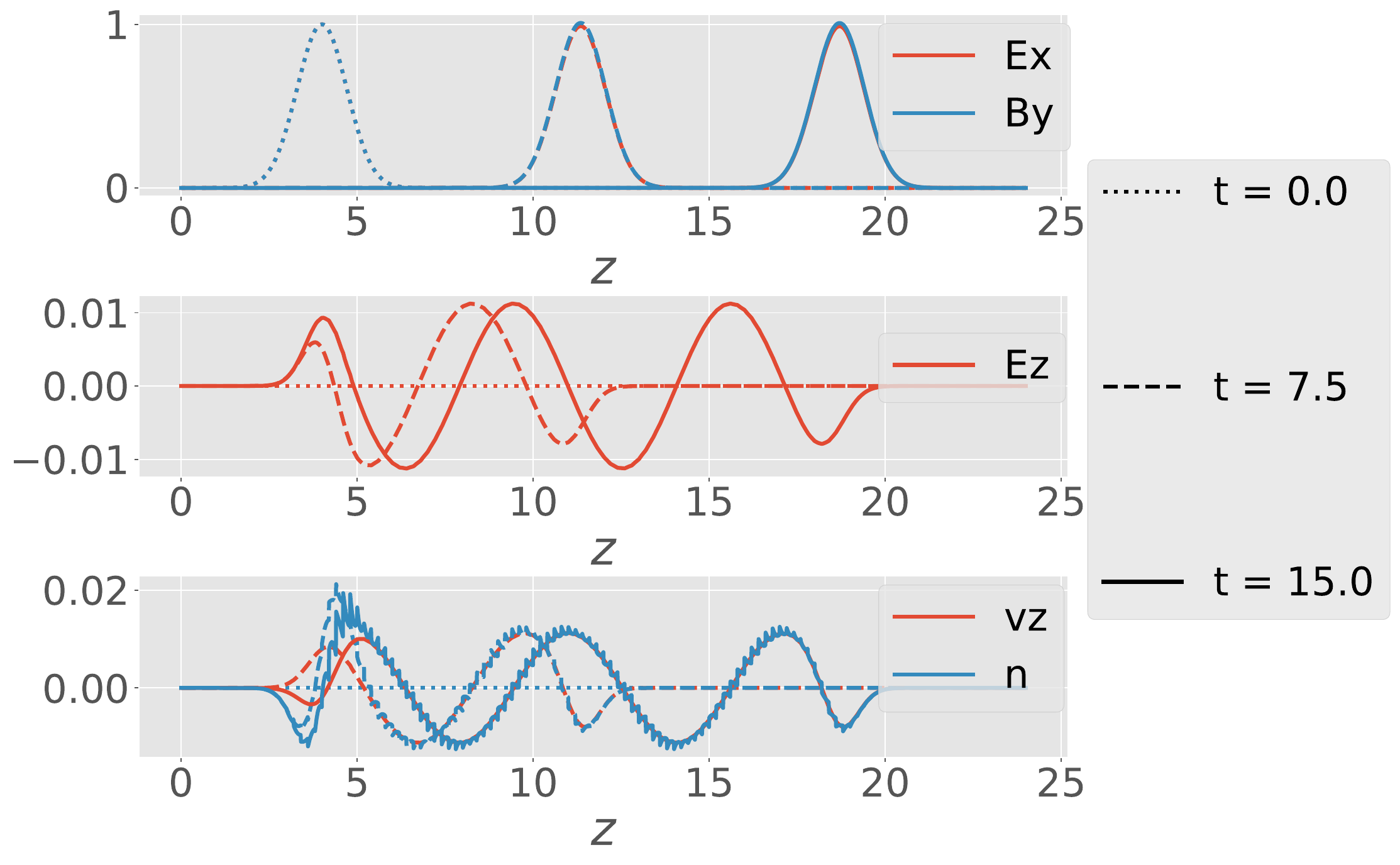}
  		\end{minipage}
		\caption{Broken-FEEC; Temporal order: $O(\Delta t^2)$; Spatial order: $3^{rd}$ for $0$-forms and $2^{nd}$ for $1$-forms}
	\end{subfigure}
	
	\begin{subfigure}{\textwidth}
    		\begin{minipage}{0.47\textwidth}
   			\centering
			\includegraphics[width=\linewidth]{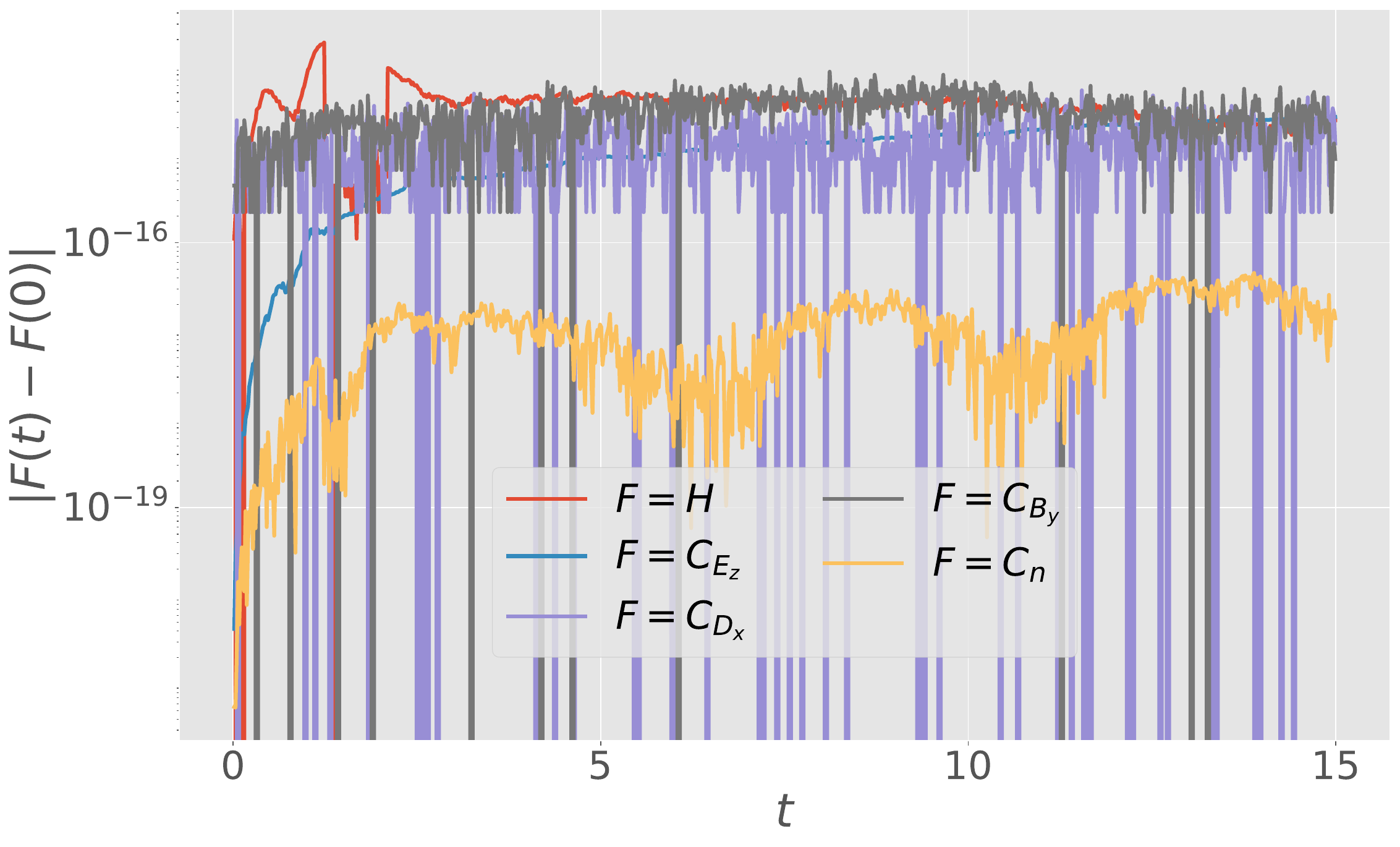}  	
		\end{minipage}\hfill
  		\begin{minipage}{0.47\textwidth}
   			\centering
        			\includegraphics[width=\linewidth]{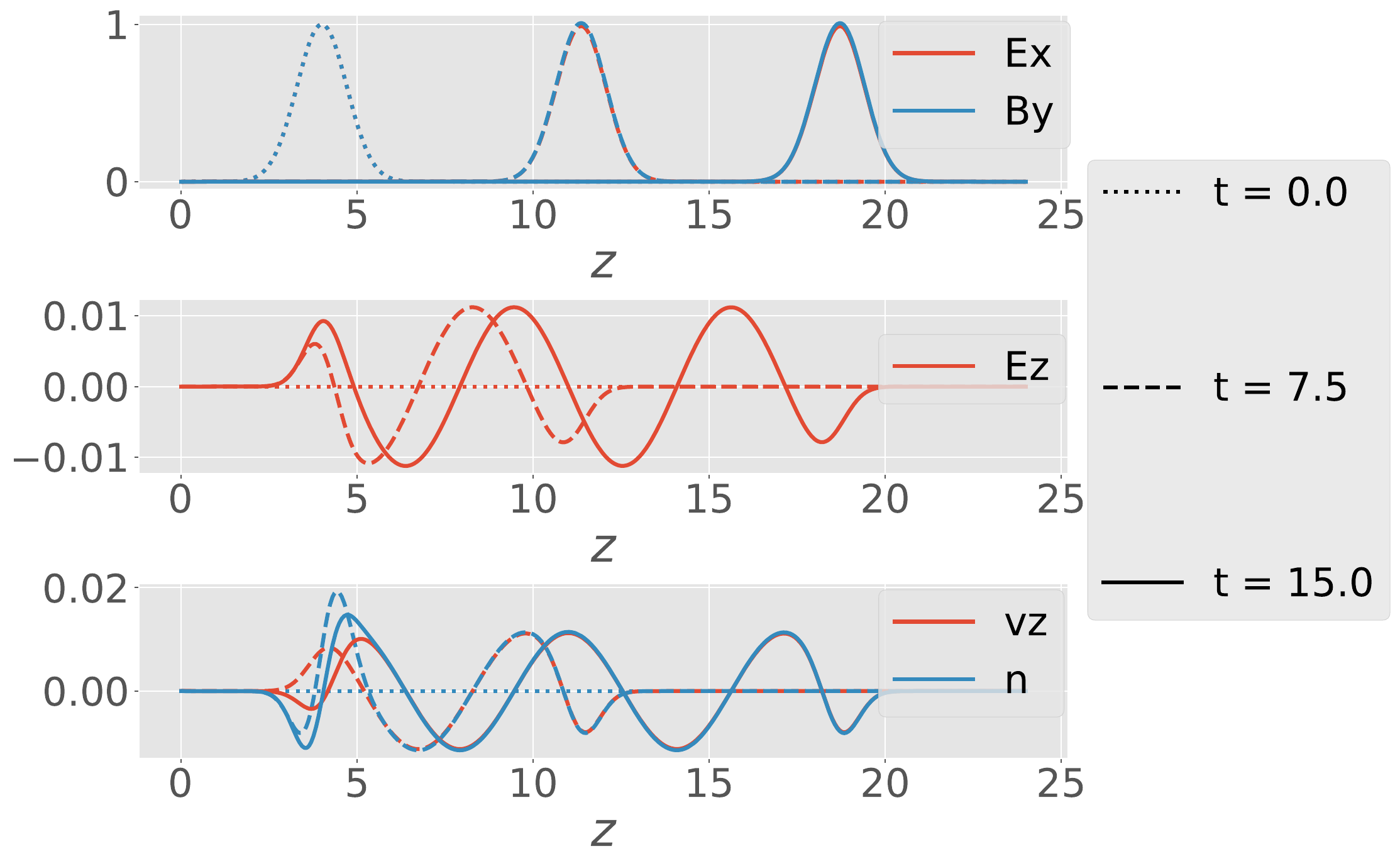}
  		\end{minipage}
		\caption{Broken-FEEC; Temporal order: $O(\Delta t^6)$; Spatial order: $6^{th}$ for $0$-forms and $5^{th}$ for $1$-forms}
	\end{subfigure}
    
    	\caption{Weakly nonlinear regime: $(\omega_p/\omega_0, \omega_c/\omega_0) = (1/5,-1/5)$. The left and right columns plot the conservation law errors and the solution respectively. \ref{sec:disc_cons_laws} defines the discrete Casimir invariants. (a) and (b) are the conforming low and high order methods respectively; (c) and (d) are the low and high order broken methods respectively.}
	\label{fig:weakNL}
	
\end{figure*}

Fig.~\ref{fig:weakNL} summarizes the results of the numerical experiments in this weakly nonlinear regime. The first column plots the error in the energy and Casimir invariants as a function of time. One can see that in each case, the Casimir invariants are conserved up to machine precision while the error in the energy does not grow and has magnitude determined by the order of the splitting method. Symplectic integrators do not exactly conserve the Hamiltonian of the continuous system, but rather that of another system which is a perturbation of the original system. The relationship between the true and perturbed systems may be determined with a technique known as backward error analysis \cite{geometric_numerical_integration}. As a result of this property of symplectic integrators, the computed value of the Hamiltonian oscillates with finite amplitude about the initial value of the energy (the amplitude of this oscillation decreases as the order of the method increases). With second order splitting, the error in the Hamiltonian is found to be $\sim 10^{-7}$. With sixth order splitting, the Hamiltonian is conserved very near machine precision likely because the amplitude of the oscillation is smaller than machine precision. 

The second column shows the solution at three consecutive times. One can see that each solution qualitatively agrees in the behavior of the system, but that the low-order broken-FEEC scheme exhibits spurious oscillations. In general, it was found that the broken-FEEC scheme required higher resolution to suppress these oscillations. The solution behaves as one might anticipate. At $t=7.5$ in the second column of Fig.~\ref{fig:weakNL}, one can see that the transverse field has induced a ponderomotive force pointing in the longitudinal direction and that this force induced a charge separation in the longitudinal direction which in turn excited a longitudinal electrostatic field. Notice, in the first column of the same figures, this longitudinal field satisfies an exact charge conservation law due to a Casimir invariant of the discrete system. The longitudinal field is relatively low amplitude because of the weakness of the nonlinearity in this system. Moreover, the transverse fields remain largely undeformed because the nonlinear polarization is so weak.

\subsubsection{Strongly nonlinear regime} \label{sec:nonlinear_regime}
We now consider a strongly nonlinear regime in which $(\omega_p/\omega_0, \omega_c/\omega_0) = (2/5, - 9/10)$. These parameters are somewhat larger than the asymptotic derivation allows, however we include these parameters in this numerical study because they more strongly activate the nonlinearities in the model. This serves as a stiffer test for the numerical method and allows us to examine its conservative properties in a more interesting context. 

\begin{figure*}
	\centering
	
	\begin{subfigure}{\textwidth}
    		\begin{minipage}{0.47\textwidth}
   			\centering
			\includegraphics[width=\linewidth]{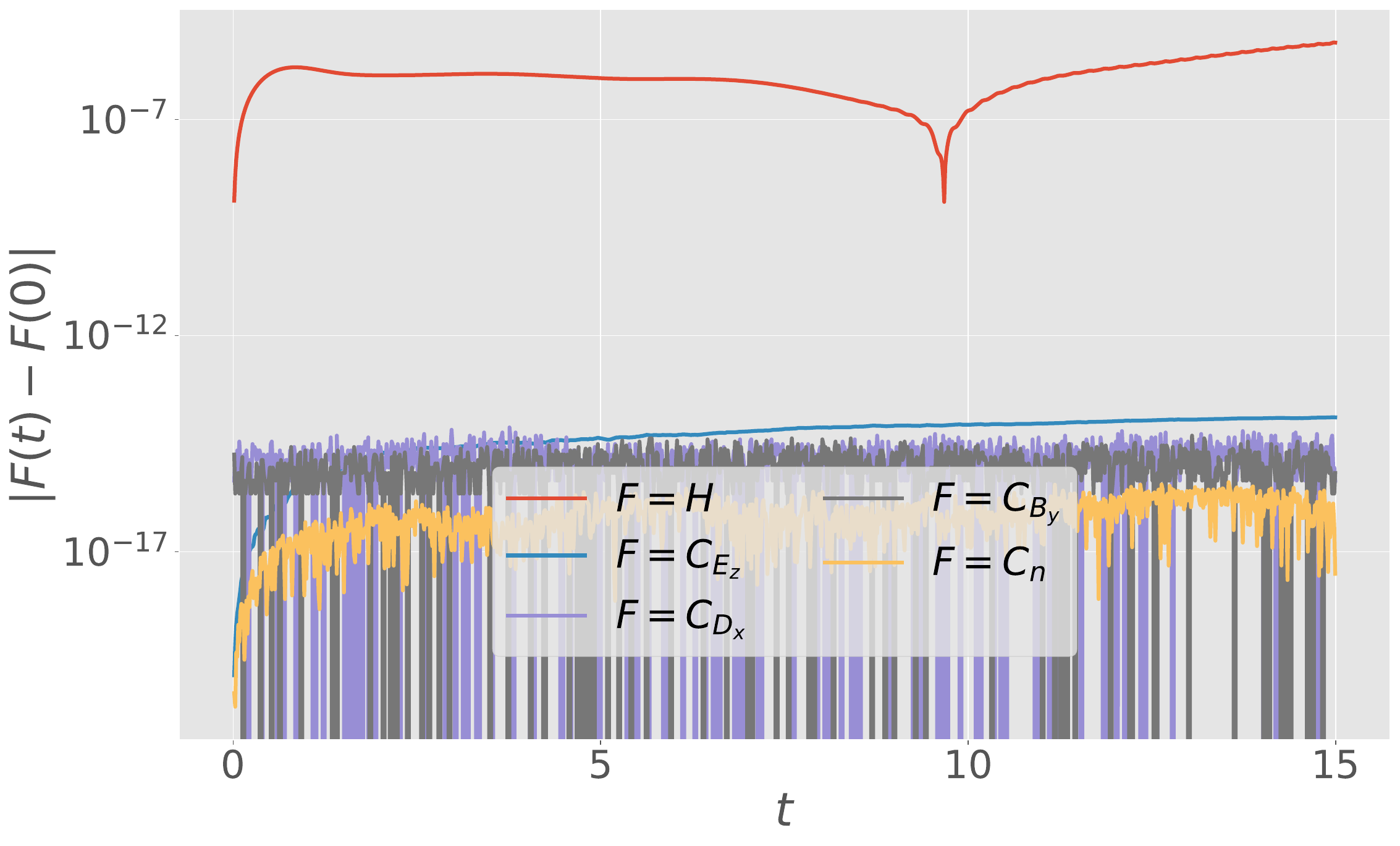}  	
		\end{minipage}\hfill
  		\begin{minipage}{0.47\textwidth}
   			\centering
        			\includegraphics[width=\linewidth]{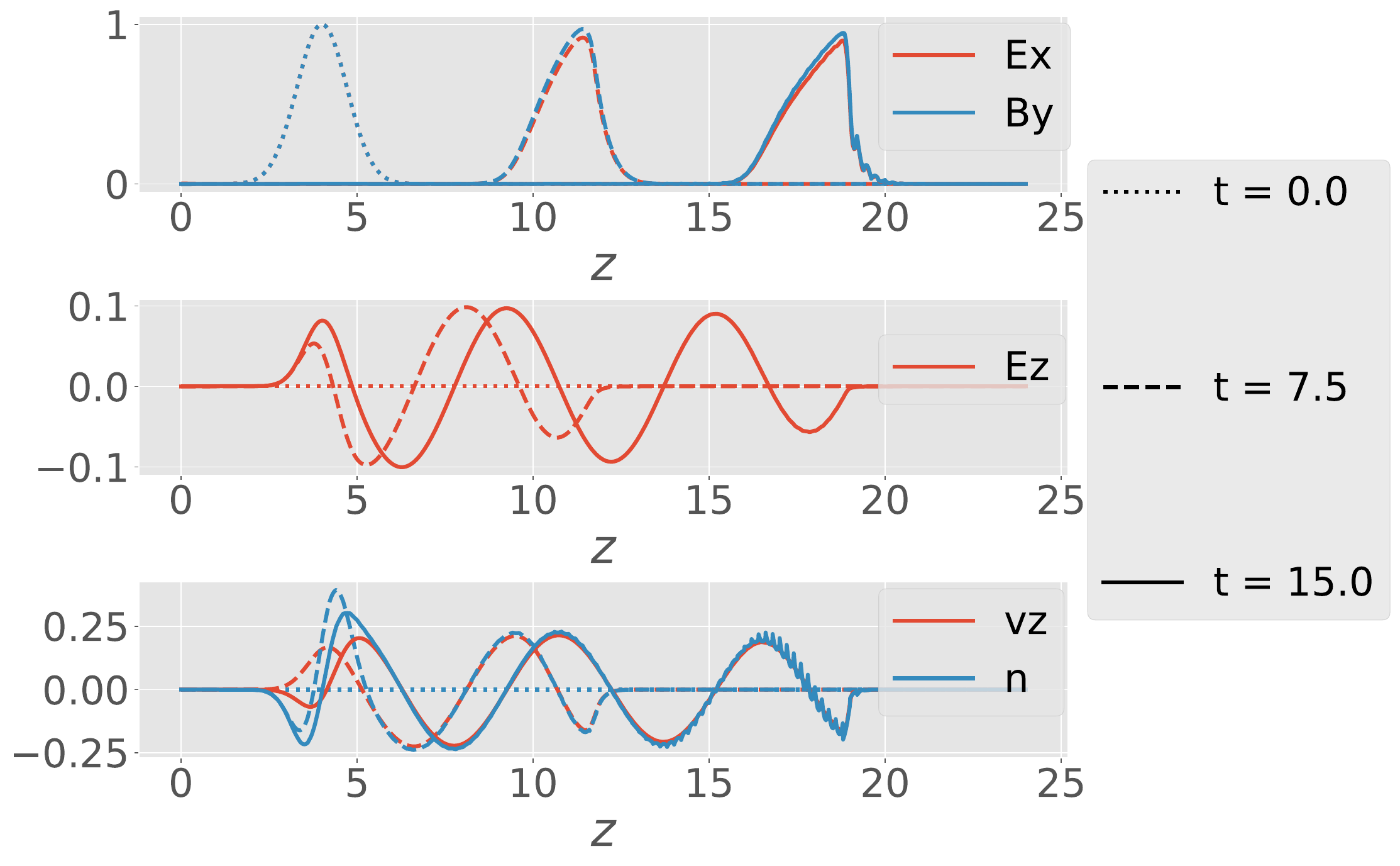}
  		\end{minipage}
		\caption{Conforming-FEEC; Temporal order: $O(\Delta t^2)$; Spatial order: $3^{rd}$ for $0$-forms and $2^{nd}$ for $1$-forms}
	\end{subfigure}
	
	\begin{subfigure}{\textwidth}
    		\begin{minipage}{0.47\textwidth}
   			\centering
			\includegraphics[width=\linewidth]{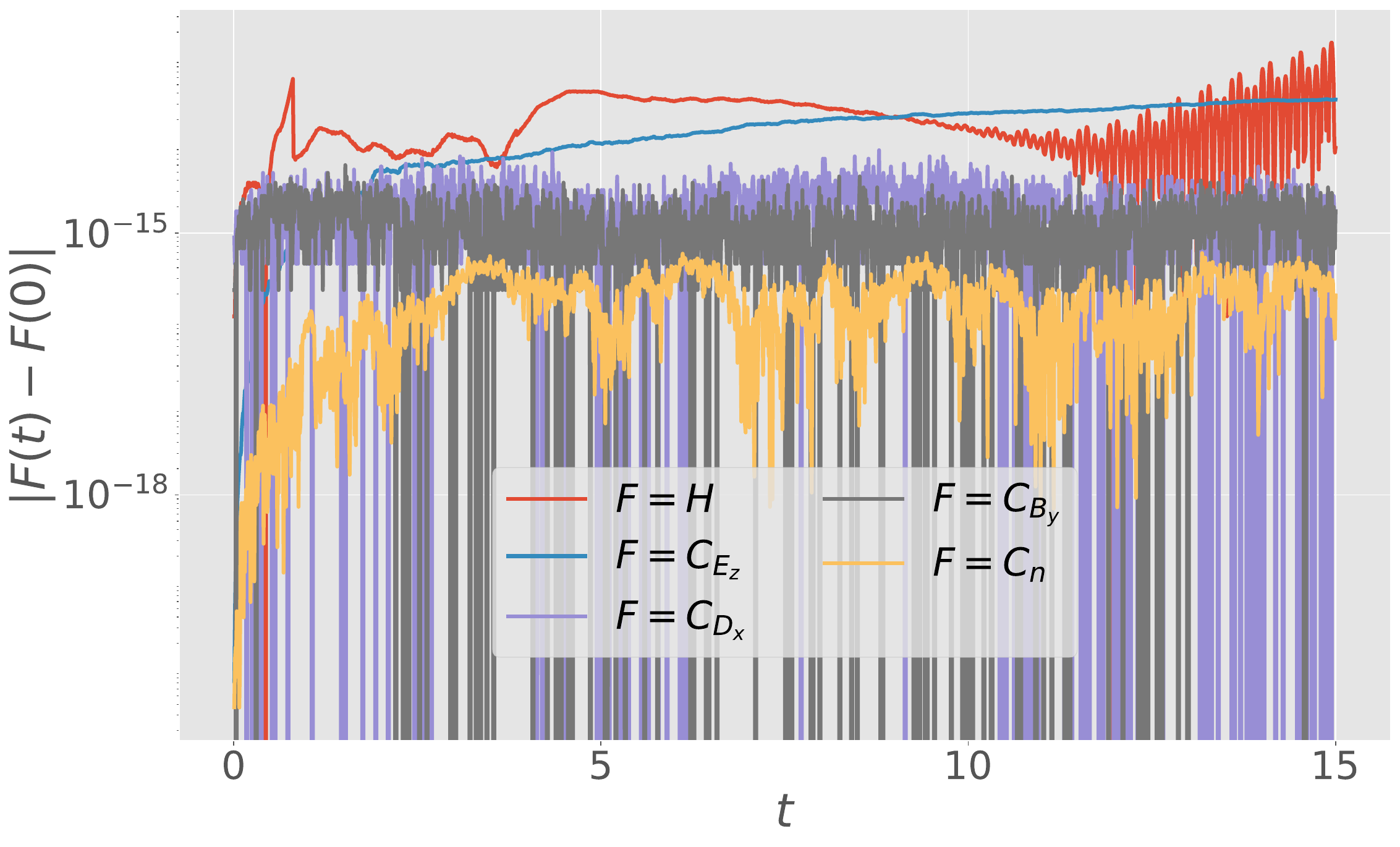}  	
		\end{minipage}\hfill
  		\begin{minipage}{0.47\textwidth}
   			\centering
        			\includegraphics[width=\linewidth]{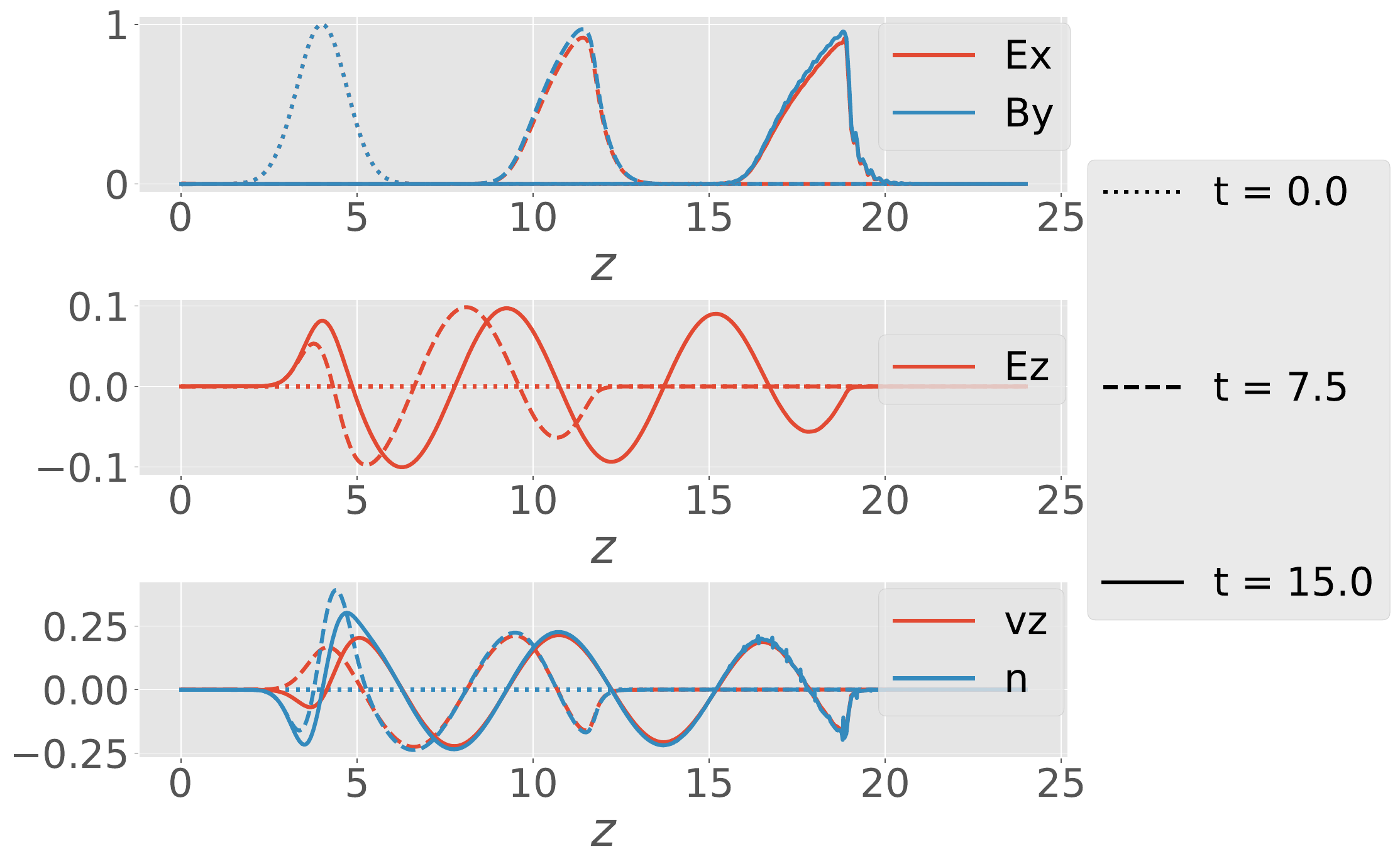}
  		\end{minipage}
		\caption{Conforming-FEEC; Temporal order: $O(\Delta t^6)$; Spatial order: $6^{th}$ for $0$-forms and $5^{th}$ for $1$-forms}
	\end{subfigure}

	\begin{subfigure}{\textwidth}
    		\begin{minipage}{0.47\textwidth}
   			\centering
			\includegraphics[width=\linewidth]{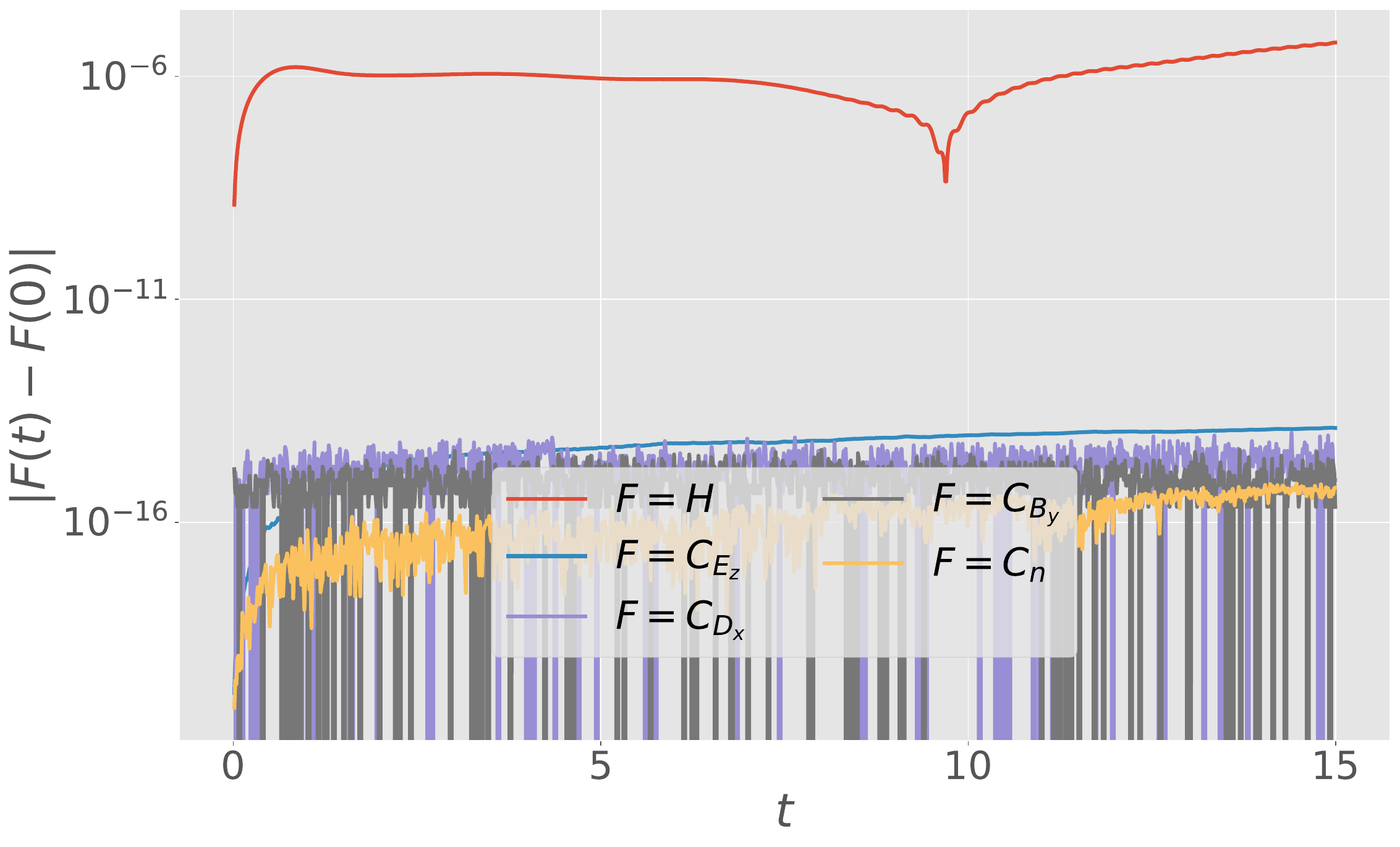}  	
		\end{minipage}\hfill
  		\begin{minipage}{0.47\textwidth}
   			\centering
        			\includegraphics[width=\linewidth]{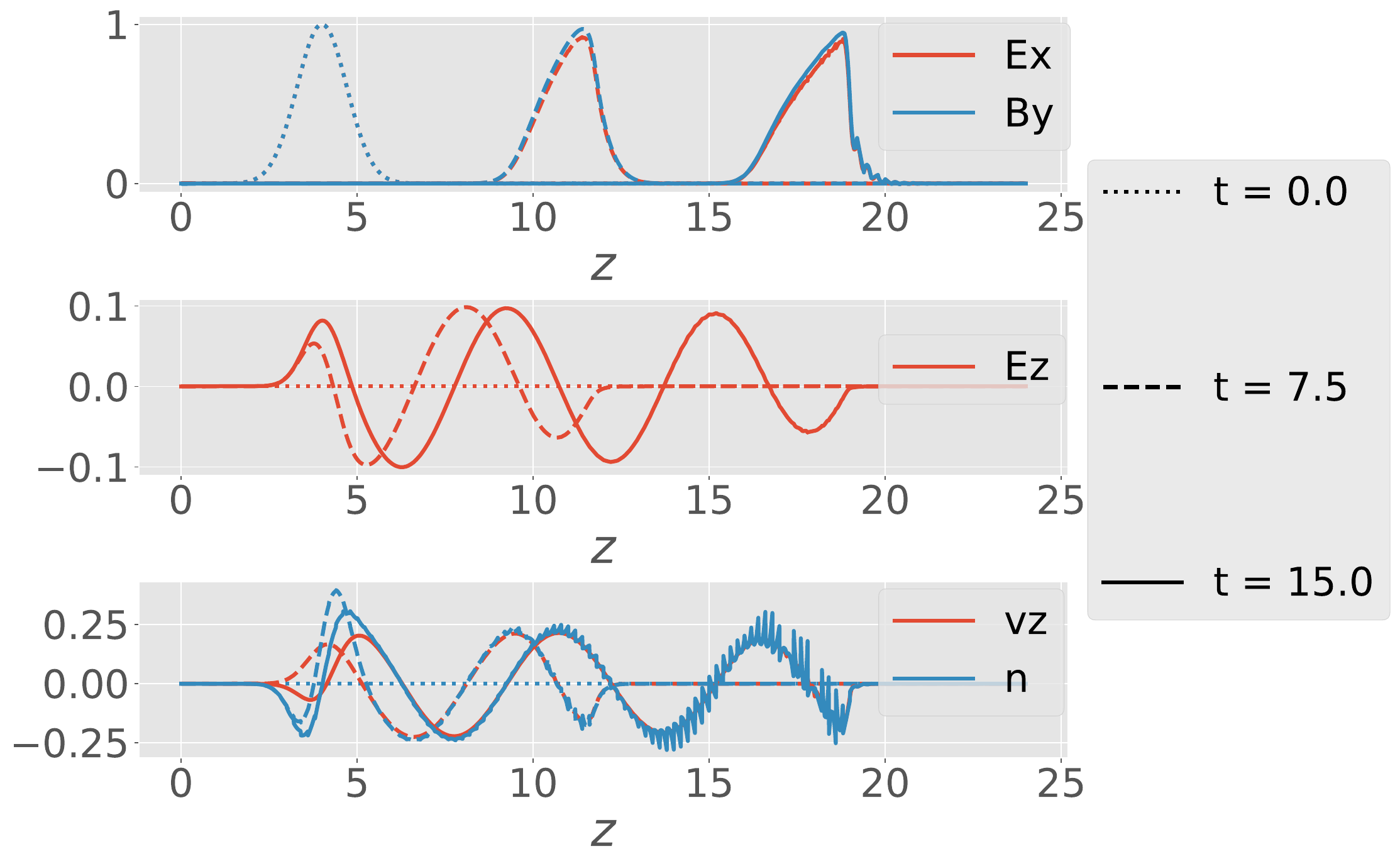}
  		\end{minipage}
		\caption{Broken-FEEC; Temporal order: $O(\Delta t^2)$; Spatial order: $3^{rd}$ for $0$-forms and $2^{nd}$ for $1$-forms}
	\end{subfigure}
	
	\begin{subfigure}{\textwidth}
    		\begin{minipage}{0.47\textwidth}
   			\centering
			\includegraphics[width=\linewidth]{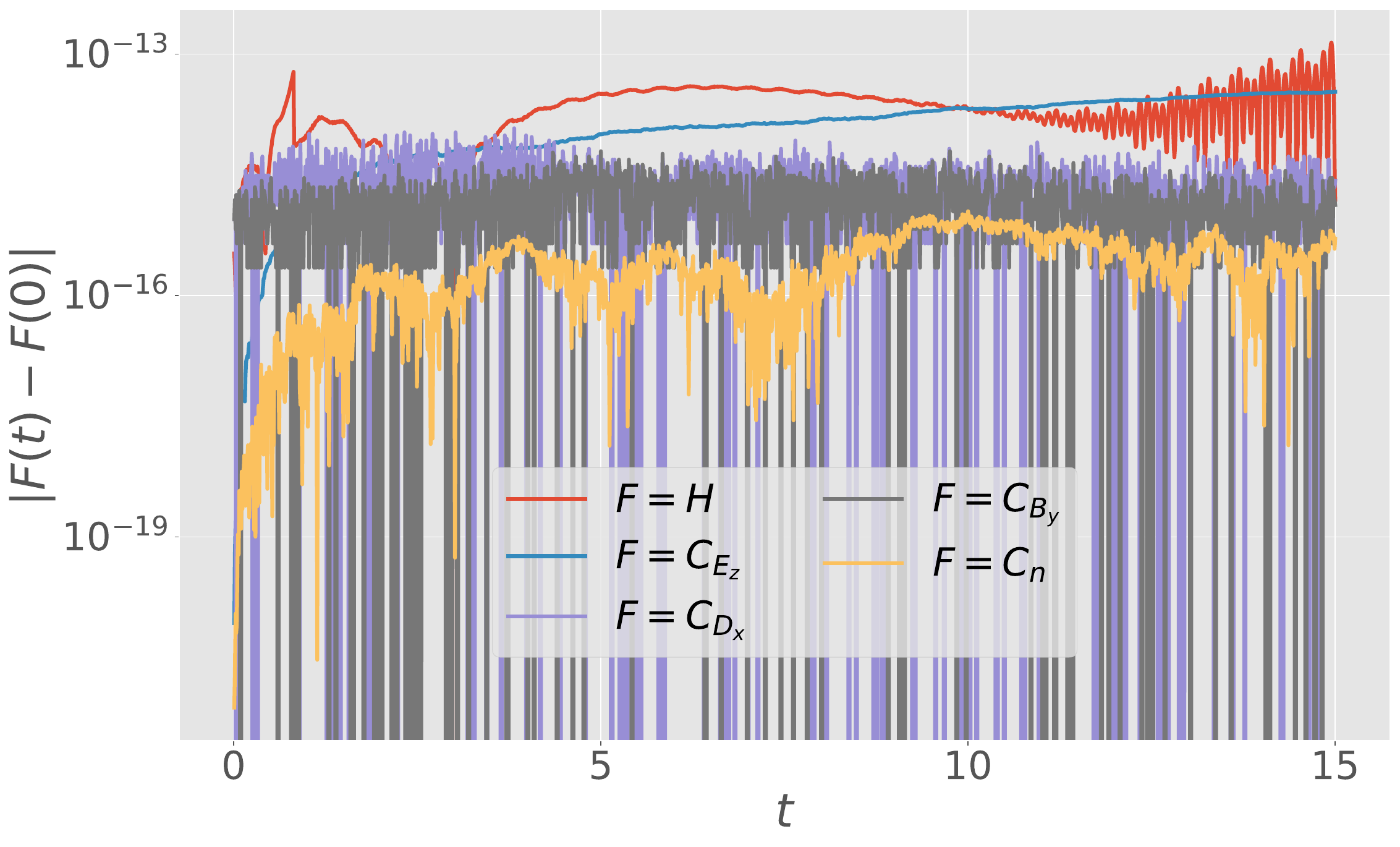}  	
		\end{minipage}\hfill
  		\begin{minipage}{0.47\textwidth}
   			\centering
        			\includegraphics[width=\linewidth]{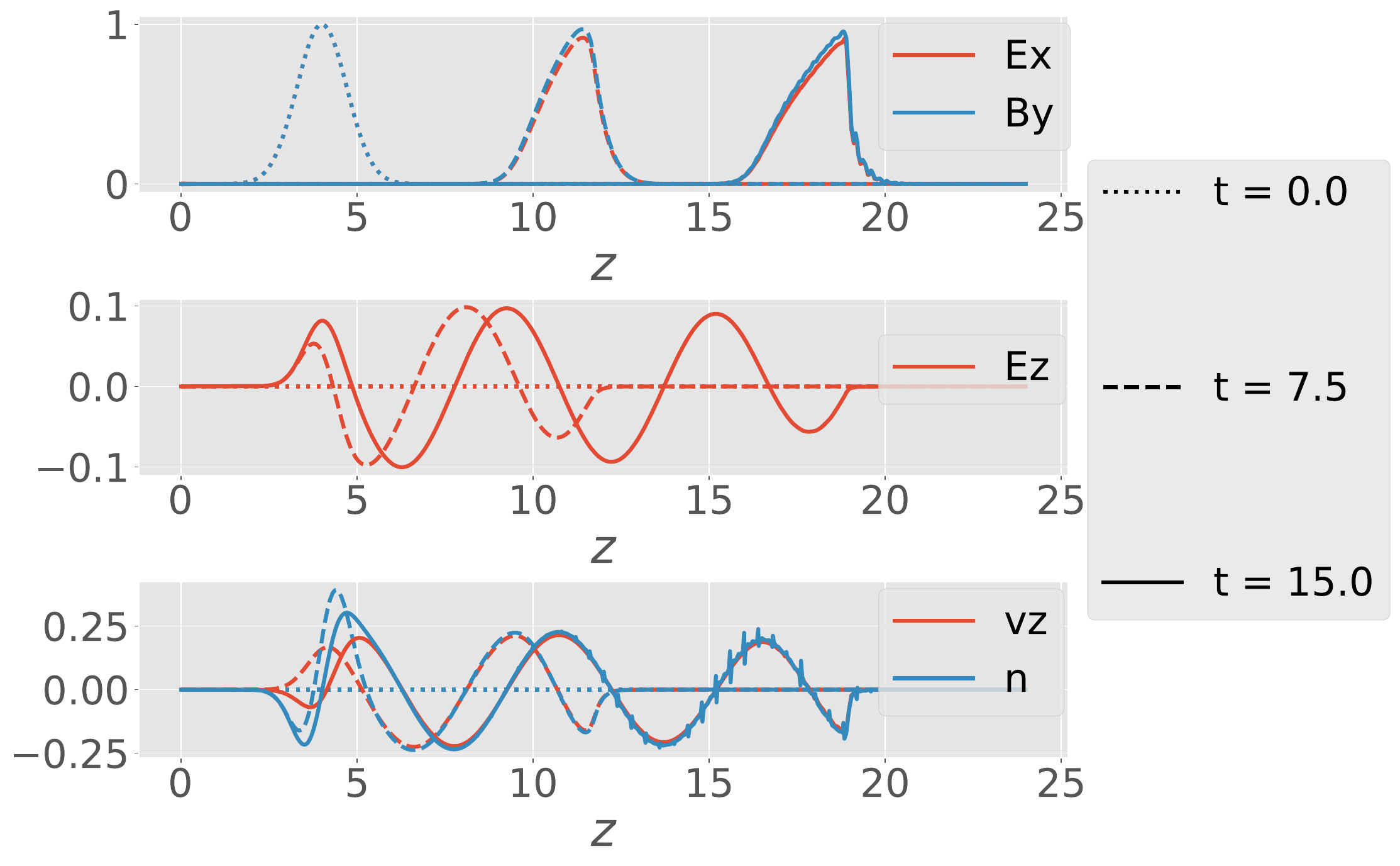}
  		\end{minipage}
		\caption{Broken-FEEC; Temporal order: $O(\Delta t^6)$; Spatial order: $6^{th}$ for $0$-forms and $5^{th}$ for $1$-forms}
	\end{subfigure}
    
    	\caption{Strongly nonlinear regime: $(\omega_p/\omega_0, \omega_c/\omega_0) = (2/5, -9/10)$. The left and right columns plot the conservation law errors and the solution respectively. \ref{sec:disc_cons_laws} defines the discrete Casimir invariants. (a) and (b) are the conforming low and high order methods respectively; (c) and (d) are the low and high order broken methods respectively.}
	\label{fig:strongNL}
	
\end{figure*}

Fig.~\ref{fig:strongNL} summarizes the results of the numerical experiments in this strongly nonlinear regime. Once again, the error in the conservation laws is in the first column while the computed solutions are shown in the second column. One can see that in each case, all Casimir invariants are conserved up to machine precision. The charge conservation Casimir invariant $C_{E_z}$ very slowly grows in magnitude, possibly due to the round-off errors incurred by the nonlinear solver in each step (which has been set to $10^{-13}$). Despite this slow drift, the charge conservation error remains near machine precision through the whole simulation. The Hamiltonian remains well conserved (the magnitude of the error being determined by the order of the splitting method) until a shock begins forming, discussed below, around $t=9$ causing spurious oscillations.
After this time, the error in the Hamiltonian begins growing presumably because of increased roundoff error as the regularity of the solution breaks down. 

In the strongly nonlinear regime, one can see that the transverse wave steepens over time, see the second column of Fig.~\ref{fig:strongNL}. This eventually leads to the formation of a shock which our scheme does not adequately capture. This is because high order polynomial interpolation exhibits Runge's phenomenon when gradients become too large. Hence, in this strongly nonlinear regime, a lower order method with a larger number of cells yields better results, see section \ref{sec:fd_method}. While the solution does exhibit spurious oscillations, it continues to conserve the invariants of motion to high precision. This issue serves as a reminder that conservative integrators, while beneficial in their ability to preserve the invariants of motion, are not immune to the challenges and failures of standard numerical methods. In addition to increasing grid resolution in regions where the gradient is large, for example by using an adaptive mesh, there are various remedies one might try in order to better resolve the shocks, i.e.~shock capturing methods, but how such techniques might be integrated into a structure preserving framework is beyond the scope of this paper. 

\subsubsection{Sensitivity to tolerance of nonlinear solver}
Each time the field $\bmsf{D}_x$ is updated (in the $\msf{H}_{Bv}$ sub-step when one calls $\texttt{updateEN}$), a new value of $\bmsf{E}_x$ must be computed using an iterative nonlinear solver. With infinite precision arithmetic, the time advance of the entire system would be a Poisson integrator exactly conserving the Casimir invariants by keeping the solution on the proper symplectic leaf and exactly conserving the symplectic two-form on that leaf \cite{geometric_numerical_integration}. However, the need to perform a nonlinear solve after each $\msf{H}_{Bv}$ sub-step incurs an error which takes the solution off of the symplectic leaf. Therefore, the algorithm is not truly a Poisson integrator, but only approximately so. In this section, we examine the sensitivity of the conservation of the Casimir invariants to the specified tolerance of the nonlinear solver. 

\begin{figure*}
	\centering
	
	\begin{minipage}{0.49\textwidth}
   			\centering
			\includegraphics[width=\linewidth]{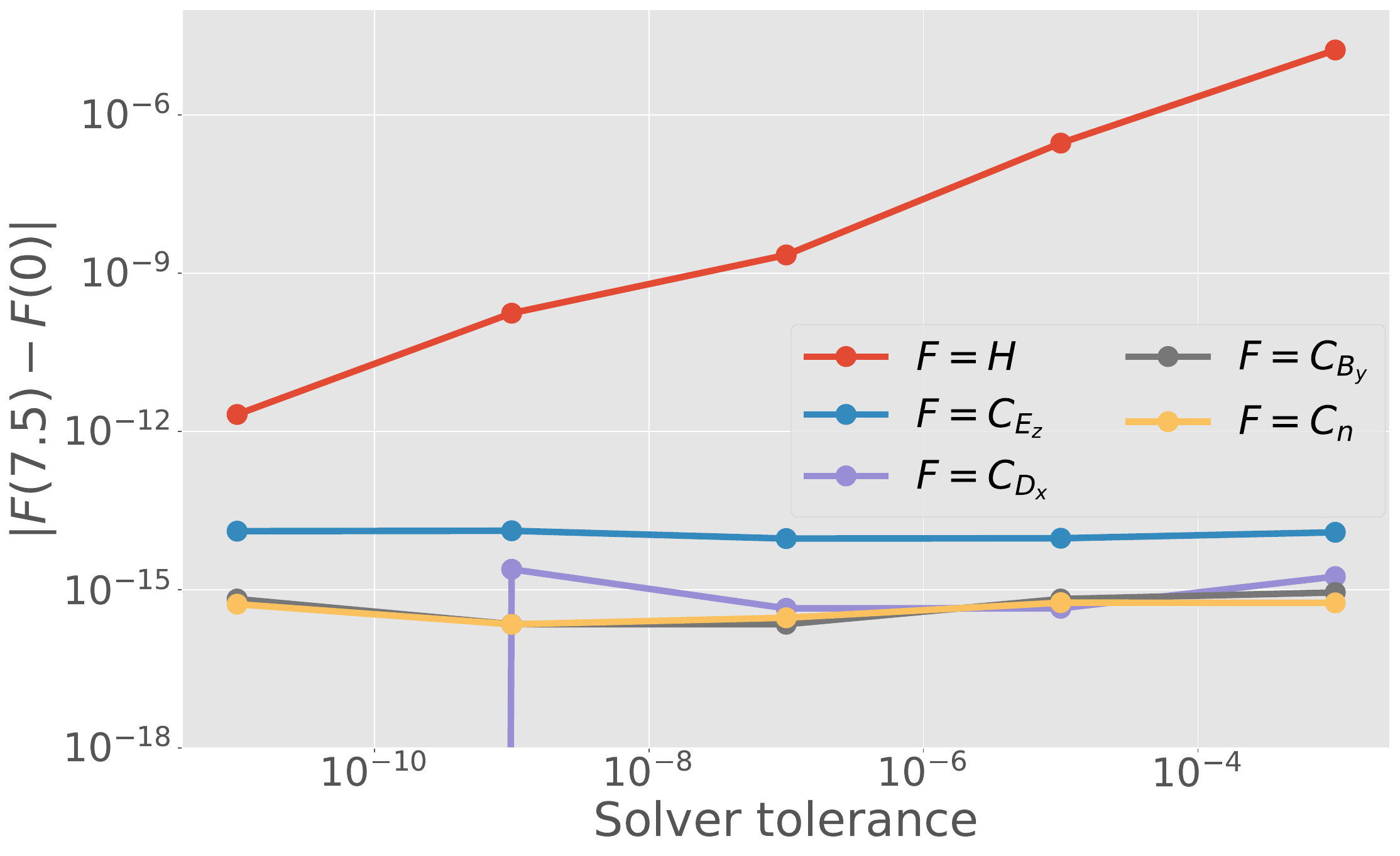}  	
		\end{minipage}\hfill
  		\begin{minipage}{0.49\textwidth}
   			\centering
        			\includegraphics[width=\linewidth]{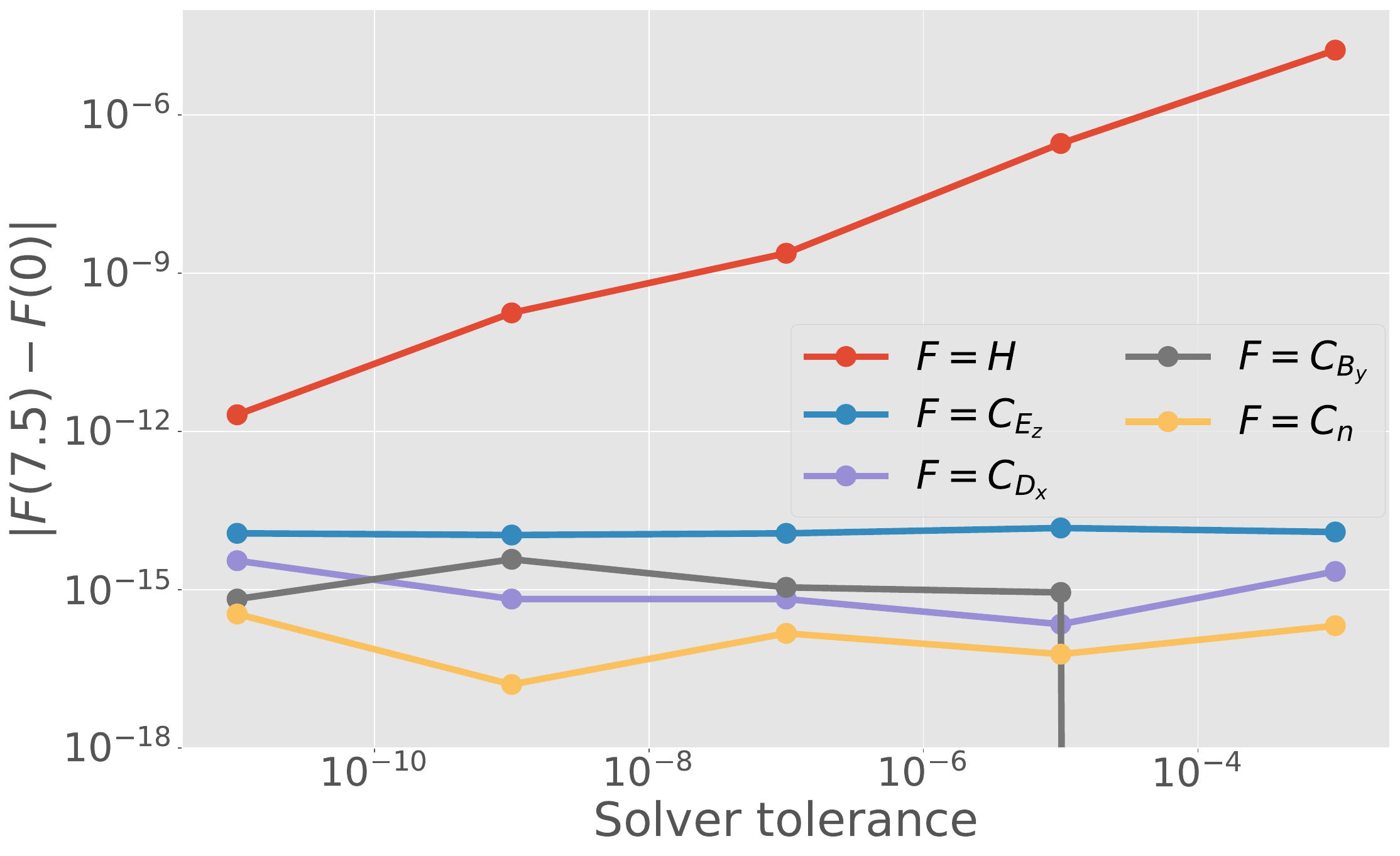}
  	\end{minipage}
	\caption{Numerical test of the sensitivity of discrete conservation laws to nonlinear solver tolerance: conforming-FEEC on the left and broken-FEEC on the right. See section \ref{sec:disc_cons_laws} for definition of the discrete Casimir invariants.}
	\label{fig:sensitivity_test}
\end{figure*}

In order to see the effects of the tolerance of the nonlinear solver in isolation, we run high-resolution ($N=6$, $K=80$ cells, and $6^{th}$-order splitting), conforming and broken FEEC tests in the strongly nonlinear regime with the same parameters as before, but only up to $t=7.5$ to avoid pollution of the results from the known issue the solvers face when a shock forms. In both the conforming and broken FEEC cases, $dt \approx 0.021$. In the conforming case, both the $0$-forms and $1$-forms have about $480$ DOFs while in the broken FEEC case, the $0$-forms have $560$ DOFs (the $1$-forms are the same in conforming and broken FEEC). The results for the weakly nonlinear regime are omitted because conservation in the strongly nonlinear regime is a much stiffer test. See Fig.~\ref{fig:sensitivity_test} for the results. One can see that the Casimirs are conserved up to machine precision regardless of the precision of the nonlinear solver. However, conservation of the Hamiltonian is dependent on the tolerance of the nonlinear solver. 

\subsection{Hamiltonian finite difference/lowest order conforming-FEEC scheme} \label{sec:fd_method}
As a point of comparison with the FEEC spectral element methods previously discussed, it is worth briefly considering how a similar Hamiltonian finite difference scheme might be derived. We compare the finite element approximation with a Hamiltonian structure preserving finite difference approximation because the obvious finite difference scheme obtained from replacing all derivatives in equations (\ref{eq:ponder_maxwell_1})-(\ref{eq:ponder_maxwell_3}) with first order finite difference stencils results in an unstable approximation. We find that the Hamiltonian structure preserving finite difference approximation is a special case of finite element method in which one uses uniform grids and lowest order interpolating polynomials. Despite this, it is still instructive to show how the method might be directly derived as a finite difference approximation. 

Suppose the domain is subdivided into an equispaced grid $\{ z_i \}_{i=1}^N$. The Hamiltonian is discretized via trapezoidal rule which, on a periodic domain, is simply
\begin{widetext}
\begin{equation}
	\msf{H}_{FD} = \frac{\Delta z}{2} \sum_{i=1}^N \left[ 
		\frac{\omega_p^2}{\omega_c^2} \msf{v}_{z,i}^2 + \msf{E}_{z,i}^2 
			+ \frac{1}{2} \left( \left(1 + \frac{\omega_p^2}{\omega_0^2} \left( 1 + \msf{n}_i 
				- \frac{3 \omega_c^2}{16 \omega_0^2} \msf{E}_{x,i}^2 \right) \right) \msf{E}_{x,i}^2 
					+ \msf{B}_{y,i}^2 \right) \right] \,,
\end{equation}
\end{widetext}
where, if $A(z)$ is a continuous field, $\msf{A}_i = A(z_i)$. We define the relationship between $\bmsf{D}_x$ and $\bmsf{E}_x$ by identifying values at the collocation points:
\begin{equation} \label{eq:const_law_dx}
	\msf{D}_{x,i} = \left[ 1 + \frac{\omega_p^2}{\omega_0^2} 
		\left( 1 + \msf{n}_i - \frac{\omega_c^2}{8 \omega_0^2}  \left( \msf{E}_{x,i} \right)^2 \right) \right] \msf{E}_{x,i}.
\end{equation}
Finally, taking derivatives of the Hamiltonian proceeds much as in the case of the finite element method. Letting $\overline{H}(\bmsf{D}_x, \bmsf{B}_y, \bmsf{v}_z, \bmsf{n}, \bmsf{E}_z) = \bmsf{H}(\bmsf{E}_x, \bmsf{B}_y, \bmsf{v}_z, \bmsf{n}, \bmsf{E}_z)$, we obtain
\begin{multline}
    \frac{\partial \overline{\msf{H}}_{FD}}{\partial \msf{D}_{x,i}} = \frac{\Delta z \msf{E}_{x,i}}{2},
	\quad
    \frac{\partial \overline{\msf{H}}_{FD}}{\partial \msf{B}_{y,i}} = \frac{\Delta z \msf{B}_{y,i}}{2}, \\
    \frac{\partial \overline{\msf{H}}_{FD}}{\partial \msf{E}_{z,i}} = \Delta z \msf{E}_{z,i}, 
        \quad
    \frac{\partial \overline{\msf{H}}_{FD}}{\partial \msf{v}_{z,i}} = \Delta z \frac{\omega_p^2}{\omega_c^2} \msf{v}_{z,i}, \\
    \text{and} \quad
    \frac{\partial \overline{\msf{H}}_{FD}}{\partial \msf{n}_{i}} = - \Delta z \frac{\omega_p^2}{4 \omega_0^2} \msf{E}_{x,i}^2.
\end{multline}

Define $\mathbbm{d}_{FD}$ to be the circulant matrix with stencil 
\begin{equation}
	(\mathbbm{d}_{FD})_{ij} = 
	\begin{cases}
		1, & j = i+1 \\
		-1, & j = i \\
		0, & \text{else}.
	\end{cases}
	\vspace{0.1em}
\end{equation}
Notice, this derivative matrix is precisely that of the conforming spectral element FEEC method. 

Before discretizing the Poisson bracket, we must briefly explain how functional derivatives discretize in this finite difference setting. Letting $\left. (\delta F/\delta A) \right|_{z = z_i}$ be the functional derivative evaluated at $z = z_i$, we have that
\begin{equation}
	\int_\Omega \frac{\delta F}{\delta A} \delta A \mathsf{d} z \approx \sum_i \left( \frac{\delta F}{\delta A} \right)_{z = z_i} \delta A_i \Delta z = \sum_i \frac{\partial \msf{F}}{\partial A_i} \delta A_i \,,
\end{equation}
where the integral is approximated with trapezoidal rule. Hence, we find
\begin{equation}
	\left( \frac{\delta F}{\delta A} \right)_{z = z_i} = \frac{1}{\Delta z} \frac{\partial \msf{F}}{\partial A_i}.
\end{equation}

The discrete Poisson bracket is then obtained by again replacing the integral with its trapezoidal rule approximation and the derivatives by first order finite differences. One finds
\begin{widetext}
\begin{equation}
	\begin{aligned}
		[ \msf{F}, \msf{G}]_{FD} &=  
                - \frac{\omega_c^2}{\omega_p^2} \sum_{i,j=1}^N \bigg[
				\frac{\partial \msf{F}}{\partial \msf{n}_i}
				(\mathbbm{d}_{FD})_{ij}
				\frac{\partial \msf{G}}{\partial \msf{v}_{z,j}}
				-
				\frac{\partial \msf{G}}{\partial \msf{n}_i}
				(\mathbbm{d}_{FD})_{ij}
				\frac{\partial \msf{F}}{\partial \msf{v}_{z,j}}
				\bigg] \frac{1}{\Delta z^2} 
                - \frac{\omega_c}{\omega_p} \sum_{i=1}^N \bigg[
				\frac{\partial \msf{F}}{\partial \msf{E}_{z,i}}
				\frac{\partial \msf{G}}{\partial \msf{v}_{z,i}}
				- 
				\frac{\partial \msf{G}}{\partial \msf{E}_{z,i}}
				\frac{\partial \msf{F}}{\partial \msf{v}_{z,i}}
				\bigg] \frac{1}{\Delta z} \\
			    &\quad
			- 2 \sum_{i,j=1}^N \left[ \frac{\partial \msf{F}}{\partial \msf{B}_{y,i}}
				(\mathbbm{d}_{FD})_{ij}
				\frac{\partial \msf{G}}{\partial \msf{D}_{x,j}}
				-
				\frac{\partial \msf{G}}{\partial \msf{B}_{y,i}}
				(\mathbbm{d}_{FD})_{ij}
				\frac{\partial \msf{F}}{\partial \msf{D}_{x,j}} \right] \frac{1}{\Delta z^2}.
	\end{aligned}
\end{equation}
\end{widetext}
The scaling of the discrete Poisson bracket by $\Delta z^{-2}$ comes from appropriately treatment of the discrete functional derivatives. However, this scaling may also be understood via an appropriate interpretation of the variables as living on the primal/dual de Rham complex. Scaling a vector by $\Delta z^2$ may be interpreted as applying the diagonal matrix $\Delta z^2 \mathbbm{I}$. The finite difference method in fact is identical to the spectral element FEEC method on a uniform grid using lowest order interpolating polynomials. In that method, the mass matrix for the $1$-forms is precisely $\Delta z^2 \mathbbm{I}$. As the bracket obtained with finite differences is similar to the FEEC spectral element Poisson bracket, it possesses an analogous set of Casimir invariants:
\begin{multline} \label{eq:fd_casimirs}
	C_{D_x} = \mathbbm{1}^T \bmsf{D}_x \Delta z, 
	\quad
	C_{B_y} = \mathbbm{1}^T \bmsf{B}_y \Delta z, 
	\quad
	C_n =  \mathbbm{1}^T \bmsf{n} \Delta z, \\
	\text{and} \quad
	C_{E_z} = \left\| \Delta z^{-1} \mathbbm{d}_{FD} \bmsf{E}_z 
                    - \frac{\omega_p}{\omega_c} \bmsf{n} \right\|_2
\end{multline}
which are the trapezoidal rule approximations of the integrals of the continuous fields (note, $\| \cdot \|_2$ is the vector $2$-norm). 

Using the fact that $\dot{F} = [F,\msf{H}_{FD}]_{FD}$, we obtain the equations of motion:
\begin{equation}
	\begin{split}
		\dot{\bmsf{D}}_{x} &= \Delta z^{-1} \mathbbm{d}_{FD}^T \bmsf{B}_y \\
		\dot{\bmsf{B}}_{y} &= - \Delta z^{-1} \mathbbm{d}_{FD} \bmsf{E}_x \\
		\dot{\bmsf{E}}_{z} &= - \frac{\omega_p}{\omega_c} \bmsf{v}_{z}
	\end{split}
	\hspace{1em}
	\begin{split}
		\dot{\bmsf{v}}_{z} &= \frac{\omega_c}{\omega_p} \bmsf{E}_z 
			- \Delta z^{-1} \frac{\omega_c^2}{4 \omega_0^2} \mathbbm{d}_{FD}^T \left( \bmsf{E}_{x}^2 \right) \\
		\dot{\bmsf{n}} &= - \Delta z^{-1} \mathbbm{d}_{FD} \bmsf{v}_z
	\end{split}
\end{equation}
which are completed with the constitutive relation for $\bmsf{D}_x(\bmsf{E}_x, \bmsf{n})$ in equation \eqref{eq:const_law_dx}. Notice, this is precisely what is obtained from simply applying a finite difference approximation to the spatial derivatives in equations (\ref{eq:ponder_maxwell_1})-(\ref{eq:ponder_maxwell_3}). 

Hence, we found that the Hamiltonian structure preserving spatial discretization using finite differences or lowest order finite elements is simply the finite difference method one might naively guess without all of this extra theoretical machinery. However, in addition to a deeper understanding of the dynamics of the discrete system and its relationship to the continuous system, what is gained from this Hamiltonian perspective is an appropriate operator splitting method to approximate temporal flow. A simple finite difference approximation to the time derivatives yields an unstable method. However, the Hamiltonian is separable and, when appropriately split, we obtain two exactly integrable subsystems:
\begin{multline}
	\begin{cases}
		\dot{\bmsf{D}}_{x} = \Delta z^{-1} \mathbbm{d}_{FD}^T \bmsf{B}_y \\
		\dot{\bmsf{E}}_{z} = - \frac{\omega_p}{\omega_c} \bmsf{v}_{z} \\
		\dot{\bmsf{n}} = - \Delta z^{-1} \mathbbm{d}_{FD} \bmsf{v}_z
	\end{cases}
	\\
        \text{and} \quad
	\begin{cases}
		\dot{\bmsf{B}}_{y} = - \Delta z^{-1} \mathbbm{d}_{FD} \bmsf{E}_x \\
		\dot{\bmsf{v}}_{z} = \frac{\omega_c}{\omega_p} \bmsf{E}_z 
			- \Delta z^{-1} \frac{\omega_c^2}{4 \omega_0^2} \mathbbm{d}_{FD}^T \left( \bmsf{E}_{x}^2 \right)
	\end{cases}
\end{multline}
which possess exact solutions
\begin{multline} \label{eq:updateBV_FD}
    \begin{cases}
        \bmsf{D}_{x}(t + \Delta t) = \bmsf{D}_{x}(t) 
            + \frac{\Delta t}{\Delta z} \mathbbm{d}_{FD}^T \bmsf{B}_y(t) \\
        \bmsf{E}_{z}(t + \Delta t) = \bmsf{E}_{z}(t) - \Delta t \frac{\omega_p}{\omega_c} \bmsf{v}_{z}(t) \\
        \bmsf{n}(t + \Delta t) = \bmsf{n}(t) - \frac{\Delta t}{\Delta z} \mathbbm{d}_{FD} \bmsf{v}_z(t)
    \end{cases}
    \\ 
    \text{and}
    \\
    \begin{cases}
        \bmsf{B}_{y}(t + \Delta t) = \bmsf{B}_{y}(t) - \frac{\Delta t}{\Delta z} \mathbbm{d}_{FD} \bmsf{E}_x(t) \\
        \bmsf{v}_{z}(t + \Delta t) = \bmsf{v}_{z}(t) + \Delta t \frac{\omega_c}{\omega_p} \bmsf{E}_z(t) \\
            \hspace{7em} - \frac{\Delta t}{\Delta z} \frac{\omega_c^2}{4 \omega_0^2} \mathbbm{d}_{FD}^T \left( \bmsf{E}_{x}^2(t) \right).
    \end{cases}
\end{multline}
These may then be composed together as in section \ref{section:temporal_discretization} to obtain approximations of the full flow. 

\begin{figure*}
	\centering
	
	\begin{minipage}{0.49\textwidth}
   			\centering
			\includegraphics[width=\linewidth]{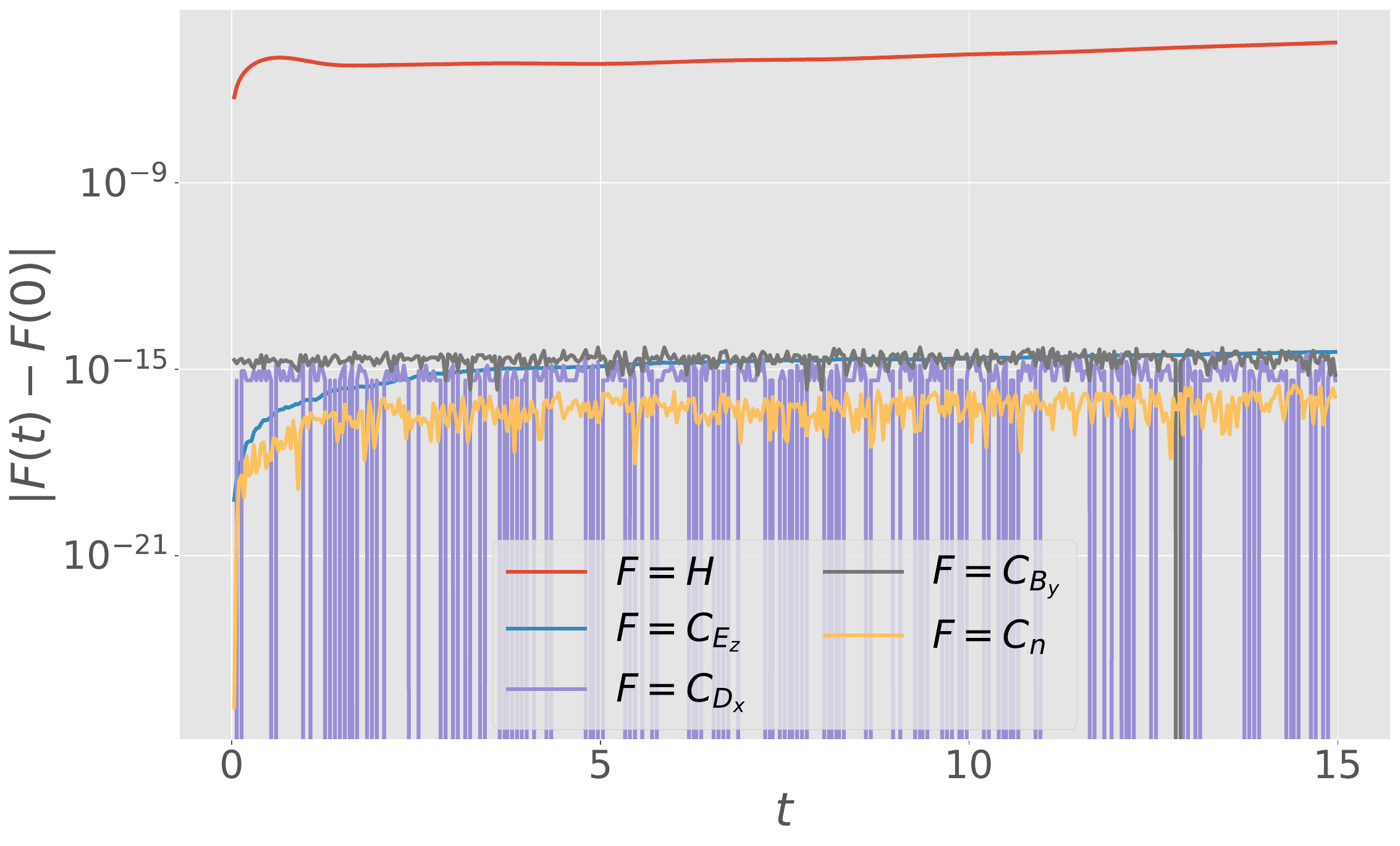}  	
		\end{minipage}\hfill
  		\begin{minipage}{0.49\textwidth}
   			\centering
        			\includegraphics[width=\linewidth]{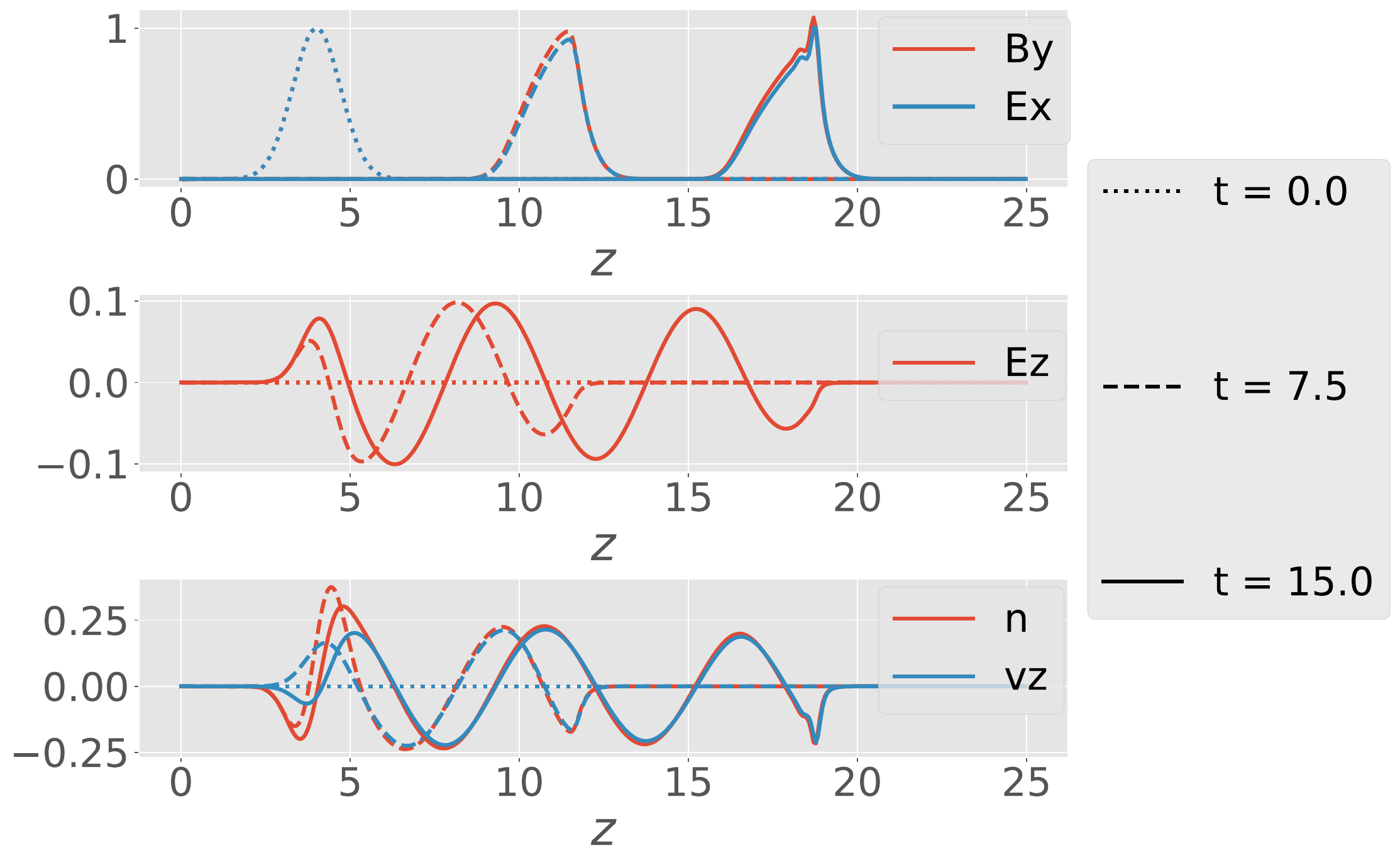}
  	\end{minipage}
	\caption{Numerical tests for strongly nonlinear regime using the finite difference/lowest order FEM solver: $(\omega_p/\omega_0, \omega_c/\omega_0) = (1/2, -1)$. See equation \eqref{eq:fd_casimirs} for the definition of the discrete Casimir invariants. The figures on the left and right are the errors in the conservation laws and a visualization of the solution respectively.}
	
	\label{fig:fd_approx}
	
\end{figure*}

We run a simulation of the 1D ponderomotive Maxwell system in the strongly nonlinear regime with the same parameters as in section \ref{sec:nonlinear_regime}. We use $2^{nd}$ order splitting and $N=600$ as this is the same number of degrees of freedom as was used in the previous spectral element FEEC simulations. The time-step is $dt \approx 0.033$. See Fig.~\ref{fig:fd_approx} for the results of this simulation. One can see that a qualitatively similar solution is obtained, but it does not possess the spurious oscillations of the previous simulations. This is because high order methods suffer from Runge's phenomenon when gradients are too large, and, in such situations, a larger number of cells is preferable to high order elements.

\section{Conclusion}


The objectives of this paper are twofold: to derive a self-consistent Hamiltonian model of the ponderomotive force, and to apply the general discretization procedure to study this model. The simple structure of the 1D ponderomotive Maxwell bracket is due to the fact that the model linearizes the fluid equations about a quiescent, uniform density background. The bracket thus becomes extraordinarily simple with all of the nonlinearity in the model arising from the complexity of the Hamiltonian. The asymptotic nature of the 1D ponderomotive Maxwell system causes it to miss certain physically significant features of the true dynamics it approximates. Moreover, even in parameter regimes where it is valid, the validity holds only on a limited time horizon. A high resolution simulation of the physics of the full, non-asymptotic system was previously studied \cite{10.1063/1.3124185}. One can see that the initial phase of the evolution of the 1D ponderomotive Maxwell system captures some of the relevant features. The oscillatory transverse electromagnetic field induces charge separation in the longitudinal direction exciting a longitudinal electric field. Moreover, the transverse Gaussian pulse becomes deformed as it propagates initially experiencing pulse steepening. However, the subsequent pulse elongation as well as the frequency modulation cannot be captured by our model as the asymptotic model omits phase dynamics. 

We also neglected the thermodynamic pressure in the simulations. This is because this term would be $O(c_S^2/c^2)$ where $c_S$ is the sound speed. In many cases, this term is asymptotically small, and its effects would be negligible over the short time horizon where this asymptotic regime remains valid. Including it would introduce sound waves to the model. These acoustic waves were eliminated in the simulations so that we could more clearly see the behavior of the nonlinear wave induced by the ponderomotive force. To include sound waves in the numerical method, one simply includes the appropriate pressure term on the righthand side of the continuity equation timestep in equation \eqref{eq:updateBV} (or equation \eqref{eq:updateBV_FD} for the finite difference implementation). 


The numerical method is not without its own limitations. We saw that the methods using high order interpolation in space underperformed in the strongly nonlinear regime with respect to discretizations using an equivalent number of degrees of freedom with lower order polynomial interpolation (which may be interpreted as a low order finite difference method) because of Runge's phenomenon. Likewise, the broken-FEEC approximations typically required greater resolution than the conforming-FEEC approximations in order to suppress spurious oscillations. Despite these current limitations, we have reason to believe that the numerical techniques introduced in this paper might provide a foundation for future powerful methods for simulating electrodynamics in nonlinearly polarized media. 


In order to avoid spurious oscillations in regions of large gradient while maintaining the efficiency of high order methods in smoother regions, several approaches might be taken. First, one might investigate adaptive mesh refinement in a structure preserving context. The problem of refinement in a structure preserving algorithm becomes that of finding a suitable projection operator from one mesh to another which does not significantly alter the Casimir invariants or the energy: i.e. each time the mesh is refined, the data is mapped to a new Hamiltonian system of a different dimension. The performance of the broken-FEEC method might be enhanced by investigating conforming projections other than the simplest averaging procedure used in this work. The FEEC approach is preferable to the finite differences because (1) it more easily accommodates general boundary conditions, (2) it more easily generalizes to higher dimensions and more complex geometries, (3) it scales to higher order discretizations. While this flexibility comes at the cost of algorithmic complexity, it has greater potential to yield a high performance nonlinear electrodynamics solver. Further, the broken-FEEC approach localizes the basis functions to each element thus allowing for a scalable, parallel implementation. Particularly notable is that this localizes the nonlinear solve, the largest bottleneck in each time-step. As a final note, the approach taken in this paper to discretize Maxwell's equation in a nonlinearly polarized medium (the ponderomotive Maxwell system is such a model) might be adapted to apply to a broad class of models in nonlinear optics. Subsequent papers will investigate this further application of the results of this paper in detail. 

\begin{acknowledgments}
We gratefully acknowledge the support of U.S. Dept. of Energy Contract \# DE-FG05-80ET-53088, NSF Graduate Research Fellowship \# DGE-1610403, and the Humboldt foundation. We would also like to thank the three anonymous referees for their attentive reading of this manuscript and for their insightful feedback. 
\end{acknowledgments}

\appendix
\section{The emergence of macroscopic fields from the Legendre transform in the electromagnetic Lagrangian} \label{appendix:legendre_transform_EM}
In this appendix, we briefly consider the reason for the particular form of the nonlinear constitutive relations in the Hamiltonian formulation of electromagnetism. In Gaussian units, $\bm{D}$ and $\bm{A}$ are are canonically conjugate up to the constant scaling factor $4 \pi c$. We may define the electromagnetic ``kinetic energy'' and ``potential energy'' as
\begin{equation}
	T = \frac{c^2}{8 \pi} \int | \dot{\bm{A}} |^2 \mathsf{d}^3 \bm{x}
	\quad \text{and} \quad
	V = \frac{1}{8\pi} \int | \nabla \times \bm{A} |^2 \mathsf{d}^3 \bm{x}.
\end{equation}
respectively. The Lagrangian is then given by
\begin{equation}
	L = T - V. 
\end{equation}
We then define the canonical momentum via a Legendre transform: $\bm{D} / 4 \pi c = \delta L/\delta \dot{\bm{A}}$, and the Hamiltonian is given by
\begin{multline}
	H = \frac{1}{4 \pi c} \int \bm{D} \cdot \dot{\bm{A}} \mathsf{d}^3 \bm{x} - L \\
		= \frac{1}{4 \pi c} \int \bm{D} \cdot \dot{\bm{A}} \mathsf{d}^3 \bm{x} -
			 \frac{1}{8 \pi} \int \left(  c^2 | \dot{\bm{A}} |^2 - | \nabla \times \bm{A} |^2  \right) \mathsf{d}^3 \bm{x}
\end{multline}
Letting $\bm{E} = c^{-1} \dot{\bm{A}}$ and $\bm{B} = \nabla \times \bm{A}$, we find
\begin{equation}
	H = \int \bm{D} \cdot \bm{E} \mathsf{d}^3 \bm{x} 
		- \frac{1}{8 \pi} \int \left( | \bm{E} |^2 - | \bm{B} |^2 \right) \mathsf{d}^3 \bm{x}
\end{equation}
which is the usual electromagnetic energy in a vacuum since $\bm{D} = \bm{E}$ for this choice of $T$. Hence, we recover the vacuum Maxwell equations. 

We now generalize to nonlinear media in which the refractive index is a perturbation from the identity. If we let
\begin{equation}
	T = \frac{c^2}{8 \pi} \int | \dot{\bm{A}} |^2 \mathsf{d}^3 \bm{x} - K,
\end{equation}
where $K$ is an arbitrary functional of the fields, we find
\begin{multline}
	H = \frac{1}{4 \pi c} \int \bm{D} \cdot \dot{\bm{A}} \mathsf{d}^3 \bm{x} + K \\
	- \frac{1}{8 \pi} \int \left( c^2 | \dot{\bm{A}} |^2 
        - | \nabla \times \bm{A} |^2  \right) \mathsf{d}^3 \bm{x}
\end{multline}
where, since $\bm{E} = c^{-1} \dot{\bm{A}}$, 
\begin{equation}
	\frac{\bm{D}}{4 \pi c} = \frac{\delta L}{\delta \bm{\dot{A}}} 
        = \frac{c^2 \dot{\bm{A}}}{4 \pi} - \frac{\delta K}{\delta \dot{\bm{A}}}
        = \frac{\bm{E}}{4 \pi c} - \frac{1}{c} \frac{\delta K}{\delta \bm{E}} \,,
\end{equation}
and
\begin{multline}
	H = \frac{1}{4 \pi} \int \bm{D} \cdot \bm{E} \mathsf{d}^3 \bm{x} + K -
			 \frac{1}{8 \pi} \int \left( | \bm{E} |^2 - | \bm{B} |^2  \right) \mathsf{d}^3 \bm{x} \\
	 = K - 4 \pi \int \frac{\delta K}{\delta \bm{E}} \cdot \bm{E} \mathsf{d}^3 \bm{x} 
        + \frac{1}{8 \pi} \int \left( | \bm{E} |^2 
        + | \bm{B} |^2 \right) \mathsf{d}^3 \bm{x}.
\end{multline}
This explains the particular form of the electromagnetic energy in a perturbed nonlinear medium. 

We can see that the (scaled) canonical bracket is simply
\begin{equation}
	\{F, G\} = 4 \pi c \int \left( \frac{\delta F}{\delta \bm{D}} \cdot \frac{\delta G}{\delta \bm{A}} 
		- \frac{\delta G}{\delta \bm{D}} \cdot \frac{\delta F}{\delta \bm{A}} \right) \mathsf{d}^3 \bm{x}. 
\end{equation}
Hence, in the variables $(\bm{D}, \bm{B})$, the bracket is 
\begin{equation}
	\{F, G\} = 4 \pi c \int \left( \frac{\delta F}{\delta \bm{D}} \cdot \nabla \times \frac{\delta G}{\delta \bm{B}} 
		- \frac{\delta G}{\delta \bm{D}} \cdot \nabla \times \frac{\delta F}{\delta \bm{B}} \right) \mathsf{d}^3 \bm{x}. 
\end{equation}
Assuming the Lagrangian is hyperregular, because the Legendre transform which maps $\dot{\bm{A}}$ to $\bm{D}$ is involutive, we have that $\delta H/\delta \bm{D} = (4 \pi c)^{-1} \dot{\bm{A}} = (4 \pi)^{-1} \bm{E}$. Finally, we simply define $\bm{H} = 4 \pi \delta H/ \delta \bm{B} = \bm{B} + 4 \pi \delta K/\delta \bm{B}$. We therefore recover the macroscopic Maxwell equations. 

Hence, an asymptotic procedure which introduces changes to the electromagnetic energy must likewise induce the appropriate compensating polarization and magnetization prescribed by the formalism discussed herein. Moreover, even in linear media, $(\bm{D}, \bm{A})$ are the naturally canonically conjugate variables with $\bm{D}$ and $\bm{E}$ being related to each other in the same manner in which the momentum and velocity are related in classical mechanics. We further note that this framework facilitates building theories of nonlinear electrodynamics which are not perturbations of the linear theory. 

\section{1D spectral element FEEC method} \label{appendix:1d_spectral_element_feec}
Here, we present a brief overview of the conforming/broken FEEC method \cite{campos-pinto_and_guclu_2021, güçlü2022broken} and urge the interested reader to consult the numerical analysis literature for the full details. The FEEC method we employ is a spectral element method which uses an interpolation/histopolation approach \cite{10.1007/978-3-642-15337-2_17}. 

\subsection{Construction of the FEM basis}
In a FEEC method, different kinds of quantities are approximated with a different set of basis functions in order to better represent the physics at a discrete level. In one spatial dimension, there are only scalar fields, however we may distinguish between two different varieties of scalars transformation rules. Suppose $F\colon \hat{\Omega} \to \Omega$ is a diffeomorphism. A $0$-form transforms as
\begin{equation}
    \phi(z) = \phi(F(\hat{z})) = \hat{\phi}(\hat{z})
\end{equation}
where $\hat{\phi} = \phi \circ F$. Alternatively, a $1$-form transforms as
\begin{equation}
    \rho(z) \mathsf{d} z = \hat{\rho}(\hat{z}) \mathsf{d} \hat{z} \implies \hat{\rho}(\hat{z}) = F'(\hat{z}) \rho(F(\hat{z})) \,.
\end{equation}
The vector space of $0$-forms is defined
\begin{equation}
    V^0 \coloneqq H^1 \Lambda^0 (\Omega) \coloneqq \{ \phi \in L^2 \Lambda^0 (\Omega) \colon \partial_z \phi \ \mathsf{d} z \in L^2 \Lambda^1 (\Omega) \} \,.
\end{equation}
This is the space of square integrable functions with square integrable derivatives which transform like differential $0$-forms. The vector space $V^1 \coloneqq L^2 \Lambda^1 (\Omega)$ is the space of square integrable functions which transform as $1$-forms. These spaces are the usual spaces $H^1$ and $L^2$ studied in functional analysis augmented with the transformation rules of $0$-forms and $1$-forms. 

As alluded to before, the finite element spaces which approximate $V^0$ and $V^1$ use different basis functions for reasons we will discuss subsequently. We denote the finite element spaces by $V^0_h$ and $V^1_h$ and the basis functions (also called shape functions) are denoted $\{ \Lambda_{0,j} \}_{j=1}^{N_0}$ and $\{ \Lambda_{1,j} \}_{j=1}^{N_1}$. Hence, if $\phi_h \in V^0_h$ and $\rho_h \in V^1_h$, we have
\begin{equation}
    \phi_h(z) = \sum_{j=1}^{N_0} \upphi_j \Lambda_{0,j}
    \quad \text{and} \quad
    \rho_h(z) = \sum_{j=1}^{N_1} \uprho_j \Lambda_{1,j} 
\end{equation}
where $\bm{\upphi} \in \mathcal{C}_0$ and $\bm{\uprho} \in \mathcal{C}_1$ are the Galerkin coefficients in the finite element expansions and $\mathcal{C}_0 \subset \mathbb{R}^{N_0}$ and $\mathcal{C}_1 \subset \mathbb{R}^{N_1}$ are the vector spaces of Galerkin coefficients. We shall uniformly use upright/sans-serif notation to denote the coefficient variables. 

The interpolation operators are simply the operators which take a vector of Galerkin coefficients and return the interpolated field in the finite element space:
\begin{equation}
    \begin{aligned}
    \phi_h &= \mathcal{I}^0 \bm{\upphi} = \sum_{j=1}^{N_0} \upphi_j \Lambda_{j,0}(z)
    \quad \text{and} \quad \\
    \rho_h &= \mathcal{I}^1 \bm{\uprho} = \sum_{j=1}^{N_1} \uprho_j \Lambda_{j,1}(z).
    \end{aligned}
\end{equation}
We define the degrees of freedom operators $\bm{\sigma}^0\colon V^0 \to \mathcal{C}_0$ and $\bm{\sigma}^1\colon V^1 \to \mathcal{C}_1$ such that 
\begin{equation}
    \sigma^0_i( \Lambda_{0,j} ) = \delta_{ij} 
    \quad \text{and} \quad
     \sigma^1_i( \Lambda_{1,j} ) = \delta_{ij}.
\end{equation}
Hence, applying the degrees of freedom operator to a field returns the degrees of freedom, the Galerkin coefficients, of that field in the appropriate finite element space, e.g. $\bm{\sigma}^1(\rho_h) = \bm{\uprho}$. Finally, we define the projection operators $\Pi^0\colon V^0 \to V^0_h$ and $\Pi^1\colon V^1 \to V^1_h$ to be the composition of the degrees of freedom and interpolation operators:
\begin{equation}
    \Pi^0 \phi = \sum_{j=1}^{N_0} \sigma^0_j(\phi) \Lambda_{j,0}(z)
    \quad \text{and} \quad
    \Pi^1 \rho = \sum_{j=1}^{N_1} \sigma^1_j(\rho) \Lambda_{j,1}(z).
\end{equation}

We may concisely summarize the spaces and operators making up our 1D FEEC method by the following commuting diagram:
\begin{equation} \label{eq:disc_derham_diagram}
	\begin{tikzcd}
	V^0 = H^1 \Lambda^0 (\Omega) \arrow{r}{\partial_z} \arrow{d}{\bm{\sigma}^{0}} \arrow[bend right=35,swap]{dd}{\Pi^{0}} & 
			V^{1} = L^2 \Lambda^1 (\Omega) \arrow[swap]{d}{\bm{\sigma}^{1}} \arrow[bend left=35]{dd}{\Pi^{1}} \\
	\mathcal{C}_{0} \arrow{r}{\mathbbm{d}_0} \arrow{d}{\mathcal{I}^{0}} & 
			\mathcal{C}_{1} \arrow[swap]{d}{\mathcal{I}^{1}} \\
	V^{0}_h \arrow{r}{\partial_{z,h} = \left. \partial_z \right|_{V^0_h} } & 
			V^{1}_h
	\end{tikzcd}
\end{equation}
We define $\mathbbm{d}_0$ below in definition \ref{def:discrete_ext_deriv}. Commutativity of the diagram means
\begin{equation}
	\mathbbm{d}_0 \bm{\sigma}^{0} = \bm{\sigma}^{1} \partial_z, 
	\quad 
	\partial_z \mathcal{I}^{0} = \mathcal{I}^{1} \mathbbm{d}_0, 
	\quad \text{and} \quad
	\partial_{z,h} \Pi^{0} = \Pi^{1} \partial_z.
\end{equation}
The reason why we utilize this FEEC approach is to ensure the consistent treatment of derivatives in our discretized models:
\begin{equation}
    \phi_h \in V^0_h \implies \partial_z \phi_h \in V^1_h
\end{equation}
and
\begin{equation}
    \bm{\sigma}^0(\phi_h) = \bm{\upphi} \in \mathcal{C}_0 \implies 
    \bm{\sigma}^1(\partial_z \phi_h) = \mathbbm{d}_0 \bm{\upphi} \in \mathcal{C}_1.
\end{equation}
The framework just described gives us these desirable properties. The remainder of this section is provides a brief overview of the definition of the finite element de Rham complex in one spatial dimension. 

To begin, let $\Omega = [0,L]$ be the physical domain. We divide this into $K$ cells which we denote $\{ \Omega_k \}_{k=1}^K$ such that $\bigcup_k \Omega_k = \Omega$. We first describe the construction on the reference element $\hat{\Omega} \coloneqq [-1,1]$ which is then mapped to each $\Omega_k$ by a diffeomorphism $F_k\colon \hat{\Omega} \to \Omega_k$. We shall denote all operators defined on the reference element with a hat to distinguish these from their mapped counterparts. 

The finite element shape functions used to interpolate the $0$-forms are simply Lagrange interpolating polynomials with the Gauss-Lobatto quadrature points as nodes. 
\begin{definition}
We denote the Gauss-Lobatto points in $[-1,1]$ by $\{ \xi_i \}_{i=0}^N$ and we denote the Lagrange interpolating polynomials over this grid by $\hat{\Lambda}_{0,i}$. We let there be $N+1$ quadrature points so that the order of the interpolating polynomials be $N$. The \textbf{$0$-form degrees of freedom operators} are defined to be
\begin{equation}
	\hat{\sigma}_{0,i}(\hat{v}) = \hat{v}(\xi_i) \implies \hat{\sigma}_{0,i}( \hat{\Lambda}_{0,j} ) = \delta_{ij}
\end{equation}
and \textbf{$0$-form interpolation} is simply $\hat{\mathcal{I}}^0 \bmsf{v} = \sum_i \msf{v}_i \hat{\Lambda}_{0,i}$. 
\end{definition}

\begin{definition}
The \textbf{degrees of freedom for the $1$-forms} are cell-wise integrals:
\begin{equation}
	\hat{\sigma}_{1,i}(v) = \int_{\xi_i}^{\xi_{i+1}} v(\xi) \mathsf{d} \xi \quad \text{for} \quad i = 0, 2, \hdots, N-1.
\end{equation}
The \textbf{$1$-form interpolating polynomials} are
\begin{multline}
	\hat{\Lambda}_{1,i}(\xi) = - \sum_{k=0}^{i} \hat{\Lambda}_{0,k}'(\xi) = \sum_{k = i+1}^{N} \hat{\Lambda}_{0,k}'(\xi) \\
    		= \frac{1}{2} \left[  \sum_{k = i+1}^{N} \hat{\Lambda}_{0,k}'(\xi) - \sum_{k=0}^{i} \hat{\Lambda}_{0,k}'(\xi) \right] \,,
\end{multline}
where the equality of these three expressions follows from the fact that
\begin{equation}
	\sum_{i=0}^N \hat{\Lambda}_{0,i}(\xi) = 1 \implies \sum_{i=0}^N \hat{\Lambda}_{0,i}'(\xi) = 0
\end{equation}
since the Lagrange interpolating polynomials form a partition of unity. 
\end{definition}
\noindent The $1$-form basis functions were selected such that the following is true:
\begin{equation}
\begin{aligned}
  \hat\sigma_{1,i}\bigl(\hat{\Lambda}_{1,j}\bigr)
  &= \int_{\xi_i}^{\xi_{i+1}} \hat{\Lambda}_{1,j}(\xi) d\xi \\
  &= \sum_{k=j+1}^N \left[ \delta_{i+1,k} - \delta_{i,k} \right] \\
  &= \delta_{i,j} + \sum_{k=j+1}^{N-1} \left(\delta_{i,k} - \delta_{i,k}\right) 
    - \delta_{i,N} \\
  &= \delta_{i,j} - \delta_{i,N} \,,
\end{aligned}
\end{equation}
where we made use of the identity $\delta_{i+c,k+c} = \delta_{i,k}\ \forall (i,k,c) \in \mathbb{Z}^3$. On a grid with $N$ elements we have that $i = 0, \dots, N-1$, and therefore $\delta_{i,N} = 0$ always. This yields the final result
\begin{equation}
  \hat\sigma_{1,i}\bigl(\hat{\Lambda}_{1,j}\bigr) = \delta_{i,j} \,.
\end{equation}

\begin{definition} \label{def:discrete_ext_deriv}
We define the \textbf{discrete exterior derivative matrix} to be 
\begin{equation}
	(\mathbbm{d}_0)_{ij} = \hat{\sigma}_{1,i}( \partial_\xi \hat{\Lambda}_{0,j}(\xi)).
\end{equation}
\end{definition}
\noindent One may show that
\begin{equation}
	\hat{\sigma}_{1,i}( \partial_\xi v ) = \int_{\xi_i}^{\xi_{i+1}} \partial_\xi v \mathsf{d} \xi 
		= v(\xi_{i+1}) - v(\xi_i) = (\mathbbm{d}_0)_{ij} \hat{\sigma}_{0,j}(v).
\end{equation}
Hence, we find that the derivative matrix is simply the adjacency matrix of the grid:
\begin{equation}
	\mathbbm{d}_0\colon \hat{\mathcal{C}}_0 \to \hat{\mathcal{C}}_1
	\quad \text{such that} \quad
	(\mathbbm{d}_0)_{ij} = 
	\begin{cases}
		1, & j = i+1 \\
		-1, & j = i \\
		0, & \text{else}.
	\end{cases}
\end{equation}
It is further possible to show $\partial_{\xi,h} \hat{\Pi}^0 = \hat{\Pi}^1 \partial_\xi$ which establishes the commutativity of the diagram. 

This defines all of the discrete operators on the reference element. We then define maps to each element in the physical domain, $F_k\colon \hat{\Omega} \to \Omega_k$, and pullback all quantities to the reference element in order to perform all computation. The manner by which $0$-forms and $1$-forms transform differ. 
\begin{definition}
We define the \textbf{push-forward operators}:
\begin{equation}
    \begin{aligned}
        \mathcal{F}^0_k \colon \hat{u} \mapsto u &\coloneqq \hat{u} \circ F^{-1}_k 
	\quad \text{and} \quad \\
        \mathcal{F}^1_k \colon \hat{u} \mapsto u &\coloneqq \frac{1}{\partial_{\hat{\xi}} F_k(\hat{\xi})} \hat{u} \circ F^{-1}_k.
    \end{aligned}
\end{equation}
\end{definition}
\noindent We see that $0$-forms transform like scalars while $1$-forms transform like densities. 
\begin{prop}
The derivative commutes with the push-forward:
\begin{equation}
	\partial_x \mathcal{F}^0_k \hat{u} = \mathcal{F}^1_k \partial_\xi \hat{u} 
	\quad \forall \hat{u} = \hat{u}(\hat{\xi}).
\end{equation}
\end{prop}
\noindent \textit{Proof:} This is simply the chain rule. \qed
\begin{definition}
The \textbf{local mapped finite element spaces} are simply $V_{h,k}^\ell = \mathcal{F}^\ell_k \hat{V}_h^\ell$. We define the \textbf{global finite element space} by
\begin{equation}
	V_h^\ell = \sum_k V_{h,k}^\ell \,,
\end{equation}
where we have extended functions outside of their local elements by zero. We then define the \textbf{global degrees of freedom} and \textbf{global interpolation} operators on the mapped domain by
\begin{equation}
	\sigma_{\ell, i, k} (u) \coloneqq \hat{\sigma}_{\ell, i}( (\mathcal{F}^\ell_k)^{-1} u)
	\quad \text{and} \quad
	\Lambda_{\ell,i, k} = \mathcal{F}^\ell_k \hat{\Lambda}_{\ell, i}. 
\end{equation}
\end{definition}
\noindent That the commuting diagram property is satisfied on the mapped domain follows from simple algebraic manipulation. When computing the degrees of freedom or inner products with the shape functions, one first pulls back to the reference element. 

Finally, in this 1D context, the distinction between conforming- and broken-FEEC methods is especially simple. Neighboring mapped elements will have their endpoints overlapping. That is, if we have $F_1\colon [-1,1] \to [0,1/2]$ and $F_2\colon [-1,1] \to [1/2, 1]$ map to two distinct elements, $F_1(1) = F_2(-1) = 1/2$. In a conforming FEEC method, we collapse the two overlapping endpoints into one single degree of freedom. In broken-FEEC, we keep these degrees of freedom distinct to allow for discontinuity across element boundaries. We then define a conforming projection operator which maps from the discontinuous space into the continuous space in order to perform derivatives. A simple choice is to average the degrees of freedom on the element boundaries. From the point of view of one using a broken-FEEC method, the primary difference then is that the discrete derivative operator must be composed with this (sparse) projection operator. 

The presence of adjoint differential operators in our Hamiltonian formulation of the 1D ponderomotive Maxwell model requires that we consider a dual de Rham complex (the complex for the adjoint differential operators) in our discretization scheme. Once the primal de Rham complex has been defined (as above), it is straightforward to define the dual complex in a subsidiary manner entirely in terms of quantities previously defined. 

\begin{definition}
The \textbf{dual degrees of freedom} $\bm{\sigma}_{\ell}^*\colon (V^{\ell})^* \to \mathcal{C}_{\ell}^*$ are defined such that 
\begin{equation}
	\sigma_{{\ell},i}^*(u) = \left\langle u, \Lambda_{{\ell},i} \right\rangle
	\quad \text{for} \quad
	u \in (V^\ell)^* \,,
\end{equation} 
where $\langle \cdot, \cdot \rangle\colon (V^\ell)^* \times V^\ell \to \mathbb{R}$ denotes the evaluation pairing of a vector space with its dual. We define the \textbf{dual basis}, $\{ \Lambda_{{\ell},i}^*\}_{i = 0}^{N_\ell}$, such that $\left( \Lambda_{{\ell},i}, \Lambda_{{\ell},j}^* \right) = \delta_{i, j}$. The \textbf{dual interpolation operator} $\mathcal{I}_{\ell}^*\colon \mathcal{C}_{\ell}^* \to (V_h^{\ell})^*$ is defined such that
\begin{equation}
	\mathcal{I}_{\ell}^* \bmsf{c}_* = \sum_{i = 0}^{N_\ell} (\msf{c}_*)_i \left( \Lambda_{{\ell},i}^*, \cdot \right) \in (V^\ell_h)^*.
\end{equation}
That is, $\mathcal{I}_{\ell}^* \bmsf{c}_*$ is a functional which acts on elements of $V^\ell_h$ such that
\begin{equation}
	\mathcal{I}_{\ell}^* \bmsf{c}_*[v_h] = \sum_{i = 0}^{N_\ell} (\msf{c}_*)_i \left( \Lambda_{{\ell},i}^*, v_h \right).
\end{equation}
The \textbf{mass matrix} is defined $(\mathbb{M}_{\ell})_{i, j} = ( \Lambda_{{\ell},i}, \Lambda_{{\ell},j} )\colon \mathcal{C}_\ell \to \mathcal{C}_\ell^*$. The \textbf{dual projection operator} is defined such that 
\begin{equation}
	\Pi_{\ell}^* = \mathcal{I}_{\ell}^* \circ \bm{\sigma}_{\ell}^*\colon (V^{\ell})^* \to (V^{\ell}_h)^*.
\end{equation}
One can see that this is simply the $L^2$ projection. 
\end{definition}

\begin{prop}
If $\bmsf{c}_* \in \mathcal{C}^*_\ell$ and $v_h \in V^\ell_h$ such that $\bm{\sigma}^\ell(v_h) = \bmsf{v} \in \mathcal{C}_\ell$, then
\begin{equation}
	\mathcal{I}_{\ell}^* \bmsf{c}_*[v_h] = \mathcal{I}_{\ell}^* \bmsf{c}_*[ \mathcal{I}^\ell \bmsf{v} ]  = \bmsf{c}_*^T \bmsf{v}.
\end{equation}
\end{prop}
\noindent \textit{Proof:} This simply follows from the fact that $\left( \Lambda_{{\ell},i}, \Lambda_{{\ell},j}^* \right) = \delta_{i, j}$. \qed

\begin{prop} \label{prop:change_of_basis}
The mass matrix provides the change of basis from $\mathcal{C}_{\ell}$ to $\mathcal{C}_{\ell}^*$. That is, if $u \in V^\ell_h$ and we denote its corresponding element in the dual space $u^*_h = (u_h, \cdot) \in (V^\ell_h)^*$, then
\begin{equation}
	\mathbb{M}_\ell \bm{\sigma}_\ell(u_h) =  \bm{\sigma}_\ell^*(u^*_h).
\end{equation}
\end{prop}
\noindent \textit{Proof:} 
\begin{equation}
	\begin{aligned}
		\bm{\sigma}_\ell^*(u^*_h) 
			&= \left\langle u^*_h, \Lambda_{{\ell},i} \right\rangle
				= \left( u_h,  \Lambda_{{\ell},i} \right) \\
			&= \sum_{j} \left( \Lambda_{\ell, i}, \Lambda_{\ell, j} \right) \msf{u}_{j}
				= \mathbb{M}_\ell \bm{\sigma}_\ell(u_h).
	\end{aligned}
\end{equation}
\qed

The primal and dual de Rham complexes together are frequently called the double de Rham complex. The discrete double de Rham complex may be schematically represented by
\begin{equation}
	\begin{tikzcd}[column sep=4em, row sep=2em]
		V^0 \arrow{d}{\partial_z} \arrow{r}{\Pi^0}
			& V^0_h \arrow{d}{ \partial_{z,h}} \arrow{r}{ \mathcal{R}_{0} }
				& (V^0_h)^* \arrow[shift left=1.5]{l}{ \mathcal{R}_{0}^{-1} } 
					& (V^0)^* \arrow{l}{ \Pi_0^* }
		\\
		V^{1} \arrow{r}{\Pi^1} 
			& V^1_h \arrow{r}{ \mathcal{R}_1}
				& (V^1_h)^* \arrow[swap]{u}{ \partial_{z,h}^* }  \arrow[shift left=1.5]{l}{ \mathcal{R}_1^{-1}} 
					& (V^1)^* \arrow[swap]{u}{ \partial_z^* } \arrow{l}{ \Pi_1^*} 
	\end{tikzcd} \\
\end{equation}
where $\mathcal{R}_\ell\colon V^\ell_h \to (V^\ell_h)^*$ is the isomorphism guaranteed by the Reisz representation theorem, or, at the coefficient level,
\begin{equation}
	\begin{tikzcd}[column sep=4em, row sep=2em]
		V^0 \arrow{d}{\partial_z} \arrow{r}{\bm{\sigma}^0}
			& \mathcal{C}_0 \arrow{d}{ \mathbbm{d}_0} \arrow{r}{ \mathbb{M}_0 }
				& \mathcal{C}_0^* \arrow[shift left=1.5]{l}{ \mathbb{M}_0^{-1} } 
					& (V^0)^* \arrow{l}{ \bm{\sigma}_{0}^* }
		\\
		V^1 \arrow{r}{\bm{\sigma}^1} 
			& \mathcal{C}_1  \arrow{r}{ \mathbb{M}_1 } 
				& \mathcal{C}_1^* \arrow[swap]{u}{ \mathbbm{d}_0^* }  \arrow[shift left=1.5]{l}{ \mathbb{M}_1^{-1}} 
					& (V^1)^* \arrow[swap]{u}{ \partial_z^* } \arrow{l}{ \bm{\sigma}_1^*} 
	\end{tikzcd}
\end{equation}
where each diagram commutes. We can see that the discrete integration by parts formula is given by 
\begin{equation}
	(\mathbbm{d}_0^*)^T \mathbb{M}_0 = \mathbb{M}_1 \mathbbm{d}_0.
\end{equation} 

Finally, it is worth briefly discussing the advantages of broken-FEEC. The advantage of broken-FEEC is that the basis functions are localized to each element. This gives the mass matrix a block diagonal structure and also localizes the nonlinear solves required by this method. Hence, at the cost of increasing the number of degrees of freedom in a broken-FEEC method, one localizes many operations allowing for greater parallelism. The one-dimensional problem considered in this paper does not require such fancy computational techniques, however similar problems in nonlinear electrodynamics in higher dimensions might benefit from these considerations. By introducing conforming projection operators \cite{campos-pinto_and_guclu_2021, güçlü2022broken} which project fields from the broken spaces into appropriate spaces in the conforming finite element de Rham complex, one can construct a commuting diagram of vector spaces like equation \eqref{eq:disc_derham_diagram} (and generalizations thereof in 2D and 3D). Hence, this version of FEEC which utilizes broken elements retains the structure preserving properties celebrated in conforming FEEC discretizations. 

\subsection{Integrals of functions in FEM space} \label{appendix:casimir_lemmas}
It is useful to know how to compute integrals of functions in the FEM spaces over the entire domain. The ponderomotive Maxwell system possesses several conservation laws which are integrals of fields over the entire domain. The following lemma tells us how such integrals may be computed. In the following, we combine the indices for the DOF within an element, and the element index into one single multiindex. 

\begin{lemma}
Suppose $v_h \in V^0_h$ and $u_h \in V^1_h$. Then if we denote $v_h = \sum_{\bm{\alpha}} \Lambda_{0,\bm{\alpha}} \msf{v}_{\bm{\alpha}}$ and $u_h = \sum_{\bm{\alpha}} \Lambda_{1,\bm{\alpha}} \msf{u}_{\bm{\alpha}}$, it follows that 
\begin{multline}
    \int_{0}^L v_h \mathsf{d} z = \sum_{ \bm{\alpha} , \bm{\beta} } (\mathbb{M}_0)_{\bm{\alpha} \bm{\beta}} \msf{v}_{\bm{\beta}} = \mathbbm{1}^T \mathbb{M}_0 \bmsf{v}
    \\ 
    \text{and} \quad
    \int_{0}^L u_h \mathsf{d} z = \sum_{\bm{\alpha}} \msf{u}_{\bm{\alpha}} = \mathbbm{1}^T \bmsf{u} \,,
\end{multline}
where $\mathbbm{1}_{\bm{\alpha}} = 1 \ \forall \bm{\alpha}$. 
\end{lemma}

\noindent \textit{Proof:} To begin, we consider the integral of $1$-forms in $V^1_h$. The proof proceeds by direct calculation and application of the definitions:
\begin{equation}
	\begin{aligned}
		\int_{0}^L u_h \mathsf{d} z 
			&= \sum_{i, k} \int_{\Omega_k} \Lambda_{1,i,k}(z) \msf{u}_{i,k} \mathsf{d} z \\
			&= \sum_{i, k} \left( \int_{\hat{\Omega}} \hat{\Lambda}_{1,i}( \xi ) \mathsf{d} \xi \right) \msf{u}_{i,k} \\
			&= \sum_{i, j, k} \left( \int_{\xi_j}^{\xi_{j+1}} \hat{\Lambda}_{1,i}(\xi) \mathsf{d} \xi \right) \msf{u}_{i,k} \\
			&= \sum_{i, j, k} \delta_{ij} \msf{u}_{i,k} 
                    = \sum_{i, k} \msf{u}_{i,k} 
                    = \sum_{\bm{\alpha}} \msf{u}_{\bm{\alpha}}
	\end{aligned}
\end{equation}
where we have used the fact that $\Lambda_{1,i,k}(z) \mathsf{d} z = \hat{\Lambda}_{1,i}(\xi) \mathsf{d} \xi$ where $F_k(\xi) = z$ and $F_k(\hat{\Omega}) = \Omega_k$. 

We proceed similarly with the $0$-forms: 
\begin{equation}
	\begin{aligned}
		\int_{0}^L v_h \mathsf{d} z 
			&= \sum_{i, k} \int_{\Omega_k} \Lambda_{0,i,k}(z) \msf{v}_{i,k} \mathsf{d} z \\
			&= \sum_{i, k} \left( \int_{\hat{\Omega}} \hat{\Lambda}_{0,i}( \xi ) \partial_{\xi} F_k(\xi) \mathsf{d} \xi \right) \msf{v}_{i,k}.
	\end{aligned}
\end{equation}
Recall, the $0$-form shape functions are a partition of unity: $\sum_l \hat{\Lambda}_{0,l}(\xi) = 1$. Hence,
\begin{multline}
    \int_{\hat{\Omega}} \hat{\Lambda}_{0,i}(\xi) \partial_{\xi} F_k(\xi) \mathsf{d} \xi 
	\\
    = \sum_l \int_{\hat{\Omega}} \hat{\Lambda}_{0,i}(\xi) \hat{\Lambda}_{0,l}(\xi) \partial_{\xi} F_k(\xi) \mathsf{d} \xi 
        \\
    = \sum_l (\mathbb{M}_0^{(k)})_{li} \,,
\end{multline}
where $\mathbb{M}_0^{(k)}$ represents the $0$-form mass matrix for the $k^{th}$ element. The global mass matrix $\mathbb{M}_0$ is related to each local mass matrix according to
\begin{equation}
	\sum_i (\mathbb{M}_0^{(k)})_{li} \msf{v}_{i,k} = (\mathbb{M}_0 \bmsf{v})_{l,k} \,,
\end{equation}
where care must be taken in how the degrees of freedom at element boundaries are handled (this differs between conforming and broken FEEC methods). Hence, it follows that
\begin{equation}
	\int_{0}^L v_h \mathsf{d} z = \sum_{i, k} (\mathbb{M}_0 \bmsf{v})_{i,k} = \mathbbm{1}^T \mathbb{M}_0 \bmsf{v}
\end{equation}
which establishes the result. \qed

\section{On discretizing functional derivatives} \label{appendix:disc_functional_derivatives}
This appendix considers the mathematically rigorous details in discretizing functional derivatives. Much of the content in this appendix assumes a background in functional analysis and may be skipped if one is only interested in the results of the paper and not the mathematical details. 

\subsection{Discrete functional derivatives}
As Hamiltonian field theories are stated in terms of functional derivatives, in order to discretize the Hamiltonian structure directly, it is necessary to consider what happens when a functional is restricted to a finite dimensional subspace (e.g.\ a finite element subspace). What we find is a convenient approach to discretize the Poisson bracket and Hamiltonian. 

Let $(U, \| \cdot \|_U)$ be a normed vector space. For any functional $K\colon U \to \mathbb{R}$, we define its Fr{\'e}chet derivative at $u \in U$ (if it exists) to be the linear functional $DK[u]$ such that
\begin{equation} \label{eq:functional_derivative}
	\frac{ \| K[u + \delta u] - K[u] - DK[u] \delta u \|_U }{ \| \delta u \|_U } = O( \| \delta u \|_U ).
\end{equation}
We shall assume in this paper that we restrict our attention to those functionals which are Fr{\'e}chet differentiable. When $U = \mathbb{R}^n$, this reduces to the usual definition of the Jacobian matrix. Hence, this definition generalizes the notion of derivatives to arbitrary normed vector spaces (e.g. function spaces). We shall use the two common notations for the functional derivative $DK[u]$ and $\delta K/\delta u$ interchangeably. 

Derivatives with respect to functions which are Galerkin expansions in our finite element framework reduce to finite dimensional derivatives with respect to the Galerkin coefficients. 
\begin{theorem} \label{thm:disc_func_deriv}
Let $K\colon V^{\ell} \to \mathbb{R}$ be an arbitrary functional on the continuous ${\ell}$-forms and let $\msf{K} \coloneqq K \circ \mathcal{I}^{\ell}\colon \mathcal{C}_{\ell} \to \mathbb{R}$ represent the discrete analog of the functional $K$. Moreover, define $\bmsf{u} = \bm{\sigma}^{\ell}(u)$ and $\bmsf{v} = \bm{\sigma}^{\ell}(v)$. Then 
\begin{equation}
	\begin{aligned}
		&\left\langle D(K \circ \Pi^{\ell})[u], v \right\rangle_{(V^{\ell})^*, V^{\ell}} \\
			&\hspace{2em}= \left\langle D(K \circ \mathcal{I}^{\ell} )[ \bm{\sigma}^{\ell}(u) ], \bm{\sigma}^{\ell}(v) \right\rangle_{\mathcal{C}_{\ell}^*, \mathcal{C}_{\ell}} \\
			&\hspace{2em}= \left\langle D\msf{K}[ \bmsf{u} ], \bmsf{v} \right\rangle_{\mathcal{C}_{\ell}^*, \mathcal{C}_{\ell}} 
				=\left( \frac{\partial \msf{K}}{\partial \bmsf{u}} \right)^T \bmsf{v}.
	\end{aligned}
\end{equation}
Similarly, if we have a functional of the dual de Rham sequence, $K_*\colon (V^{\ell})^* \to \mathbb{R}$, defining $\msf{K}_* = K \circ \mathcal{I}_{\ell}^*$, $\bmsf{u}_* = \bm{\sigma}_{\ell}^*(u)$, and $\bmsf{v}_* = \bm{\sigma}_{\ell}^*(v)$, we have
\begin{equation}
	\begin{aligned}
		&\left\langle D(K_* \circ \Pi_{\ell}^*)[u], v \right\rangle_{V^{\ell}, (V^{\ell})^*} \\ 
			&\hspace{2em}= \left\langle D(K_* \circ \mathcal{I}_{\ell}^* )[ \bm{\sigma}_{\ell}^*(u) ], 
				\bm{\sigma}_{\ell}^*(v) \right\rangle_{\mathcal{C}_{\ell}, \mathcal{C}_{\ell}^*} \\
			&\hspace{2em}= \left\langle D\msf{K}_*[ \bmsf{u}_* ], \bmsf{v}_* \right\rangle_{\mathcal{C}_{\ell}, \mathcal{C}_{\ell}^*} 
				= \left( \frac{\partial \msf{K}_*}{\partial \bmsf{u}_*} \right)^T \bmsf{v}_*.
	\end{aligned}
\end{equation}
\end{theorem}
\noindent \textit{Proof:} The result follows from the chain rule and the linearity of the degrees of freedom operator. By the chain rule, we have
\begin{multline}
	\left\langle D(K \circ \Pi^{\ell})[u], v \right\rangle_{(V^{\ell})^*, V^{\ell}} \\
			= \left\langle D(K \circ \mathcal{I}^{\ell} )[ \bm{\sigma}^{\ell}(u) ] \circ D \bm{\sigma}^\ell[u], v \right\rangle_{(V^\ell)^*, V^{\ell}}.
\end{multline}
But by the linearity of $\bm{\sigma}^\ell$, $D \bm{\sigma}^\ell[u] = \bm{\sigma}^\ell$. Hence, we find
\begin{multline}
    \left\langle D(K \circ \Pi^{\ell})[u], v \right\rangle_{(V^{\ell})^*, V^{\ell}} \\
		=  \left\langle D(K \circ \mathcal{I}^{\ell} )[ \bm{\sigma}^{\ell}(u) ], \bm{\sigma}^{\ell}(v) \right\rangle_{\mathcal{C}_{\ell}^*, \mathcal{C}_{\ell}}.
\end{multline}
The remainder of the result follows from simple notational manipulation. The result for $K_*$ follows from an entirely analogous argument.
\qed

Hence, one may directly translate continuous functional derivatives into discrete functional derivatives. Moreover, by appropriate interpretation of the discrete representation of the variables on the dual sequence, we may also prescribe discrete versions of functional derivatives with respect to variables in the dual space. 

\subsection{Adjoint of degrees of freedom operator}
In this section, we prove a technical result which is needed to discretize the Poisson bracket. This may safely be skipped if one is only interested in the results and not the mathematical substructure of the numerical method. In the following, we use $\dagger$ to denote operator adjoints. 

\begin{lemma} \label{lemma:disc_to_cont_duality}
Suppose $\bmsf{u}_* \in \mathcal{C}^*_\ell$ and $\bmsf{v} \in \mathcal{C}^\ell$. Then
\begin{equation}
	\left\langle \bmsf{u}_*, \bmsf{v} \right\rangle_{C^*_\ell, C^\ell} 
		= \left\langle \mathcal{I}_\ell^* \bmsf{u}_*, \mathcal{I}^\ell \bmsf{v} \right\rangle_{(V^\ell)^*, V^\ell}.
\end{equation}
\end{lemma}
\noindent \textit{Proof:} The result follows by simple application of the definitions:
\begin{equation}
	\begin{aligned}
	       &\left\langle \mathcal{I}_\ell^* \bmsf{u}_*, \mathcal{I}^\ell \bmsf{v} \right\rangle_{(V^\ell)^*, V^\ell} \\
			&\hspace{2em}= \left\langle \sum_{i=1}^{N_{\ell}} (\msf{u}_*)_i \left( \Lambda_{{\ell},i}^*, \cdot \right), 
				\sum_{j=1}^{N_{\ell}} \msf{v}_j \Lambda_{{\ell},j} \right\rangle_{(V^\ell)^*, V^\ell} \\
			&\hspace{2em}= \sum_{i,j=1}^{N_{\ell}} (\msf{u}_*)_i \msf{v}_j \left( \Lambda_{{\ell},i}^*, \Lambda_{{\ell},j} \right) \\
			&\hspace{2em}= \bmsf{u}_*^T \bmsf{v} \coloneqq \left\langle \bmsf{u}_*, \bmsf{v} \right\rangle_{C^*_\ell, C^\ell}.
	\end{aligned}
\end{equation}
\qed

\begin{theorem}
With respect to the natural evaluation duality pairing, $\left( \left. \bm{\sigma}^\ell \right|_{V^\ell} \right)^\dagger = \mathcal{I}_\ell^*$. Moreover, 
\begin{multline}
    \frac{ \left| \left( [ (\bm{\sigma}^\ell)^\dagger - \mathcal{I}_\ell^* ] \bmsf{u}_*, v \right) \right|}{ \| \bmsf{u}_* \|_{C^*_\ell} \| v \|_{V^\ell}} 
		\leq \| I - \Pi_\ell \| \| \mathcal{I}_\ell^* \| \\
    \forall \bmsf{u} \in \mathcal{C}^*_\ell \text{ and } \forall v \in V^\ell.
\end{multline}
Hence, $\mathcal{I}_\ell^*$ approximates $\bm{\sigma}_\ell^\dagger$ in the following sense:
\begin{multline}
	\| (\bm{\sigma}^\ell)^\dagger - \mathcal{I}_\ell^* \| 
		\coloneqq \sup_{ \substack{\| \bmsf{u} \|_{C^*_\ell} \leq 1 \\ \bmsf{u}_* \in \mathcal{C}^*_\ell } } 
		    \sup_{ \substack{\| v \|_{V^\ell} \leq 1 \\ v \in V^\ell}} 
		    \frac{ \left| \left\langle ( \bm{\sigma}_\ell^\dagger - \mathcal{I}_\ell^*) \bmsf{u}_*, v \right\rangle \right|}{ \| \bmsf{u}_* \|_{C^*_\ell} \| v \|_{V^\ell}} 
				\\
    = O(h^{p+1}).
\end{multline}
An analogous result holds for $\mathcal{I}^\ell$ and $\bm{\sigma}_\ell^*$.
\end{theorem}
\noindent \textit{Proof:} Let $v_h \in V^\ell_h$. Then by the previous lemma, 
\begin{multline}
    \left\langle \bmsf{u}_*, \bm{\sigma}^\ell(v_h) \right\rangle_{C^*_\ell, C^\ell} 
		= \left\langle \mathcal{I}_\ell^* \bmsf{u}_*, \mathcal{I}^\ell \circ \bm{\sigma}^\ell(v_h) \right\rangle_{(V^\ell)^*, V^\ell} \\
    = \left\langle \mathcal{I}_\ell^* \bmsf{u}_*, v_h \right\rangle_{(V^\ell)^*, V^\ell}
\end{multline}
since $\mathcal{I}^\ell \circ \bm{\sigma}^\ell = \Pi^\ell = I$ on $V^\ell_h$. This proves the first claim. 

Now, let $v \in V^\ell$ be arbitrary. Then
\begin{equation}
	\left\langle \bmsf{u}_*, \bm{\sigma}^\ell(v) \right\rangle_{C^*_\ell, C^\ell} 
		= \left\langle \mathcal{I}_\ell^* \bmsf{u}_*, \Pi^\ell v \right\rangle_{(V^\ell)^*, V^\ell}.
\end{equation}
Hence,
\begin{multline}
    \left\langle \bmsf{u}_*, \bm{\sigma}^\ell(v) \right\rangle_{C^*_\ell, C^\ell} - \left\langle \mathcal{I}_\ell^* \bmsf{u}_*, v \right\rangle_{(V^\ell)^*, V^\ell} \\
    = - \left\langle \mathcal{I}_\ell^* \bmsf{u}_*, (I - \Pi^\ell) v \right\rangle_{(V^\ell)^*, V^\ell}
\end{multline}
which implies
\begin{equation}
	\begin{aligned}
		&\left| \left\langle \bmsf{u}_*, \bm{\sigma}^\ell(v) \right\rangle_{C^*_\ell, C^\ell} 
			- \left\langle \mathcal{I}_\ell^* \bmsf{u}_*, v \right\rangle_{(V^\ell)^*, V^\ell} \right| \\
			&\hspace{2em}
            = \left| \left\langle \mathcal{I}_\ell^* \bmsf{u}_*, (I - \Pi^\ell) v \right\rangle_{(V^\ell)^*, V^\ell} \right| \\
			&\hspace{2em}
            \leq \| I - \Pi^\ell \| \| \mathcal{I}_\ell^* \bmsf{u}_* \|_{(V^\ell)*} \| v \|_{V^\ell} \\
			&\hspace{2em}
            \leq \| I - \Pi^\ell \| \| \mathcal{I}_\ell^* \| \| \bmsf{u}_* \|_{(V^\ell)*} \| v \|_{V^\ell} 
	\end{aligned}
\end{equation}
by repeated application of the triangle inequality. Since $\mathcal{I}_\ell^*\colon C_\ell^* \to (V^\ell_h)^*$ is a map between finite dimensional spaces, it is bounded. Moreover, by assumption, $\| I - \Pi^\ell \| = O(h^{p+1})$. The result follows. \qed

This combined with the notion of discrete functional derivatives in the previous section gives us the following result.
\begin{theorem} \label{thm:disc_func_deriv_2}
Let $K\colon V^\ell \to \mathbb{R}$. Define $\mathsf{K} \coloneqq K \circ \mathcal{I}^\ell \colon \mathcal{C}_\ell \to \mathbb{R}$. Then
\begin{equation}
	\frac{\delta (K \circ \Pi^\ell) }{\delta u} = (\bm{\sigma}^\ell)^\dagger \frac{\partial \mathsf{K}}{\partial \bmsf{u} } = \mathcal{I}^*_\ell \frac{\partial \mathsf{K}}{\partial \bmsf{u}} + O(h^{p+1}) \,,
\end{equation}
where $\bmsf{u} = \bm{\sigma}^\ell(u)$. If our functional depends only on the finite dimensional space $V^\ell_h$, i.e. $K\colon V^\ell_h \to \mathbb{R}$, then
\begin{equation}
	\frac{\delta K}{\delta u_h} = \mathcal{I}^*_\ell \frac{\partial \mathsf{K} }{\partial \bmsf{u} }.
\end{equation}
A similar result holds for $K\colon (V^\ell)^* \to \mathbb{R}$. 
\end{theorem}
\noindent \textit{Proof:} As before, from the chain rule, we have
\begin{equation}
	\begin{aligned}
		&\left\langle D(K \circ \Pi^{\ell})[u], v \right\rangle_{(V^{\ell})^*, V^{\ell}} \\
		&\hspace{2em}= \left\langle D(K \circ \mathcal{I}^{\ell} )[ \bm{\sigma}^{\ell}(u) ], \bm{\sigma}^{\ell}(v) \right\rangle_{\mathcal{C}_{\ell}^*, \mathcal{C}_{\ell}} \\
		&\hspace{2em}= \left\langle (\bm{\sigma}^\ell)^\dagger \left( D \msf{K}[ \bmsf{u} ] \right), v \right\rangle_{(V^\ell)^*, V^\ell}.
	\end{aligned}
\end{equation}
Hence, simply changing notation, we find that
\begin{equation}
	\frac{\delta (K \circ \Pi^\ell) }{\delta u} = (\bm{\sigma}^\ell)^\dagger \frac{\partial \mathsf{K}}{\partial \bmsf{u} }.
\end{equation}
The result follows from the previous theorem. \qed

\bibliography{ponder_maxwell.bib}

\begin{thebibliography}{22}%
\makeatletter
\providecommand \@ifxundefined [1]{%
 \@ifx{#1\undefined}
}%
\providecommand \@ifnum [1]{%
 \ifnum #1\expandafter \@firstoftwo
 \else \expandafter \@secondoftwo
 \fi
}%
\providecommand \@ifx [1]{%
 \ifx #1\expandafter \@firstoftwo
 \else \expandafter \@secondoftwo
 \fi
}%
\providecommand \natexlab [1]{#1}%
\providecommand \enquote  [1]{``#1''}%
\providecommand \bibnamefont  [1]{#1}%
\providecommand \bibfnamefont [1]{#1}%
\providecommand \citenamefont [1]{#1}%
\providecommand \href@noop [0]{\@secondoftwo}%
\providecommand \href [0]{\begingroup \@sanitize@url \@href}%
\providecommand \@href[1]{\@@startlink{#1}\@@href}%
\providecommand \@@href[1]{\endgroup#1\@@endlink}%
\providecommand \@sanitize@url [0]{\catcode `\\12\catcode `\$12\catcode
  `\&12\catcode `\#12\catcode `\^12\catcode `\_12\catcode `\%12\relax}%
\providecommand \@@startlink[1]{}%
\providecommand \@@endlink[0]{}%
\providecommand \url  [0]{\begingroup\@sanitize@url \@url }%
\providecommand \@url [1]{\endgroup\@href {#1}{\urlprefix }}%
\providecommand \urlprefix  [0]{URL }%
\providecommand \Eprint [0]{\href }%
\providecommand \doibase [0]{http://dx.doi.org/}%
\providecommand \selectlanguage [0]{\@gobble}%
\providecommand \bibinfo  [0]{\@secondoftwo}%
\providecommand \bibfield  [0]{\@secondoftwo}%
\providecommand \translation [1]{[#1]}%
\providecommand \BibitemOpen [0]{}%
\providecommand \bibitemStop [0]{}%
\providecommand \bibitemNoStop [0]{.\EOS\space}%
\providecommand \EOS [0]{\spacefactor3000\relax}%
\providecommand \BibitemShut  [1]{\csname bibitem#1\endcsname}%
\let\auto@bib@innerbib\@empty
\bibitem [{\citenamefont {Esarey}\ \emph {et~al.}(1997)\citenamefont {Esarey},
  \citenamefont {Sprangle}, \citenamefont {Krall},\ and\ \citenamefont
  {Ting}}]{641305}%
  \BibitemOpen
  \bibfield  {author} {\bibinfo {author} {\bibfnamefont {E.}~\bibnamefont
  {Esarey}}, \bibinfo {author} {\bibfnamefont {P.}~\bibnamefont {Sprangle}},
  \bibinfo {author} {\bibfnamefont {J.}~\bibnamefont {Krall}}, \ and\ \bibinfo
  {author} {\bibfnamefont {A.}~\bibnamefont {Ting}},\ }\bibfield  {title}
  {\enquote {\bibinfo {title} {{Self-focusing and guiding of short laser pulses
  in ionizing gases and plasmas}},}\ }\href {\doibase 10.1109/3.641305}
  {\bibfield  {journal} {\bibinfo  {journal} {IEEE Journal of Quantum
  Electronics}\ }\textbf {\bibinfo {volume} {33}},\ \bibinfo {pages}
  {1879--1914} (\bibinfo {year} {1997})}\BibitemShut {NoStop}%
\bibitem [{\citenamefont {Antonsen~Jr.}\ and\ \citenamefont
  {Mora}(1993)}]{10.1063/1.860884}%
  \BibitemOpen
  \bibfield  {author} {\bibinfo {author} {\bibfnamefont {T.~M.}\ \bibnamefont
  {Antonsen~Jr.}}\ and\ \bibinfo {author} {\bibfnamefont {P.}~\bibnamefont
  {Mora}},\ }\bibfield  {title} {\enquote {\bibinfo {title} {{Self‐focusing
  and Raman scattering of laser pulses in tenuous plasmas}},}\ }\href {\doibase
  10.1063/1.860884} {\bibfield  {journal} {\bibinfo  {journal} {Physics of
  Fluids B: Plasma Physics}\ }\textbf {\bibinfo {volume} {5}},\ \bibinfo
  {pages} {1440--1452} (\bibinfo {year} {1993})},\ \Eprint
  {http://arxiv.org/abs/https://pubs.aip.org/aip/pfb/article-pdf/5/5/1440/12440575/1440\_1\_online.pdf}
  {https://pubs.aip.org/aip/pfb/article-pdf/5/5/1440/12440575/1440\_1\_online.pdf}
  \BibitemShut {NoStop}%
\bibitem [{\citenamefont {Morrison}(2013)}]{morrison_gauge_free_lifting}%
  \BibitemOpen
  \bibfield  {author} {\bibinfo {author} {\bibfnamefont {P.~J.}\ \bibnamefont
  {Morrison}},\ }\bibfield  {title} {\enquote {\bibinfo {title} {A general
  theory for gauge-free lifting},}\ }\href {\doibase 10.1063/1.4774063}
  {\bibfield  {journal} {\bibinfo  {journal} {Physics of Plasmas}\ }\textbf
  {\bibinfo {volume} {20}},\ \bibinfo {pages} {012104} (\bibinfo {year}
  {2013})},\ \Eprint {http://arxiv.org/abs/https://doi.org/10.1063/1.4774063}
  {https://doi.org/10.1063/1.4774063} \BibitemShut {NoStop}%
\bibitem [{\citenamefont {Morrison}(1982)}]{pjm82}%
  \BibitemOpen
  \bibfield  {author} {\bibinfo {author} {\bibfnamefont {P.~J.}\ \bibnamefont
  {Morrison}},\ }\bibfield  {title} {\enquote {\bibinfo {title} {{P}oisson
  {b}rackets for {f}luids and {p}lasmas},}\ }\href@noop {} {\bibfield
  {journal} {\bibinfo  {journal} {AIP Conf. Proc.}\ }\textbf {\bibinfo {volume}
  {88}},\ \bibinfo {pages} {13--46} (\bibinfo {year} {1982})}\BibitemShut
  {NoStop}%
\bibitem [{\citenamefont {Arnold}, \citenamefont {Falk},\ and\ \citenamefont
  {Winther}(2010)}]{ArnoldDouglasN2010Feec}%
  \BibitemOpen
  \bibfield  {author} {\bibinfo {author} {\bibfnamefont {D.~N.}\ \bibnamefont
  {Arnold}}, \bibinfo {author} {\bibfnamefont {R.~S.}\ \bibnamefont {Falk}}, \
  and\ \bibinfo {author} {\bibfnamefont {R.}~\bibnamefont {Winther}},\
  }\bibfield  {title} {\enquote {\bibinfo {title} {Finite element exterior
  calculus: from {H}odge theory to numerical stability},}\ }\href@noop {}
  {\bibfield  {journal} {\bibinfo  {journal} {Bulletin (new series) of the
  American Mathematical Society}\ }\textbf {\bibinfo {volume} {47}},\ \bibinfo
  {pages} {281--354} (\bibinfo {year} {2010})}\BibitemShut {NoStop}%
\bibitem [{\citenamefont {Campos~Pinto}\ and\ \citenamefont
  {Sonnendr{\"u}cker}(2017{\natexlab{a}})}]{conga_pic_1}%
  \BibitemOpen
  \bibfield  {author} {\bibinfo {author} {\bibfnamefont {M.}~\bibnamefont
  {Campos~Pinto}}\ and\ \bibinfo {author} {\bibfnamefont {E.}~\bibnamefont
  {Sonnendr{\"u}cker}},\ }\bibfield  {title} {\enquote {\bibinfo {title}
  {{Compatible Maxwell solvers with particles I: conforming and non-conforming
  2D schemes with a strong Ampere law}},}\ }\href {\doibase
  10.5802/smai-jcm.20} {\bibfield  {journal} {\bibinfo  {journal} {SMAI Journal
  of Computational Mathematics}\ }\textbf {\bibinfo {volume} {3}},\ \bibinfo
  {pages} {53--89} (\bibinfo {year} {2017}{\natexlab{a}})}\BibitemShut
  {NoStop}%
\bibitem [{\citenamefont {Campos~Pinto}\ and\ \citenamefont
  {Sonnendr{\"u}cker}(2017{\natexlab{b}})}]{conga_pic_2}%
  \BibitemOpen
  \bibfield  {author} {\bibinfo {author} {\bibfnamefont {M.}~\bibnamefont
  {Campos~Pinto}}\ and\ \bibinfo {author} {\bibfnamefont {E.}~\bibnamefont
  {Sonnendr{\"u}cker}},\ }\bibfield  {title} {\enquote {\bibinfo {title}
  {{Compatible Maxwell solvers with particles II: conforming and non-conforming
  2D schemes with a strong Faraday law}},}\ }\href {\doibase
  10.5802/smai-jcm.21} {\bibfield  {journal} {\bibinfo  {journal} {SMAI Journal
  of Computational Mathematics}\ }\textbf {\bibinfo {volume} {3}},\ \bibinfo
  {pages} {91--116} (\bibinfo {year} {2017}{\natexlab{b}})}\BibitemShut
  {NoStop}%
\bibitem [{\citenamefont {Campos~Pinto}\ and\ \citenamefont
  {Sonnendr{\"u}cker}(2016)}]{campos-pinto2016:CONGA}%
  \BibitemOpen
  \bibfield  {author} {\bibinfo {author} {\bibfnamefont {M.}~\bibnamefont
  {Campos~Pinto}}\ and\ \bibinfo {author} {\bibfnamefont {E.}~\bibnamefont
  {Sonnendr{\"u}cker}},\ }\bibfield  {title} {\enquote {\bibinfo {title}
  {{Gauss-compatible Galerkin schemes for time-dependent Maxwell equations}},}\
  }\href {https://www.jstor.org/stable/mathcomp.85.302.2651} {\bibfield
  {journal} {\bibinfo  {journal} {Mathematics of Computation}\ }\textbf
  {\bibinfo {volume} {85}},\ \bibinfo {pages} {2651--2685} (\bibinfo {year}
  {2016})}\BibitemShut {NoStop}%
\bibitem [{\citenamefont {Campos~Pinto}\ and\ \citenamefont {G{\"u}{\c
  c}l{\"u}}(2021)}]{campos-pinto_and_guclu_2021}%
  \BibitemOpen
  \bibfield  {author} {\bibinfo {author} {\bibfnamefont {M.}~\bibnamefont
  {Campos~Pinto}}\ and\ \bibinfo {author} {\bibfnamefont {Y.}~\bibnamefont
  {G{\"u}{\c c}l{\"u}}},\ }\href {\doibase 10.48550/ARXIV.2109.02553} {\enquote
  {\bibinfo {title} {{Broken-FEEC approximations of Hodge Laplace problems}},}\
  } (\bibinfo {year} {2021})\BibitemShut {NoStop}%
\bibitem [{\citenamefont {Güçlü}, \citenamefont {Hadjout},\ and\
  \citenamefont {Pinto}(2022)}]{güçlü2022broken}%
  \BibitemOpen
  \bibfield  {author} {\bibinfo {author} {\bibfnamefont {Y.}~\bibnamefont
  {Güçlü}}, \bibinfo {author} {\bibfnamefont {S.}~\bibnamefont {Hadjout}}, \
  and\ \bibinfo {author} {\bibfnamefont {M.~C.}\ \bibnamefont {Pinto}},\
  }\href@noop {} {\enquote {\bibinfo {title} {{A broken FEEC framework for
  electromagnetic problems on mapped multipatch domains}},}\ } (\bibinfo {year}
  {2022}),\ \Eprint {http://arxiv.org/abs/2208.05238} {arXiv:2208.05238
  [math.NA]} \BibitemShut {NoStop}%
\bibitem [{\citenamefont {Kraus}\ \emph {et~al.}(2017)\citenamefont {Kraus},
  \citenamefont {Kormann}, \citenamefont {Morrison},\ and\ \citenamefont
  {Sonnendr{\"u}cker}}]{GEMPIC}%
  \BibitemOpen
  \bibfield  {author} {\bibinfo {author} {\bibfnamefont {M.}~\bibnamefont
  {Kraus}}, \bibinfo {author} {\bibfnamefont {K.}~\bibnamefont {Kormann}},
  \bibinfo {author} {\bibfnamefont {P.~J.}\ \bibnamefont {Morrison}}, \ and\
  \bibinfo {author} {\bibfnamefont {E.}~\bibnamefont {Sonnendr{\"u}cker}},\
  }\bibfield  {title} {\enquote {\bibinfo {title} {{GEMPIC}: geometric
  electromagnetic particle-in-cell methods},}\ }\href@noop {} {\bibfield
  {journal} {\bibinfo  {journal} {Journal of Plasma Physics}\ }\textbf
  {\bibinfo {volume} {83}} (\bibinfo {year} {2017})}\BibitemShut {NoStop}%
\bibitem [{\citenamefont {Burby}(2017)}]{10.1063/1.4976849}%
  \BibitemOpen
  \bibfield  {author} {\bibinfo {author} {\bibfnamefont {J.~W.}\ \bibnamefont
  {Burby}},\ }\bibfield  {title} {\enquote {\bibinfo {title}
  {{Finite-dimensional collisionless kinetic theory}},}\ }\href {\doibase
  10.1063/1.4976849} {\bibfield  {journal} {\bibinfo  {journal} {Physics of
  Plasmas}\ }\textbf {\bibinfo {volume} {24}},\ \bibinfo {pages} {032101}
  (\bibinfo {year} {2017})},\ \Eprint
  {http://arxiv.org/abs/https://pubs.aip.org/aip/pop/article-pdf/doi/10.1063/1.4976849/15934438/032101\_1\_online.pdf}
  {https://pubs.aip.org/aip/pop/article-pdf/doi/10.1063/1.4976849/15934438/032101\_1\_online.pdf}
  \BibitemShut {NoStop}%
\bibitem [{\citenamefont {Kirchhoff}(1867)}]{kirchhoff_point_vortex}%
  \BibitemOpen
  \bibfield  {author} {\bibinfo {author} {\bibfnamefont {G.}~\bibnamefont
  {Kirchhoff}},\ }\href@noop {} {\emph {\bibinfo {title} {{Vorlesungen über
  Mathematische Physik}}}}\ (\bibinfo  {publisher} {Leipzig},\ \bibinfo {year}
  {1867})\ Chap.~\bibinfo {chapter} {XX}\BibitemShut {NoStop}%
\bibitem [{\citenamefont {Spencer}\ and\ \citenamefont
  {Kaufman}(1982)}]{PhysRevA.25.2437}%
  \BibitemOpen
  \bibfield  {author} {\bibinfo {author} {\bibfnamefont {R.~G.}\ \bibnamefont
  {Spencer}}\ and\ \bibinfo {author} {\bibfnamefont {A.~N.}\ \bibnamefont
  {Kaufman}},\ }\bibfield  {title} {\enquote {\bibinfo {title} {Hamiltonian
  structure of two-fluid plasma dynamics},}\ }\href {\doibase
  10.1103/PhysRevA.25.2437} {\bibfield  {journal} {\bibinfo  {journal} {Phys.
  Rev. A}\ }\textbf {\bibinfo {volume} {25}},\ \bibinfo {pages} {2437--2439}
  (\bibinfo {year} {1982})}\BibitemShut {NoStop}%
\bibitem [{\citenamefont {Burby}\ and\ \citenamefont
  {Ruiz}(2020)}]{10.1063/1.5099383}%
  \BibitemOpen
  \bibfield  {author} {\bibinfo {author} {\bibfnamefont {J.~W.}\ \bibnamefont
  {Burby}}\ and\ \bibinfo {author} {\bibfnamefont {D.~E.}\ \bibnamefont
  {Ruiz}},\ }\bibfield  {title} {\enquote {\bibinfo {title} {{Variational
  nonlinear WKB in the Eulerian frame}},}\ }\href {\doibase 10.1063/1.5099383}
  {\bibfield  {journal} {\bibinfo  {journal} {Journal of Mathematical Physics}\
  }\textbf {\bibinfo {volume} {61}},\ \bibinfo {pages} {053101} (\bibinfo
  {year} {2020})},\ \Eprint
  {http://arxiv.org/abs/https://pubs.aip.org/aip/jmp/article-pdf/doi/10.1063/1.5099383/14076683/053101\_1\_online.pdf}
  {https://pubs.aip.org/aip/jmp/article-pdf/doi/10.1063/1.5099383/14076683/053101\_1\_online.pdf}
  \BibitemShut {NoStop}%
\bibitem [{\citenamefont {Trotter}(1959)}]{trotter1959product}%
  \BibitemOpen
  \bibfield  {author} {\bibinfo {author} {\bibfnamefont {M.}~\bibnamefont
  {Trotter}},\ }\bibfield  {title} {\enquote {\bibinfo {title} {On the product
  of semi-groups of operators},}\ }\href {\doibase 10.2307/2032958} {\bibfield
  {journal} {\bibinfo  {journal} {Proceedings of the American Mathematical
  Society}\ }\textbf {\bibinfo {volume} {10}},\ \bibinfo {pages} {545--551}
  (\bibinfo {year} {1959})}\BibitemShut {NoStop}%
\bibitem [{\citenamefont {Strang}(1968)}]{doi:10.1137/0705041}%
  \BibitemOpen
  \bibfield  {author} {\bibinfo {author} {\bibfnamefont {G.}~\bibnamefont
  {Strang}},\ }\bibfield  {title} {\enquote {\bibinfo {title} {On the
  construction and comparison of difference schemes},}\ }\href {\doibase
  10.1137/0705041} {\bibfield  {journal} {\bibinfo  {journal} {SIAM Journal on
  Numerical Analysis}\ }\textbf {\bibinfo {volume} {5}},\ \bibinfo {pages}
  {506--517} (\bibinfo {year} {1968})},\ \Eprint
  {http://arxiv.org/abs/https://doi.org/10.1137/0705041}
  {https://doi.org/10.1137/0705041} \BibitemShut {NoStop}%
\bibitem [{\citenamefont {Yoshida}(1990)}]{YOSHIDA1990262}%
  \BibitemOpen
  \bibfield  {author} {\bibinfo {author} {\bibfnamefont {H.}~\bibnamefont
  {Yoshida}},\ }\bibfield  {title} {\enquote {\bibinfo {title} {Construction of
  higher order symplectic integrators},}\ }\href {\doibase
  https://doi.org/10.1016/0375-9601(90)90092-3} {\bibfield  {journal} {\bibinfo
   {journal} {Physics Letters A}\ }\textbf {\bibinfo {volume} {150}},\ \bibinfo
  {pages} {262--268} (\bibinfo {year} {1990})}\BibitemShut {NoStop}%
\bibitem [{\citenamefont {Hairer}, \citenamefont {Wanner},\ and\ \citenamefont
  {Lubich}()}]{geometric_numerical_integration}%
  \BibitemOpen
  \bibfield  {author} {\bibinfo {author} {\bibfnamefont {E.}~\bibnamefont
  {Hairer}}, \bibinfo {author} {\bibfnamefont {G.}~\bibnamefont {Wanner}}, \
  and\ \bibinfo {author} {\bibfnamefont {C.}~\bibnamefont {Lubich}},\
  }\href@noop {} {\emph {\bibinfo {title} {{Geometric Numerical
  Integration}}}},\ \bibinfo {series} {Springer Series in Computational
  Mathematics}, Vol.~\bibinfo {volume} {31}\ (\bibinfo  {publisher} {Springer
  Berlin Heidelberg},\ \bibinfo {address} {Berlin, Heidelberg})\BibitemShut
  {NoStop}%
\bibitem [{\citenamefont {Anderson}(1965)}]{10.1145/321296.321305}%
  \BibitemOpen
  \bibfield  {author} {\bibinfo {author} {\bibfnamefont {D.~G.}\ \bibnamefont
  {Anderson}},\ }\bibfield  {title} {\enquote {\bibinfo {title} {Iterative
  procedures for nonlinear integral equations},}\ }\href {\doibase
  10.1145/321296.321305} {\bibfield  {journal} {\bibinfo  {journal} {J. ACM}\
  }\textbf {\bibinfo {volume} {12}},\ \bibinfo {pages} {547–560} (\bibinfo
  {year} {1965})}\BibitemShut {NoStop}%
\bibitem [{\citenamefont {Shadwick}, \citenamefont {Schroeder},\ and\
  \citenamefont {Esarey}(2009)}]{10.1063/1.3124185}%
  \BibitemOpen
  \bibfield  {author} {\bibinfo {author} {\bibfnamefont {B.~A.}\ \bibnamefont
  {Shadwick}}, \bibinfo {author} {\bibfnamefont {C.~B.}\ \bibnamefont
  {Schroeder}}, \ and\ \bibinfo {author} {\bibfnamefont {E.}~\bibnamefont
  {Esarey}},\ }\bibfield  {title} {\enquote {\bibinfo {title} {{Nonlinear laser
  energy depletion in laser-plasma acceleratorsa)}},}\ }\href {\doibase
  10.1063/1.3124185} {\bibfield  {journal} {\bibinfo  {journal} {Physics of
  Plasmas}\ }\textbf {\bibinfo {volume} {16}},\ \bibinfo {pages} {056704}
  (\bibinfo {year} {2009})},\ \Eprint
  {http://arxiv.org/abs/https://pubs.aip.org/aip/pop/article-pdf/doi/10.1063/1.3124185/14033624/056704\_1\_online.pdf}
  {https://pubs.aip.org/aip/pop/article-pdf/doi/10.1063/1.3124185/14033624/056704\_1\_online.pdf}
  \BibitemShut {NoStop}%
\bibitem [{\citenamefont {Gerritsma}(2011)}]{10.1007/978-3-642-15337-2_17}%
  \BibitemOpen
  \bibfield  {author} {\bibinfo {author} {\bibfnamefont {M.}~\bibnamefont
  {Gerritsma}},\ }\bibfield  {title} {\enquote {\bibinfo {title} {Edge
  functions for spectral element methods},}\ }in\ \href@noop {} {\emph
  {\bibinfo {booktitle} {Spectral and High Order Methods for Partial
  Differential Equations}}},\ \bibinfo {editor} {edited by\ \bibinfo {editor}
  {\bibfnamefont {J.~S.}\ \bibnamefont {Hesthaven}}\ and\ \bibinfo {editor}
  {\bibfnamefont {E.~M.}\ \bibnamefont {R{\o}nquist}}}\ (\bibinfo  {publisher}
  {Springer Berlin Heidelberg},\ \bibinfo {address} {Berlin, Heidelberg},\
  \bibinfo {year} {2011})\ pp.\ \bibinfo {pages} {199--207}\BibitemShut
  {NoStop}%
\end{thebibliography}%

\end{document}